\documentclass[11pt,a4paper]{article}
\usepackage[utf8]{inputenc}
\usepackage{hyperref}
\usepackage[english]{babel}
\usepackage[T1]{fontenc}
\usepackage{amsmath}
\usepackage{amsfonts}
\usepackage{amssymb}
\usepackage{caption}
\newcommand{\Tr}{\mathrm{Tr}}
\newcommand{\tr}{\mathrm{tr}}

\newcommand{\inv}{\mathrm{inv}}
\newcommand{\ad}{\mathrm{Ad}}
\newcommand{\rk}{\mathrm{rk}}
\newcommand{\sign}{\mathrm{Sign}}
\usepackage{array,multirow,makecell}
\setcellgapes{1pt}
\makegapedcells
\newcolumntype{R}[1]{>{\raggedleft\arraybackslash }b{#1}}
\newcolumntype{L}[1]{>{\raggedright\arraybackslash }b{#1}}
\newcolumntype{C}[1]{>{\centering\arraybackslash }b{#1}}
\newtheorem{theorem}{Theorem}
\newtheorem{definition}{Definition}
\newtheorem{proposition}{Proposition}
\newtheorem{lemma}{Lemma}
\newtheorem{corollary}{Corollary}
\usepackage{multicol}
\pdfoutput=1
\usepackage{graphicx}
\usepackage[left=2cm,right=2cm,top=2cm,bottom=2cm]{geometry}
\begin{document}

\begin{center}
\begin{Large}
\textbf{Renormalization of a tensorial field theory  \\ on the homogeneous space $SU(2)/U(1)$}\\
\end{Large}
\end{center}

\begin{center}
\vspace{20pt}
Vincent Lahoche \footnote{vincent.lahoche@th-psud.fr; LPT-UMR 8627, Universit\'e Paris 11, 91405 Orsay Cedex, France, EU.},
Daniele Oriti\footnote{daniele.oriti@aei.mpg.de, Max Planck Institute for Gravitational Physics, Albert Einstein Institute, Am M\"uhlenberg 1, 14476, Potsdam, Germany.}
\begin{abstract}
\textit{We study the renormalization of a general field theory on the homogeneous space $(SU(2)/$ $U(1))^{\times d}$ with tensorial interaction and gauge invariance under the diagonal action of $SU(2)$. We derive the power counting for arbitrary $d$. For the case $d=4$, we prove perturbative renormalizability to all orders via multi-scale analysis, study both the renormalized and effective perturbation series, and establish the asymptotic freedom of the model. We also outline a general power counting for the homogeneous space $(SO(D)/SO(D-1))^{\times d}$, of direct interest for quantum gravity models in arbitrary dimension, and point out the obstructions to the direct generalization of our results to these cases.}
\end{abstract}
\end{center}
%\bigskip
%\tableofcontents

%\newpage 
%\begin{multicols}{2}
\setlength{\columnseprule}{1pt}
\setlength{\columnsep}{30pt}
%\vspace{10pt}
\section{Introduction}
Group field theories \cite{GFT} (GFTs) are a candidate formalism for the fundamental degrees of freedom  of quantum spacetime (the \lq atoms of space\rq), and an approach to quantum gravity which merges insights and mathematical structures from  loop quantum gravity and spin foam models \cite{LQG}, simplicial quantum gravity and tensor models \cite{tensor}, which have all achieved remarkable progress in recent years.  

They share with loop quantum gravity the general structure of quantum states, associated to graphs labeled by group-theoretic data, and indeed can be seen as a 2nd quantized, Fock space-based reformulation of both its kinematics and its operator dynamics \cite{GFT-LQG}. And they encode and complete the covariant definition of the same quantum dynamics, formulated in terms of spin foam models, which in fact appear generically as GFT Feynman amplitudes \cite{mike-carlo}. The same amplitudes can be recast in the form of simplicial gravity path integrals \cite{GFT-noncomm, GFT-Immirzi}, clarifying their discrete geometric content, and substantiating further the analysis of the quantum geometry of loop quantum gravity states and spin foam amplitudes \cite{LQG-quantgeom}. At the same time, they are based on the same combinatorial structures (in their action, Feynman graphs and transition amplitudes) of tensor models, which they enrich by adding group-theoretic data. The hope is that this nice interplay between combinatorics and algebra, in a quantum field theory setting, will prove powerful enough to explain from first principles the emergence of spacetime and geometry from more fundamental entities, i.e. the dynamical realization in the full quantum theory of a regime where the fundamental degrees of freedom of the theory, which are generically not interpretable in geometric terms and at best can be associated to piecewise-flat geometries, can be approximated well by smooth manifolds and a smooth geometric field, governed by (a possibly modified form of) the equations of General Relativity  \cite{emergence}. 

Indeed, not only they merge the key elements  of these related approaches (and thus most results obtained in them), e.g. the quantum states and variables of loop quantum gravity, the amplitudes of spin foam models, the combinatorial structures of tensor models, but group field theories offer a promising mathematical context for tackling some of their outstanding open issues, thanks to QFT methods, most notably renormalization. In particular, they allow to identify stringent criteria for: constraining spin foam model building, controlling quantisation ambiguities in both spin foam and canonical formulations of loop quantum gravity, and ensuring consistency of the resulting quantum dynamics. These issues, in fact, translate into the problem of proving perturbative renormalizability of their GFT reformulation, since the GFT action encodes the choice of operator spin network dynamics and the GFT Feynman amplitudes coincide with spin foam models. The issue of controlling the sum over spin foam complexes, which completes the definition of spin foam models, and of defining the full quantum spin network dynamics, i.e. going beyond the perturbative treatment and summing all spin foam diagrams, encoded in a projection operator onto physical states or in their partition function, on the other hand, translates into the problem of making sense of the corresponding non-perturbative GFT dynamics and of unravelling the macroscopic phase diagram (and interesting phase transitions) of the theory. This is the problem of the continuum limit of the theory, which is in many ways {\it the} outstanding issue of the whole approach (alongside the physical issue of extracting the effective dynamics of the theory in the same continuum limit). Again, QFT tools become available thanks to the GFT reformulation, be it in the form of constructive renormalisation or of functional renormalisation group techniques. 

It is here that the input from tensor models has proven most relevant, in particular, the large body of recent results on colored tensor models \cite{jimmyrazvan}, where the use of colour labels on combinatorial structures ensures a greater control over their topology, and an analytic understanding of their scaling limits. In turn, this led to important results about the universality classes of tensor models, and to a precise suggestion for the class of allowed tensor interactions: those satisfying a \lq tensor invariance\rq criterion, which can be seen as the tensor analogue of the notion of locality in standard quantum field theory on flat spacetime. 

This becomes particularly relevant for group field theories and their renormalisation analysis. In fact, by treating GFT fields as quantum geometrically-enriched tensors, one has then a prescription for the relevant theory space that the renormalisation group flow should explore. This defines the class of GFTs known as {\it tensorial group field theories} (TGFTs), where most work on renormalisation has been carried out (after the very first step in this research direction \cite{laurent-razvan-daniele}).

Beside constructive analysis \cite{constructiveTGFT} and the first FRG studies \cite{TGFT-FRG}, most developments up to now concerned perturbative renormalizability of TGFT models. Such development can be seen as progressively approaching TGFT models for 4d quantum gravity, as developed in the spin foam context. The first models \cite{vincent-joseph, joseph} that have been shown to be renormalizable to all orders in perturbation theory were Abelian ones, with the TGFT field defined on several copies of a $U(1)$ group manifold (the number of copies matching the dimensionality of the cellular complex arising in their Feynman expansion). Next \cite{COR, fabien} came Abelian models incorporating a gauge invariance condition in their amplitudes, which turns them into lattice gauge theories and proper spin foam models, and gives the states of the theory the structure of spin networks. Then came the first proof of perturbative renormalizability at all orders of a non-abelian model, based on $SU(2)$, with the same gauge invariance \cite{COR2}. For many of these models, the renormalizability analysis was completed by the computation of the beta functions, with very interesting results on their asymptotic freedom (or safety) \cite{TGFT-beta}.\footnote{The focus of these renormalizability analysis has been to identify just-renormalizable models. Super-renormalizable models (and finite models) are equally well defined from the point of view of perturbative QFT, and thus would be equally good TGFT candidates for a fundamental formulation of quantum gravity. However, experience from standard QFT suggests that just-renormalizable models have a more interesting RG flow and a richer phase diagram, thus possibly a more interesting range of effective physics at different scales. Heuristically, one  imagine that just-renormalizable TGFT models, therefore, have a higher chance to reproduce at the effective level the rich continuum physics we expect from a theory of quantum gravity. Obviously, any heuristic motivation will have to be substantiated by explicit analysis.}

A bulk of solid work and understanding has therefore already accumulated. The stage is now set for tackling full-blown 4d quantum gravity models, as developed in the spin foam context (recall that the topological dimension of the simplicial structures generated by the TGFT perturbative expansion corresponds to the rank of the TGFT field, thus 4d gravity models require TGFT fields of rank $d=4$). Some results on radiative corrections in the simplicial setting (where more is known also in the 3d case \cite{josephBoulatov}) are available \cite{4dRadiative}, but we lack any systematic analysis, like the ones mentioned in the TGFT setting. Beside a better geometric understanding of the \lq tensor invariance\rq $\,$condition, this requires a generalisation to higher-dimensional non-abelian groups, i.e. $SO(4)$ or the even more interesting non-compact Lorentz group $SO(3,1)$, and, most important, the imposition of additional constraints on the amplitudes, the so-called \lq simplicity constraints\rq (see \cite{LQG, GFT-Immirzi} and references therein). 

The most developed strategy for model-building in 4d, in fact, is based on the so-called Plebanski-Holst formulation of classical General Relativity in the continuum. Here, the basic fields are a 2-form valued in the Lie algebra of the Lorentz group $SO(3,1)$ (or its euclidean counterpart $SO(4)$), usually indicated as $B$ and a 1-form connection field $A$ valued in the same algebra. These are the same field variables of a topological field theory of BF-type in 4d. The dynamics of the gravitational theory is defined by an action that add to the BF action a set of constraints on the $B$ field (dependent on an additional parameter called \lq Immirzi parameter\rq, which also plays a crucial role in canonical loop quantum gravity). These are called \lq simplicity constraints\rq. They have the effect of forcing the constrained 2-forms $B$ to be functions of a tetrad field. Inserting such solutions of the constraints back into the action one obtains a Palatini formulation of gravity in terms of the tetrad field and the connection $A$, plus a topological term (not affecting the classical theory) dependent on the Immirzi parameter. Current spin foam models, and the corresponding GFT models, follow a similar procedure at the quantum level, after the discrete counterpart of the simplicity constraints has been identified, so to be applied to the discrete variables that correspond to the $B$ field and are assigned to a simplicial complex. One then starts with a quantum formulation of topological BF theory and imposes a quantum counterpart of the simplicity constraints, to get a model for quantum 4d gravity. For more details, see \cite{LQG, GFT-Immirzi} and references therein. 

Depending on the exact model considered (i.e. the chosen way of imposing the simplicity constraints and the value of the so-called Immirzi parameter), these have the effect of reducing the initial domain of the GFT fields from the Lorentz group (or its euclidean counterpart $SO(4)$) to its homogeneous space $SO(3,1)/SO(3)$ (or $SO(4)/SO(3)\simeq \mathcal{S}_3$), or to another sub- manifold of the same group. This is the main physical reason why we are interested in (T)GFT models based on such domain manifolds. A renormalizability analysis of TGFT models of 4d quantum gravity requires therefore an extension of the known results and techniques from simple group manifolds to these more complicated domains, starting with homogeneous spaces. 

In the present paper, we perform one more step towards establishing the renormalizability of 4d quantum gravity TGFT models, by studying the renormalization of a TGFT model on the homogeneous space $\left(SU(2)/U(1)\right)^d$, endowed with the additional gauge invariance conditions characterising spin foam models. This is the simplest model that still combines the three technical challenges required for a full analysis of TGFT models for 4d gravity: restriction to a sub-manifold of the original group manifold, gauge invariance, and non-abelian character. The imposition of the constraints reducing the field variables to the homogeneous space is obtained in a covariant manner, using the formalism developed in \cite{GFT-Immirzi}. By rigorous multi-scale analysis, we prove renormalizability to all orders in perturbation theory of the model for $d=4$ (in $d=3$ our results imply super-renormalizability). For the same model, we also compute both the renormalised and effective perturbative series, analyse the 2-point and 4-point correlation functions, compute the beta function and establish asymptotic freedom at one-loop order.
Moreover, we generalise several of our results to arbitrary homogeneous spaces of the type $SO(D)/SO(D-1)\simeq \mathcal{S}_{D-1}$; in particular we establish a general Abelian power counting and classify such models in terms of their potential renormalizability, as seen from the Abelian power counting, for various choices of $D$ and $d$. However, we also discuss why this can be a misleading classification, since the exact power counting of other non-abelian models may deviate from the Abelian one, and what aspects of the analysis need to be carried out in detail for these cases in order to really prove (or disprove) their perturbative renormalizability. 

The model is defined in detail in Section 2. In Section 3 we provide an equivalent definition of the same model in terms of projections onto the homogeneous space, which is more elegant and lends itself immediately to the higher-dimensional generalisation. We then set-up the multi-scale analysis of the model, in Section 4, and obtain the Abelian power counting. The analysis of perturbative renormalizability of the model is performed in Section 5, while in Section 6 we go beyond this to study the full renormalisation flow of the model, computing also the renormalised and effective series. In Section 7, we report the study of the beta function at one-loop, and the proof of asymptotic freedom to the same order.

\section{Preliminary, Tensorial field theories on $\mathcal{S}_2^{\times d}$}

\subsection{Definition}
\noindent
We consider a tensorial quantum field theory on $d$ copies of the homogeneous space $SU(2)/U(1)$, which is isomorphic to the two dimensional sphere $\mathcal{S}_2$. The phase space of the theory is the cotangent bundle $\left( \mathcal{T}^*\mathcal{S}_2\right)^{\times d}\cong \left( \mathcal{S}_2\times \mathbb{R}^2\right)^{\times d}$. The complex field $\psi \in L_2(\mathcal{S}_2^{\times d})$, assumed to be square-integrable, is defined as 
\begin{align*}
\psi : \left[ SU(2)/U(1)\right]^d &\rightarrow \mathbb{C}\,, \\
(x_1,...,x_d)\in \left[ SU(2)/U(1)\right]^d &\rightarrow \psi (x_1,...,x_d)\,,\\
\int_{\mathcal{S}_2^{\times d}} \prod_{i=1}^d dx_i \bar{\psi}(x_1,...,x_d)&\psi(x_1,...,x_d)<\infty\,.
\end{align*}
The quantum dynamics is defined by the partition function
\begin{equation}
\mathcal{Z}=\int d\mu_{C_0}(\psi,\bar{\psi})e^{-S_{int}(\psi,\bar{\psi})},
\end{equation}
where the Gaussian measure with \textit{covariance} $C_0$, $d\mu_{C_0}(\psi,\bar{\psi})$ encode the kinetic part of the classical action, and define the free 2-point function, %to be detailed below.
and the interaction part $S_{int}$ of the action is constructed with all the \textit{trace invariant} contractions
\begin{equation}
S_{int}=\sum_{b}\lambda_{b} \Tr_{b}(\psi,\bar{\psi})\,.
\end{equation}
These traces are labeled by a $d$-colored bipartite regular graph (i.e. strictly $d$-valent, with links colored with $d$ colors at each node), called \textit{bubbles}, whose some examples are pictured on Figure \ref{fig1} below. Each black and white nodes correspond respectively to the fields $\psi$ and $\bar{\psi}$, the $d$ half-lines hooked to a black (resp. white) node picture the $d$ variable of the corresponding field $\psi$ (resp. $\bar{\psi}$), and the connectivity of the graph give the pattern of contraction between each fields. For instance, the bubble on Figure \ref{fig1}a corresponds to the following interaction:
\begin{equation}
\Tr_{b_{\mathrm{Fig} \ref{fig1}a}}(\psi,\bar{\psi})=\int \prod_{i=1}^3 dx_idx_i^{\prime} \psi(x_1,x_2,x_3)\bar{\psi}(x_1^{\prime},x_2,x_3)\psi(x_1^{\prime},x_2^{\prime},x_3^{\prime})\bar{\psi}(x_1,x_2^{\prime},x_3^{\prime})\,.
\end{equation}
This trace or \textit{tensorial invariant} provides a well characterization of the theory space. Moreover, the tensorial structure of the interaction allows to organize a power-counting, and as we will see in Section \ref{sectionren}, it provides a well definition of locality, essential for the definitions of counter-terms and renormalization. 
\begin{center}
\includegraphics[scale=1]{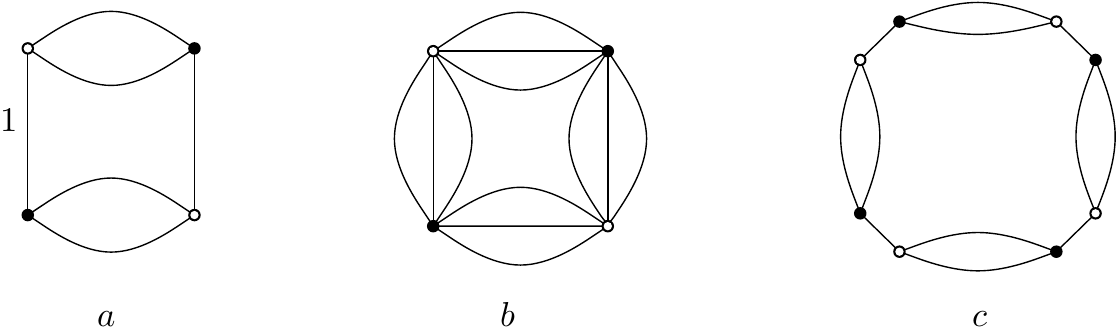} 
\captionof{figure}{Example of interaction bubbles of valence $4$, $4$ and $8$, and of rank $d=3, 6$ and $3$ respectively.}\label{fig1}
\end{center}
If the kinetic action is properly defined (i.e. if the propagator is invertible), the Gaussian measure can be defined by the choice of a kinetic action as follows:
\begin{equation}
d\mu_{C_0}(\psi,\bar{\psi}):=e^{-S_{kin}[\bar{\psi},\psi]}d\psi d\bar{\psi}\,,
\end{equation}
with
\begin{equation}
S_{kin}[\bar{\psi},\psi]:=\int_{\mathcal{S}_2^{\times d}}\prod_{i=1}^{d} \bigg[dx_i\sqrt{|g|}\bigg]\bar{\psi}({\vec{x}})\bigg(-\sum_{i=1}^d\Delta_i+m^2\bigg)\psi({\vec{x}})\,. \label{kineticaction}
\end{equation}
where $\Delta_i$ is the Laplacian operator on the 2-sphere of unit radius, $|g|$ the determinant of the metric in coordinates $\{x_i\,,i=1,2\}$, and $m^2$ be a real parameter playing the role of a mass term.
As an integrable function on $\mathcal{S}_2^{\times d}$, the field $\psi$ can be expanded on the spherical harmonics basis $\{Y_{l,m}(\theta,\phi)\}$, which is a complete basis of $L^2$-functions on the 2-sphere. In this basis, the propagator (or covariance) $C_0(\{\theta_i,\phi_i,\theta'_i,\phi'_i\})$, defined by the kinetic action \eqref{kineticaction} writes as:
\begin{align}
C_0(\{\theta_i,\phi_i,\theta'_i,\phi'_i\})= \int d\mu_{C_0} \bar{\psi}(\{\theta_i,\phi_i\})\psi(\{\theta'_i,\phi'_i\})= \sum_{\{l_i,m_i\}}C_{0\,\{l_i,m_i\}}\prod_{i=1}^d Y^*_{l_i,m_i}(\theta_i,\phi_i)Y_{l_i,m_i}(\theta'_i,\phi'_i),
\end{align}
where the coefficients
\begin{equation}
C_{0\,\{l_i,m_i\}}:=\dfrac{1}{\sum_i l_i(l_i+1)+m^2},
\end{equation}
do not depend on the magnetic indices $m_i$. This definition of the theory is in fact highly formal, because some divergences can occur in the perturbative expansion. In order to circumvent this difficulty, we introduce an ultra-violet cut-off $\Lambda$, and define the regularized propagator using \textit{Schwinger regularization}:
\begin{align}
C_{0\,\Lambda}(\{\theta_i,\phi_i,\theta'_i,\phi'_i\})=\int_{1/\Lambda^2}^{+\infty}d\alpha e^{-\alpha m^2}\times\sum_{\{l_i,m_i\}}\prod_{i=1}^d e^{-\alpha l_i(l_i+1)} Y^*_{l_i,m_i}(\theta_i,\phi_i)Y_{l_i,m_i}(\theta'_i,\phi'_i)\,.
\end{align}
Interestingly for the computation of Feynman amplitudes, this propagator involves the heat kernel
\begin{align}\label{heatkernel}
K_{\alpha}(\{\theta,\phi,\theta',\phi'\})=\sum_{\{l,m\}} e^{-\alpha l(l+1)} Y^*_{l,m}(\theta,\phi)Y_{l,m}(\theta',\phi')\,,
\end{align}
which verifies the heat equation,
\begin{equation}
\dfrac{\partial}{\partial \alpha} K_{\alpha}=\Delta K_{\alpha}\,,\label{heat1}
\end{equation}
and boundary conditions:
\begin{equation}
K_{\alpha=0}(\theta,\phi;\theta^{\prime},\phi^{\prime})=\delta(\cos \theta-\cos \theta^{\prime})\delta(\phi-\phi^{\prime})\,.
\end{equation}
Moreover, the heat kernel satisfies the composition law:
\begin{align}
\int \sin\theta d\theta d\phi &K_{\alpha_1}(\{\theta',\phi',\theta,\phi\})K_{\alpha_2}(\{\theta,\phi,\theta'',\phi''\})\label{keyprop}=K_{\alpha_1+\alpha_2}(\{\theta',\phi',\theta'',\phi''\}).
\end{align}
\begin{center}
\includegraphics[scale=1]{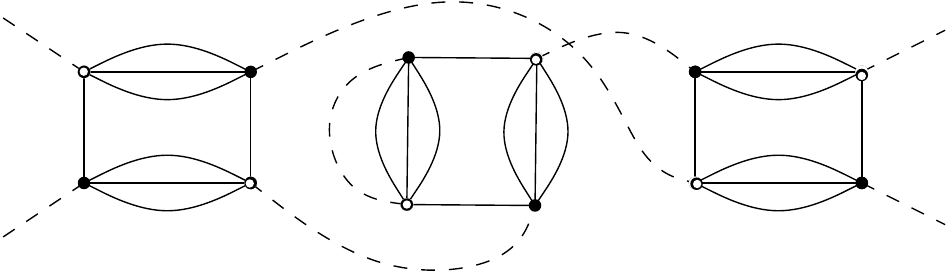} 
\captionof{figure}{Example of Feynman graph for a rank $4$ model with interactions of order $4$. The dotted lines represent Wick-contracted propagators.}\label{fig2}
\end{center}
This is in turn the key property to obtain the expression for the Feynman amplitudes entering in the perturbative expansion of the $N$-point correlation functions $S_N$. Such a function can be expanded as a sum indexed by Feynman graphs:
\begin{equation}
S_N=\sum_{G}\frac{1}{s(G)}\Big(\prod_{b\in G} -\lambda_b\Big)\mathcal{A}_{G}.
\end{equation}
where $s(G)$, the symmetry factor. Using \eqref{keyprop}, the amplitude $\mathcal{A}_{G}$ for the graph ${G}$ can be written as (see \cite{COR2}):
\begin{align}
&\mathcal{A}_{G}= {\left[ \prod_{e\in{L}(G)} \int_{1/\Lambda^2}^{\infty}{d\alpha_{e} e^{-\alpha_{e} m^{2}}}  \right]}\times\label{amplitude}\prod_{f\in F(G)}\sum_{l_{f}}(2l_f+1)e^{-\alpha(f) l_{f}(l_{f}+1)}  \\\nonumber
&\quad \times\Bigg( \prod_{f\in F_{ext}(G)} \sum_{l_f,m_f}e^{-\alpha(f) l_f(l_f+1)} Y^*_{l_{f},m_{f}}(\theta_{s(f)},\phi_{s(f)})Y_{l_f,m_f}(\theta_{t(f)},\phi_{t(f)})\Bigg),
\end{align}
where $L(G), F(G),$ and $F_{ext}(G)$ denote respectively the sets of lines, faces, and external faces of the graph $G$, and $s$ and $t$ map open faces to their boundary variables, and $\alpha(f):=\sum_{e\in\partial f}\alpha_e$, where $\partial f$ denote the set of boundary lines of $f$. Feynman graph can be depicted graphically as in Figure \ref{fig2}, with the rule that bubble vertices are depicted as in Figure \ref{fig1}, and Wick contractions with a dotted line between a black and a white vertex, both in the same bubble or not. 
\bigskip

\subsection{Closure constraint}

\noindent
For physical reasons, in which we will return in Appendix \ref{Sectiongem}, we will impose an additional condition on our field, which we call \textit{closure constraint}, and which make sense only for $d>1$. It can be understood as a gauge symmetry for the field, which reduces the manifold $\mathcal{S}_2^{\times d}$ as:
\begin{equation}
 [SU(2)/U(1)]^d \rightarrow [SU(2)/U(1)]^d/SU(2),
\end{equation}
identifying the field components up to a global $SU(2)$ group action. More precisely, if we denote the action of the group element $g\in SU(2)$ on the field $\psi$ as $\mathcal{\hat{R}}(g)\triangleright\psi$, where $\mathcal{\hat{R}}$ and $\triangleright$ are defined by the explicit group action of $SU(2)$ (see below), the closure constraint identifies, for a given $\psi$, all the elements $\mathcal{\hat{R}}(g)\triangleright\psi\, \forall g\in SU(2)$. \\

\noindent
Let us specify further the group action. We observe that the $2$-sphere admits a natural embedding in $\mathbb{R}^3$, and, using this, into $SO(3)$:
\begin{align}
\qquad \pi :\mathcal{S}_2 &\rightarrow SO(3)\\
(\theta,\phi)&\rightarrow \pi(\theta,\phi)\in SO(3)
\end{align}
with the following explicit expression in local coordinates $(\theta,\phi)$:
\begin{align}
\pi(\theta,\phi)[\hat{z}]=\vec{n}(\theta,\phi)
\end{align}
where:
\begin{align}
&\vec{n}:(\theta,\phi) \rightarrow \vec{n}(\theta,\phi):=(\sin{\theta}\cos{\phi},\sin{\theta}\sin{\phi},\cos{\theta})\in \mathbb{R}^3
\end{align}
Hence, $\pi(\theta,\phi)$ is the rotation of $SO(3)$ mapping the $\hat{z}$ axis in the direction $\vec{n}$ \footnote{This is a unique group element, up to an initial rotation around $\hat{z}$ and a final rotation around $\vec{n}$: $\pi(\theta,\phi)\sim \mathcal{R}_{\vec{n}}\pi(\theta,\phi)\mathcal{R}_{\hat{z}}$, where $ \mathcal{R}_{\vec{n}}$ denote an arbitrary rotation around the axis $\vec{n}$.}. Starting from our field on $\mathcal{S}_{2}^{\times d}$, this mapping enable us to define a new field on $SO(3)^{\times d}$, $\tilde{\psi}\in L_2(SO(3)^{\times d})$ such as $\mathbf{\pi}^*\tilde{\psi}\in L_2(\mathcal{S}_{2}^{\times d}):=\psi$, with the notation $\mathbf{\pi}^*$ defined as: $\mathbf{\pi}^*\tilde{\psi}(x_1,...,x_d):=\tilde{\psi}(\pi(x_1),...,\pi(x_d))$.\\

\noindent
For the field $\tilde{\psi}\in L_2(SO(3)^{\times d})$, there are a natural right action of the group $SO(3)$. Hence, we can define the gauge symmetry as the identification of all the fields up to a global right action of $SO(3)$. More concretely, we introduce the symmetric rotation $\hat{R}$ on $L_2(SO(3)^{\times d})$, such that $\hat{R}(g)\equiv\hat{R}(-g)\, \forall g\in SU(2)$ (it is more convenient to work with a compact simply connected group). Hence, the operator $\hat{R}$ can be understood as a function on $SO(3)\sim SU(2)/\mathbb{Z}_2$. Let $\mathcal{R}$ the map $\mathcal{R}:SU(2)\to SO(3)$, which identify $g$ and $-g$. We can therefore define the transformation law: 
\begin{align}
\hat{R}(g):L_2(SO(3)^{\times d})&\rightarrow L_2(SO(3)^{\times d})\\\nonumber
[\hat{R}(g)\tilde{\psi}](\{\pi(\theta_i,\phi_i)\})&:=\tilde{\psi}(\{\pi(\theta_i,\phi_i)\mathcal{R}(g)\}),
\end{align}
Now, we can clarify the definition of the action $\triangleright$ introduced before. More precisely, $\mathcal{\hat{R}}(g)$ acts on $\psi$ as:
\begin{align}
\mathcal{\hat{R}}(g)\triangleright\psi(\{\theta_i,\phi_i\})&:=\pi^*[\hat{R}(g)\tilde{\psi}](\{\theta_i,\phi_i\})\\\nonumber
&\,=\psi(\pi^{-1}[(\pi(\theta_i,\phi_i))\mathcal{R}(g)])\,.
\end{align}
\noindent
Then, the closure constraint, or gauge invariance, is the requirement :
\begin{equation}
\mathcal{\hat{R}}(g)\triangleright\psi=\psi\,, \qquad \forall g\in SU(2)\,.
\end{equation}
Let us consider the \textit{projector into gauge invariant fields} $\int dg \mathcal{\hat{R}}(g)$. We can think to implement the closure constraint by projection of the fields involved in the definition of the kinetic action $S_{kin}$. But in this way, we can not define easily an explicit propagator, because the kinetic kernel is not invertible on the space of fields. However, the Wick theorem states that the Gaussian measure, and with it, the perturbative expansion of the quantum theory, is well defined as long as the 2-point function at $\lambda_b=0$ (the propagator $C_{\Lambda}(\{\theta_i,\phi_i\},\{\theta'_i,\phi'_i\})$) is properly defined. As a result, we choose\footnote{Let us recall that, strictly speaking, the heat kernel is a sum of Markovian paths, and only depend on the geodesic distance between the end points on the considered manifold. As a result, it is consistent to make projection over a single field only.}: 
\begin{align}
&\int d\mu_{C}(\psi,\bar{\psi}){\psi}(\{\theta_i,\phi_i\})\bar{\psi}(\{\theta'_i,\phi'_i\})\label{propagator}=\int_{SU(2)} d{g}\int d\mu_{C_0}(\psi,\bar{\psi})\mathcal{\hat{R}}(g)\triangleright\psi(\{\theta_i,\phi_i\})\bar{\psi}(\{\theta'_i,\phi'_i\})\,,
\end{align}
or
\begin{align}
\int d\mu_{C}(\psi,\bar{\psi}){\psi}(\{\theta_i,\phi_i\})\bar{\psi}(\{\theta'_i,\phi'_i\})=\int_{SU(2)} dg \int_{1/\Lambda^2}^{\infty} d\alpha e^{-\alpha m^2}\prod_{i=1}^{d} K_{\alpha}(\{\pi^{-1}[(\pi(\theta_i,\phi_i))\mathcal{R}(g)];\theta'_i,\phi'_i\}) \,.
\end{align}
We can obtain an explicit expression for this constrained propagator. Using the expression for the heat kernel \eqref{heatkernel} and the decomposition, 
\begin{align*}
Y_{l,m}&((\pi(\theta_i,\phi_i))\mathcal{R}(g)\hat{z})=\sum_{m^{\prime}=-l}^{+l}D_{m^{\prime}m}^{(l)}[\mathcal{R}(g)^{-1}\pi(\theta_i,\phi_i)^{-1}]Y_{l,m^{\prime}}(\hat{z}),
\end{align*}
where $D^{(l)}$ is the well known Wigner matrix defined, in the usual Dirac notation for the canonical basis of angular momentum, as:
\begin{equation*}
D^{(l)}_{mm^{\prime}}[\mathcal{R}(g)]:=\langle m,l|\hat{\mathcal{R}}(g)|l,m^{\prime}\rangle\quad l\in\mathbb{N},
\end{equation*} 
we obtain, using the fact that $Y_{l,m}^{\,*}=\big[\frac{2l+1}{4\pi}\big]^{1/2} D^{(l)}_{m0}$, and that $Y_{l,m}(0,\phi)=\big[\frac{2l+1}{4\pi}\big]^{1/2}\delta_{m,0}$:
\begin{align}\label{propagatorgaugebis}
\int d\mu_{C}(\psi,\bar{\psi})\bar{\psi}(\{\theta_i,\phi_i\})\psi(\{\theta'_i,\phi'_i\})&=\int_{SU(2)} dg \int_{1/\Lambda^2}^{\infty} d\alpha e^{-\alpha m^2}\\\nonumber
&\times\sum_{\{l_i\}}\prod_{i=1}^d e^{-\alpha l_i(l_i+1)} \frac{2l+1}{4\pi}D_{00}^{(l_i)}[\mathcal{R}(g)\pi(\theta_i,\phi_i)^{-1}\pi(\theta'_i,\phi'_i)].
\end{align}
Note that the integral over the group of a product of such representation matrices defines a resolution of the identity in the space of \textit{intertwiners} of the group $SO(3)^{\otimes d}$ :
\begin{equation*}
\int dg \prod_{i} D_{m_i m^{\prime}_i}^{(l_i)}[\mathcal{R}(g)] \in \inv \big(SO(3)^{\otimes d}\big).
\end{equation*}
 
\medskip

\subsection{Regularized parametric representation of correlation functions}

We wish to obtain now the expression of the N-point correlation functions in perturbative expansion. The argument involving \eqref{keyprop} is still valid, and not affected by the closure constraint. Using the \textit{addition formula}:
\begin{equation}
\sum_{m=-l}^{+l}Y^*_{l,m}(\theta,\phi)Y_{l,m}(\theta^{\prime},\phi^{\prime})=\dfrac{2l+1}{4\pi} P_l(\vec{u}\cdot \vec{u}^{\prime}),
\end{equation}
where $\vec{u}$ (resp $\vec{u}^{\prime}$) is the unit vector pointing on the 2-sphere of radius unity in the direction $(\theta,\phi)$ (resp $(\theta^{\prime},\phi^{\prime})$), we deduce, using the explicit expression \eqref{propagatorgaugebis}, the expression for the Feynman amplitudes of the constrained theory:
%\begin{align}
%\sum_{m=-l}^{+l}Y^*_{l,m}&(\{\pi^{-1}[\mathcal{R}(g)(\pi(\theta,\phi))])Y_{l,m}%(\theta^{\prime},\phi^{\prime})\\\nonumber
%&=\sum_{m=-l}^{+l}Y^*_{l,m}(\theta,\phi)Y_{l,m}(\{\pi^{-1}[\mathcal{R}(g^{-1})%(\pi(\theta^{\prime},\phi^{\prime}))]).
%\end{align}
%\begin{align}
%&\mathcal{A}_{G}= {\Bigg[ \prod_{l\in{L}(G)} \int_{1/\Lambda^2}^{\infty}{d\alpha_{l} e^{-\alpha_{l} m^{2}}}  \int_{[SU(2)]^{|L(G)|}}\prod_{l\in L(G)}dh_l\Bigg]}\label{amplitudebis}\\\nonumber
%&\times\Bigg(\prod_{f\in F_{int}(G)} \int d^2\vec{u}_f K_{\alpha(f)}\big(\mathcal{R}\big[{\prod_{l\in \partial f}h^{\epsilon_{ef}}}\big](\vec{u}_f),\vec{u}_f\big)\Bigg) \\\nonumber
%&\times\quad \Bigg( \prod_{f\in F_{ext}(G)}K_{\alpha(f)}\big(\mathcal{R}\big[{\prod_{l\in \partial f}h^{\epsilon_{ef}}}\big](\vec{u}_{s{f}}),\vec{u}_{t(f)}\big)\Bigg).
%\end{align}
\noindent
%Using finally the explicit expression \ref{propagatorgaugebis}, we obtain :
\begin{align}
&\mathcal{A}_{G}= {\Bigg[ \prod_{e\in{L}(G)} \int_{1/\Lambda^2}^{\infty}{d\alpha_{e} e^{-\alpha_{e} m^{2}}}  \int_{[SU(2)]^{|L(G)|}}\prod_{e\in L(G)}dh_e\Bigg]}\label{amplitudebisbis}\\\nonumber
&\times\Bigg(\prod_{f\in F(G)} \sum_{l_f}(2l_f+1)D^{(l_f)}_{00}\big[\mathcal{R}\big({\vec{\Pi}_{e\in \partial f}h^{\epsilon_{ef}}}\big)\big]e^{-\alpha(f)l_f(l_f+1)}\Bigg) \\\nonumber
&\times\Bigg( \prod_{f\in F_{ext}(G)}\sum_{l_f}e^{-\alpha(f) l_f(l_f+1)}\dfrac{2l_f+1}{4\pi}D^{(l_f)}_{00}\Big(\mathcal{R}\big({\bar{\Pi}_{e\in \partial f}h^{\epsilon_{ef}}}\big)\pi(\theta_{s(f)},\phi_{s(f)})^{-1}\pi(\theta_{t(f)},\phi_{t(f)})\Big)\Bigg),
\end{align}
where the notation $|Q|$ means the cardinality of the sets $Q$, and $\epsilon_{ef}$ is the incidence matrix, which contains the information on whether a line belongs to the boundary of a face and their relative orientation: $\epsilon_{ef}=0$ if $e\notin\partial f$, $+1$ if $e\in\partial f$ and they have the same orientations, $-1$ otherwise.

%\begin{align}
%\mathcal{A}_{G}= {\Bigg[ \prod_{l\in{L}(G)} \int_{1/\Lambda^2}^{\infty}{d\alpha_{l} e^{-\alpha_{l} m^{2}}}  \int_{[SU(2)]^{|L(G)|}}\prod_{l\in L(G)}dh_l\Bigg]}\label{amplitudebisbis}\\\nonumber
%&\times\Bigg(\prod_{f\in F_{int}(G)} K_{\alpha(f)}\big(\mathcal{R}_{\prod_{l\in \partial f}h^{\epsilon_{ef}}}(\vec{e},\vec{e}\big)\Bigg) \\\nonumber
%&\times\quad \Bigg( \prod_{f\in F_{int}(G)}K_{\alpha(f)}\big(\mathcal{R}_{\prod_{l\in \partial f}h^{\epsilon_{ef}}}(\vec{u}_{s{f}},\vec{u}_{t(f)}\big)\Bigg),
%\end{align}
%where $\vec{e}$ is an arbitrary unit vector on the 2-sphere. 
%\noindent
%where $\chi^{(l)}:=\Tr D^{(l)}$ is the characteristic function in the representation $l$. 

\bigskip
\section{Covariant formulation with projections}

The problem of the above formulation, defined directly on the homogeneous space, is that we lose the explicit group structure of the group field theory. This leads to some practical difficulties in dealing with the theory, in particular in studying the divergence structure of its Feynman amplitudes and its renormalisability, following what has been done in previous works. These difficulties are mainly due to the fact that elements of the homogeneous space do not compose via multiplication to other elements of the homogeneous space. 
The way to proceed is to recast the field theory as a field theory on (several copies of) $SU(2)$, but with the fields subject to constraints effectively projecting them to the homogeneous space. This way one can perform all calculations using the standard SU(2) formalism. This is indeed well-known and already used in the GFT formulation of constrained spin foam models for $4d$ quantum gravity, in particular the BC model\cite{BCrevisited,DP-F-K-R,P-R,gluing}. An ensuing subtle point is that special care should be payed to the compatibility between the constraints projecting the field onto the homogeneous space and the gauge invariance  condition to be satisfied by the same fields. More precisely, the constraints have to be imposed {\it covariantly} with respect to the diagonal group action. This was also realised in the context of GFTs and spin foam models for 4d quantum gravity \cite{alexandrov,BCrevisited,GFT-Immirzi}, and a properly covariant construction was identified, which we now describe in some detail.
\\

\subsection{Constrained representation}
We\label{3.1} choose an element of the Lie algebra $\mathfrak{su}(2)$, $\sigma_z$ for instance, and note that the set of group elements $g\in SU(2)$, such as $g\sigma_zg^{-1}=\sigma_z$, the stabilizer group of $\sigma_z$, is isomorphic to the group $U(1)$:
\begin{equation}
U(1)_{\sigma_z}:=\{g=e^{i\theta \sigma_z}\,\,\forall \theta \in [0,2\pi)\,\} \sim U(1).
\end{equation}
\noindent
Now we can simply define a field theory for a new field $\Psi:SU(2)^{\times d}\rightarrow \mathbb{C}$ with the constraint :
\begin{align}
\Psi(g_1,...,g_d)\label{const1}=\Psi(g_1h_1,...,g_dh_d)\quad \forall (h_1,...,h_d)\in U(1)_{\sigma_z}^{\times d}.
\end{align}
\noindent
For this new field we define the partition function:
\begin{equation}
\mathcal{Z}=\int d\mu_{\tilde{C}_{\Lambda}}(\Psi,\bar{\Psi})e^{-S_{int}(\Psi,\bar{\Psi})},
\end{equation}
where, as in the previous construction, $S_{int}$ is a sum of tensorial invariants. The only difference between the two formulations is that in the first one, the fields have $2d$ local coordinates, while the new field has $3d$ local coordinates, with a constraint which reduces the number of degrees of freedom from $3d$ to $3d-d=2d$, so that we left in the end with the same degrees of freedom. \\

\noindent
The covariance $\tilde{C}_{\Lambda}$ for this model is defined as:
\begin{align}
&\int  d\mu_{\tilde{C}_{\Lambda}}\Psi(\{g_i\}) \bar{\Psi}(\{g_i'\})\label{prop2}:=\int_{U(1)_{\sigma_z}^{\times d}}\prod_{i=1}^ddh_i\int_{1/\Lambda^2}^{+\infty}d\alpha e^{-\alpha m^2}\prod_{i=1}^d K_{\alpha}(g_i h_i g_i^{\prime-1}).
\end{align}
from which we deduce the Feynman expansion of a $N$-point function $S_N$, indexed by graphs $G$:
\begin{align}
\mathcal{A}_{G}&= {\left[ \prod_{e\in {L}(G)} \int_{1/\Lambda^2}^{\infty}{d\alpha_{e} e^{-\alpha_{e} m^{2}}} \prod_{i=1}^ddh_{i_e} \right]}\label{amplitude}\\\nonumber
&\times{\left( \prod_{f\in F(G)} K_{\alpha_{(f)}} {\left( \vec{\prod}_{e\in\partial{f}}h_{i(f)_e}^{\epsilon_{ef}} \right)} \right)} \times{\left( \prod_{f\in F_{ext}(G)} K_{\alpha_{(f)}} {\left( g_{s(f)}\vec{\prod}_{e\in\partial{f}}h_{i(f)_e}^{\epsilon_{ef}}g^{-1}_{t(f)} \right)} \right)},
\end{align}
where $i(f)$ is the color of the face $f$ and $K_{\alpha}$ is the solution of the heat equation on $SU(2)$ (given by the same formula \eqref{heat1} with $\Delta$ replaced by the Laplace operator on $SU(2)$). 

We now confirm briefly the equivalence of the two constructions at the dynamical level. 
This can be seen immediately noting that the spherical harmonics are just the Wigner representation matrices for $SU(2)$ integrated over a one-dimensional subgroup isomorphic to $U(1)$. Indeed, the heat kernel is a class function on $SU(2)$ and, by virtue of the Peter-Weyl theorem, it can be expanded on the (class invariant) basis of characters as:
\begin{equation}
K_{\alpha}(g_1g_2^{-1}):=\sum_{j\in\mathbb{N}/2} (2j+1)e^{-4\alpha j(j+1)}\chi^{j}(g_1g_2^{-1}),
\end{equation}
where the characters $\chi^{j}:=\Tr_jD^{(j)}$ of the irreducible representation $j$, verify :
\begin{equation}
\Delta_{SU(2)} \chi^{j}(g)=-4j(j+1)\chi^{j}(g).
\end{equation}
Now, in the Euler angles parametrization
\begin{align*}
\chi^{j}(ge^{i\sigma_z\theta})&=\sum_m D^{(j)}_{mm}(ge^{i\sigma_z\theta})=\sum_m\langle m,j|e^{i\gamma J_z}e^{i\beta J_y}e^{i(\alpha+\theta)J_z}|j,m\rangle,
\end{align*}
and:
\begin{align*}
\int_{0}^{2\pi}\dfrac{d\theta}{2\pi} \chi^{j}(ge^{i\sigma_z\theta}) = \langle 0,j|e^{i\beta J_y}|j,0\rangle=D^{(j)}_{00}(g).
\end{align*}
Note that because $m=0$, $j$ is necessarily an integer. When applying the previous result to $\int d\theta \chi(g_1e^{i\theta \sigma_z}g_2^{-1})$, we find:
\begin{align*}
\int_{0}^{2\pi}&\dfrac{d\theta}{2\pi} \chi^{j}(g_1e^{i\sigma_z\theta}g_2^{-1})=D^{(j)}_{00}(g_2^{-1}g_1)=\sum_m D^{(j)}_{0m}(g_2^{-1})D^{(j)}_{m0}(g_1)=\sum_m D^{(j)*}_{m0}(g_2)D^{(j)}_{m0}(g_1).
\end{align*}
Hence, because of the relation : $Y_{l,m}^{\,*}=\big[\frac{2l+1}{4\pi}\big]^{1/2} D^{(l)}_{m0}$,  the equivalence between the two representations (up to a change of normalization of $\alpha$ and $m$ : $\alpha\to\alpha/4$, $m\to2m$) follows easily.\\

\medskip

We now move on to the imposition of the gauge invariance (closure) constraint in a covariant way. The aim is to combine the constraint \eqref{const1} with a global constraint of the form $\psi(g_1,...g_d)=\psi(g_1l,...,g_dl)\,\forall l\in SU(2)$. We first define the two transformations
\begin{align}\label{inv1}
\hat{T}_l:\Psi(g_1,...,g_d)&\mapsto \Psi(g_1l,...,g_dl)\\\label{inv2}
\hat{t}_{h_i}^{(i)}:\Psi(g_1,...,g_d)&\mapsto\Psi(g_1,...,g_ih_i,...,g_d),
\end{align}
satisfying : 
\begin{equation}\label{noncom}
\hat{T}_l\circ \hat{t}_{h_i}^{(i)}=\hat{t}_{l^{-1}h_il}^{(i)}\circ \hat{T}_l
\end{equation}
Hence, by defining
\begin{equation}
\Psi_{\sigma_z}(g_1,...,g_d):=\int_{U(1)_{\sigma_z}^{\times d}}\prod_{i=1}^d dh_i \hat{t}_{h_i}^{(i)}[\Psi](g_1,...,g_d),
\end{equation}
we have:
\begin{equation}
\hat{T}_l[\Psi_{\sigma_z}](g_1,...,g_d)=\Psi_{l^{-1}\sigma_zl}(g_1l,...,g_dl),
\end{equation}
with, for any $k\in \mathfrak{su}(2)$:
\begin{equation}\label{def1}
\Psi_{k}(g_1,...,g_d):=\int_{U(1)_{k}^{\times d}}\prod_{i=1}^d dh_i \hat{t}_{h_i}^{(i)}[\Psi](g_1,...,g_d).
\end{equation}
As a result, to include the closure constraint, we recast the theory in terms of the fields $\phi_k(\{g_i\}),\bar{\phi}_k(\{g_i\}):[SU(2)]^d\times \mathfrak{su}(2)\to\mathbb{C}$, with the gauge invariance condition:
\begin{equation}
\phi_k(g_1,...,g_d)=\phi_{h^{-1}kh}(g_1h,...,g_dh).
\end{equation}
Note that, because of the invariance of the Haar measure, this definition implies that the field $\psi:=\int dk \phi_k$ satisfies the standard closure constraint : $\hat{T}_h[\psi](g_1,...,g_d)=\psi(g_1,...,g_d)\forall h\in SU(2)$. In terms of this new fields $\phi_k$ and $\bar{\phi}_k$, our covariant quantum field theory is defined by the partition function:
\begin{equation}\label{partnew}
\mathcal{Z}=\int d\mu_{C}(\phi,\bar{\phi})e^{-S_{int}(\phi,\bar{\phi})},
\end{equation}
where, following \cite{GFT-Immirzi} the interaction is chosen of the form:
\begin{equation}
S_{int}(\phi,\bar{\phi})=\sum_{b}\lambda_{b}\Tr_{b}\left(\hat{P}[\phi],\hat{P}[\bar{\phi}]\right)\label{actionint},
\end{equation}
where $\hat{P}$ denotes the projector into the subset of gauge invariant fields:
\begin{equation}
\hat{P}:=\int_{\mathcal{S}_2} dk\int_{SU(2)}dl\int_{U(1)_{k}^{\times d}}\prod_{i=1}^d dh_i \hat{t}_{h_i}^{(i)}\circ\hat{T}_l.
\end{equation}
This choice allows to choose a Gaussian measure without gauge projection, and we adopt the following definition for $d\mu_{C}(\phi,\bar{\phi})$:
\begin{align}
\int d\mu_{{C}}&(\phi,\bar{\phi})\phi_k(\{g_i\})\bar{\phi}_{k^{\prime}}(\{g'_i\})\label{prop3}:=\delta_{k,k^{\prime}}\int_{1/\Lambda^2}^{+\infty}d\alpha e^{-\alpha m^2}\prod_{i=1}^d K_{\alpha}(g_ig_i^{\prime-1})\quad.
\end{align}
Our model is then completely defined, and divergence free due to the cut-off over $\alpha$ integration. Note that in the perturbative expansion, due to the Wick contractions, we can think in terms of the effective field $(\Psi,\bar{\Psi}):=(\hat{P}[\phi],\hat{P}[\bar{\phi}])$. These effective fields satisfy the closure constraints, and are associated to the effective propagator $\bar{C}$:
\begin{align}
\nonumber\int d\mu_{\bar{C}}(\Psi,\bar{\Psi})\Psi(\{g_i\})\bar{\Psi}(\{g'_i\})&=\int_{SU(2)}dl\int dk\int_{U(1)_{k}^{\times d}}\prod_{i=1}^ddh_i\int_{1/\Lambda^2}^{+\infty}d\alpha e^{-\alpha m^2}\prod_{i=1}^d K_{\alpha}(g_ilh_ig_i^{\prime-1}).
\end{align}
From the Wick theorem, using this effective propagator, we find the Feynman amplitude $\mathcal{A}_G$, whose boundary variable are projected into the gauge invariant subspace $\ker[\hat{P}-\mathbb{I}]$:
\begin{align}\label{amplitudeconst}
\mathcal{A}_G = &\Bigg[ \prod_{e \in L(G)}\int d\alpha_e e^{-\alpha_e m^2}\int dl_{e} \int dk_e \prod_{i=1}^dD^{k_e}h_{i,e}\Bigg]\\\nonumber
&\times\Bigg(\prod_{f\in F(G)}K_{\alpha(f)}\bigg(\prod_{e\in \partial f}(l_eh_{i(f),e})^{\epsilon_{ef}}\bigg)\Bigg) \\\nonumber
&\times\Bigg(\prod_{f\in F_{ext}(G)}K_{\alpha(f)}\bigg(\prod_{e\in \partial f}g_{s(f)}(l_eh_{i(f),e})^{\epsilon_{ef}}g^{-1}_{t(f)}\bigg)\Bigg)
\end{align}
where $s$ and $t$ map open faces to their boundary variables, $\epsilon$ is the incidence matrix, $i(f)$ is the ``color" of the face $f$, and 
\begin{equation}
D^{k}h_{i}:=dh_i \delta(k-h_ik(h_i)^{-1})
\end{equation}
which reduces the integration from $SU(2)^{\times d}$ to $U(1)_k^{\times d}$. Because of the integration over $k_e$, we deduce the following proposition:

\begin{proposition}\label{prop1}:
The amplitude $\mathcal{A}_G$ for a connected graph $G$ has a $SU(2)^{\times|V(G)|}$ gauge symmetry, which allows to fix variables along a spanning tree $\mathcal{T}\subset G$, such as $l_e =\mathbb{I}\,\,\forall e\in L(\mathcal{T})$.
\end{proposition}
\textit{\textbf{Proof}}: The expression \eqref{amplitudeconst} is invariant under the transformation:
\begin{equation}
l_e\rightarrow g_{t(e)}l_eg_{s(e)}^{-1}\,,\qquad k_e\rightarrow  g_{s(e)} k_e g_{s(e)}^{-1}\,,
\end{equation}
where $t(e)$ and $s(e)$ are the target and source vertex of an oriented edge $e$, and with the additional rule that one of the two group elements is the identity for open lines. Because of this invariance, $|V(G)|$ gauge variables can be freely fixed. Because there are only $|V(G)|-1$ lines in a spanning tree, the proposition is proved.
\begin{flushright}
$\square$
\end{flushright}
Moreover note that the gauge invariance at each (black or white) node allows to choose all the $k_e$ in the same direction, say $Oz$, up to a global translation for boundary variables (i.e. the variables attached to the external lines). Then, is this gauge, the integration over $k_e$ drops out of the integral. Following \cite{GFT-Immirzi}, we call \textit{time gauge} this gauge fixing.\\

\noindent
This formulation is more convenient for the study of the renormalizability of the model, and it also lends itself more easily to generalisation to other homogeneous spaces $SO(D)/SO(D-1)\simeq \mathcal{S}_{D-1}$, making clearer the role of the group manifold dimension parametrized by $D$. \\

In Appendix \ref{Sectiongem}, we give some details on the geometrical interpretation of this construction using \textit{group Fourier transform}, which also motivates its interest from a quantum gravity perspective.

\subsection{Generalization : Constrained field theory over $[SO(D)]^{\times d}$}
\label{sectiongen} 
In this paper we focus on the field theory on $\mathcal{S}_2^d$, and on its renormalisation. However, most of our construction as well as part of the renormalizability analysis,  can easily be extended to the homogeneous space $[SO(D)/SO(D-1)]^{\times d}$, using the projector formulation introduced above. In this section, we reframe the essential results obtained in the previous section for the homogeneous space $[SO(D)/SO(D-1)]^{\times d}$. The extension is straightforward, therefore we give only the essential steps, without too many details. Note that the motivation to extend the analysis to this case from the quantum gravity perspective, is that this is the basis for model building of $d$-dimensional euclidean quantum gravity models in the spin foam and discrete gravity context, via a generalised Barrett-Crane construction \cite{gluing, FK, DP-F-K-R}. \\

\noindent
Let $\{(L_{\mu\nu})_{\rho\sigma}\}$, a basis of anti-symmetric $D\times D$ matrices of the Lie Algebra $\mathfrak{so}(D)$, and $k=\{k_{\mu}\}$ an unit vector of $\mathbb{R}^D$ (the Greek indices run over $1,...,D$ and label the Euclidean coordinates on $\mathbb{R}^D$). Any element $g\in SO(D)$ can be written as (we use the Einstein convention for sums over Greek indices):
\begin{equation}
g=e^{\Omega_{\mu\nu}L_{\mu\nu}},
\end{equation}
and any element $h$ of the stabilizer group of $k$, isomorphic to $SO(D-1)$ (denoted $SO_k(D-1)$), can be written as:
\begin{equation}
h=e^{\Omega_{\mu\nu}\mathcal{P}_{\mu\mu'}^k\mathcal{P}_{\nu\nu'}^kL_{\mu'\nu'}} \in SO_k(D-1)
\end{equation}
where $\mathcal{P}^k=\mathbb{I}-k\otimes k$, $(\mathcal{P}^{k})^2=\mathcal{P}^k$ is the projector onto the subspace orthogonal to $k$. As for the $SU(2)$ case, we define a field theory on $SO(D)^{\times d}$ as a map $\Psi:SO(D)^{\times d} \to \mathbb{C}$, and we reduce the manifold to the homogeneous space $[SO(D)/SO(D-1)]^{\times d}$ imposing the constraint
\begin{equation}\label{extendconst}
\Psi(g_1,...,g_d)=\Psi(g_1h_1,...,g_dh_d)\quad \forall (h_1,...,h_d)\in [SO_k(D-1)]^{\times d} \quad .
\end{equation}
The corresponding quantum theory is defined by the choice of a partition function, or in other worlds, by the choice of an action $S_{kin}$ and of a (UV regularized) Gaussian measure $d\mu_{\bar{C}_{\Lambda}}$. As before, the action is a sum of tensorial invariants, built again as in correspondence with colored bipartite graphs (bubbles).  And the closure constraint have to be implemented in a covariant way. To this end, we define the operators $\hat{T}_l$ and $\hat{t}^{(i)}_{h_i}$
\begin{equation}
\hat{T}_l:\Psi(g_1,...,g_d)\to \Psi(g_1l,...,g_dl)
\end{equation}
\begin{equation}
\hat{t}_{h_i}^{(i)}:\Psi(g_1,...,g_d)\to \Psi(g_1,...,g_d) \quad ,
\end{equation}
satisfying again : $\hat{T}_l\circ \hat{t}_{h_i}^{(i)}=\hat{t}_{l^{-1}h_il}^{(i)}\circ \hat{T}_l$, and implying that the field
\begin{equation}
\Psi_k(g_1,...,g_d)=\int_{SO_k(D-1)^d}\prod_{i=1}^ddh_i\hat{t}_{h_i}^{(i)}[\Psi](g_1,...,g_d) \quad ,
\end{equation}
verifies:
\begin{equation}
\hat{T}_l[\Psi_k](g_1,...,g_d)=\Psi_{\mathcal{R}_{l}^{-1}[k]}(g_1,...,g_d) \quad ,
\end{equation}
where $\mathcal{R}_{l}^{-1}[k]$ is the vector $k$ rotated by $l\in SO(D)$. At this stage, all the definitions following \eqref{partnew} can be applied formally without change. We define the partition function as
\begin{equation}\label{partnew2}
\mathcal{Z}=\int d\mu_{C}(\phi,\bar{\phi})e^{-S_{int}(\phi,\bar{\phi})} \quad ,
\end{equation}
with the action
\begin{equation}
S_{int}(\phi,\bar{\phi})=\sum_{b}\lambda_{b}\Tr_{b}\left(\hat{P}[\phi],\hat{P}[\bar{\phi}]\right)\label{actionint2},
\end{equation}
and the gauge invariant propagator for the effective field $\Psi:= \hat{P}[\phi]$:
\begin{align}
\nonumber\int d\mu_{{C}}(\Psi,\bar{\Psi})\Psi(\{g_i\})\bar{\Psi}(\{g'_i\})&=\int_{SO(D)}dl\int dk\int_{SO_{k}(D-1)^{\times d}}\prod_{i=1}^ddh_i\int_{1/\Lambda^2}^{+\infty}d\alpha e^{-\alpha m^2}\prod_{i=1}^d K_{\alpha}(g_ilh_ig_i^{\prime-1}) \quad .
\end{align}
The Feynman amplitudes for the corresponding field theory take then the form \eqref{amplitudeconst}. Note that for the action \eqref{actionint2}, we have adopted the definition of \cite{GFT-Immirzi}.

\section{Abelian power counting}

In this section we explore the power counting for the divergences of the theory, in order to find renormalizability criteria that would allow to identify the renormalizable interactions. For the first time we focus on the $SU(2)/U(1)$ $SO(D)/SO(D-1)$ case, but we aim at extending the results to $SO(D)/SO(D-1)$. We begin by studying the divergences in the Abelian approximation, expected to be optimal from the results obtained recently in \cite{COR2}. We will give some additional arguments in favor of this intuition in Section \ref{sectionOB}, and we will see that the Abelian power counting becomes exact, for the $SU(2)/U(1)$ model, in the next section. We also point out why the same arguments do not generalize trivially to arbitrary dimension $D$, and what needs to be understood in order to achieve such a generalization.

\subsection{Multiscale expansion}

We move on to a systematic analysis provided by the \textit{multi-scale expansion} \cite{COR2}. It attributes a scale to each line $e \in \mathcal{L} ({G}) $ of any amplitude
of any Feynman graph ${G}$, and allows to deduce power-counting in a more systematic and rigorous way. Moreover, it renormalizes any graph in a sequence of successive steps, providing a concrete implementation of Wilson's ideas directly at the graphical level. Note that, for the rest of this section, we set $\mathbf{G}=SU(2)$.\\

\noindent
For convenience, we choose the UV-regulator $\Lambda$ so that $\Lambda=M^{-2\rho}$, and the complete effective propagator\footnote{We leave the ``tilde" out for this Section.} $C_{\Lambda}\equiv C^{\rho}$ is sliced according to
\begin{equation}
C_{\Lambda}=\sum_{i<\rho} C_i\,,
\end{equation}
where the cut-off $\Lambda$ is chosen of the form $\Lambda = M^{\rho}$, $M>1$, and the effective propagator ``in the slice i" $C_i$ is
\begin{align}
C_i &= \int dk\int_{SU(2)} dh\int_{[SU(2)]^d} \prod_{j=1}^{d} dl_j \delta(k-l_jk(l_j)^{-1})\int_{M^{-i}}^{M^{-(i-1)}} d\alpha e^{-\alpha m^2}\prod_{i=1}^{d} K_{\alpha}(g_ih l_i(g'_i)^{-1}).
\end{align}
Let us start by establishing general power counting via a multi-scale analysis, following the notations and general strategy of \cite{Rivasseau:1991ub}. \\

\noindent
The amplitude of a graph $ \mathcal{G} $, $ \mathcal {A} (\mathcal {G}) $, with fixed external momenta, is thus divided into the sum of all the scale attributions
$ \mu = \{i_{e}, e \in \mathcal{L} (  \mathcal {G} ) \} $, where $i_e$ is the scale of the momentum $p$ of line $e$:
\begin{equation}
\mathcal{A}_G=\sum_{\mu}\mathcal{A}_{G,\mu}.
\end{equation}
We will prove the following key theorem, which gives the power counting of the theory and a divergence criterion for a graph amplitude, and is the first step of the perturbative renormalizability analysis at all orders. 
\begin{theorem}
Let a Feynman graph $G$ with sliced amplitude $\mathcal{A}_{G,\mu}$ and scale assignment $\mu=\{i_{l_1},...,i_{l_{|L(G)|}}\}\, l_i\in L(G)$. This amplitude satisfies the uniform bound:
\begin{equation}
|\mathcal{A}_{G,\mu}|\leq K^{|L(G)|}\prod_i\prod_{k=1}^{\rho(i)}M^{\omega(G_i^\rho)},
\end{equation}
where $G_i^\rho$ is the $\rho$-th connected component of the sub-graph $G_i \subset G$, which contains only the lines of the graph $G$ with a slice $i_l\geq i$, and where the divergence degree $\omega(G_i^\rho)$ is given by:
\begin{equation}
\omega(G_i^\rho)=-2|L(G_i^\rho)|+2(|F(G_i^\rho)|-R(G_i^\rho)).
\end{equation}
\end{theorem}
\noindent
\textit{\textbf{Proof}}:\\
\noindent
The first step is to bound the heat kernel. The heat kernel $K_{\alpha}(g)$ on $SU(2)$ has a complicated expression (see for example \cite{camporesi}). However, it can be approximated in the UV regime, i.e. for large representation labels, by the following uniform bound:
\begin{align}\label{boundprop}
C_i(\{g_j\},\{g^{\prime}_j\}) \leq K M^{(3d-2)i}\int dk\int_{SU(2)} dh\int_{[U(1)_k]^d} \prod_{j=1}^{d} dl_j e^{-\delta M^i\sum_{j=1}^{d}|g_jhl_jg^{\prime\,-1}_j|}.
\end{align}
where here $|g_1g_2^{-1}|$ indicates the geodesic distance (using the standard metric on $SU(2)\simeq \mathcal{S}_3$) between the two group elements $g_1$ and $g_2$, and $\delta$, $K$ are two positive constants which can be precisely computed (the values of these constants do not affect the proof).\\

This result allows us to bound the (multi-)scale decomposition $\mathcal{A}_{G, \mu}$ of the amplitude. The first step is to rewrite in a suitable manner the term $\prod_{l\in L(G)}M^{(3d-2)i_l}$. To this end, note that, trivially: $M^i=\prod_i M$. This allows to rewrite the product over the lines of the graph so that $\prod_{l\in L(G)}M^{(3d-2)i_l}=\prod_{l\in L(G)}\prod_{i=1}^{i_l}M^{(3d-2)}$. Now, we wish to invert the order of the double product. Selecting a scale-assignment $i$, and a subset of lines in $G$ so that, for each of these lines, the scale assignment is higher than or equal to $i$,  we define the subgraph $G_i$ of $G$. It follows that
\begin{align*}
\prod_{l\in L(G)}M^{(3d-2)i_l}=\prod_{l\in L(G)}\prod_{i=1}^{i_l}M^{(3d-2)}=\prod_i\prod_{l\in L(G_i)}M^{(3d-2)} \quad .
\end{align*}
Because the graph ${G}_i$ is not necessarily connected, we introduce the notation ${G}_i^{\rho}$ for its connected components, so that ${G}_i=\cup_{\rho=1}^{\rho(i)}{G}_i^{\rho}$. It follows that the previous decomposition becomes
\begin{align*}
\prod_{l\in L(G)}M^{(3d-2)i_l}=\prod_i\prod_{l\in L(\cup_{\rho=1}^{k(i)}G_i^\rho)}M^{(3d-2)}=\prod_i\prod_{\rho=1}^{\rho(i)}\prod_{l\in L(G_i^\rho)}M^{(3d-2)}=\prod_i\prod_{\rho=1}^{\rho(i)}M^{(3d-2)L(G_i^\rho)} \quad .  
\end{align*}
The second step is to bound the contributions of the internal faces. Using the same trick to reorganize the products and the compactness of the group $U(1)$, we obtain the following contribution for the internal faces:
\begin{equation}
\prod_i\prod_{\rho=1}^{\rho(i)}M^{-3d|L(G_i^\rho)|+3|F(G_i^\rho)|} \quad .
\end{equation}
Note that, to obtain this formula, we have chosen an optimal tree in each face on which we perform the integrations over the angle variables. This result, combined with the first one provided by the factors $M^{(3d-2)i}$ gives
\begin{equation}\label{firstbound}
\prod_i\prod_{\rho=1}^{\rho(i)}M^{-2|L(G_i^\rho)|+3|F(G_i^\rho)|} \quad .
\end{equation}
The third and last contribution comes from the remaining integrals
\begin{equation}\label{gaugeremain}
\int\prod_e dh_e dk_e\prod_{f}dl_{k_e}^{c(f)} e^{-\delta M^{i(f)}|\prod_{e\in\partial f}(h_el_{k_e}^{c(f)})^{\epsilon_{ef}}|} \quad ,
\end{equation}
where $i(f):=\inf{\{i_l,\,l\in \partial f\}}$ and $c(f)$ is the color of the face $f$. It is at this point that the Abelian approximation intervenes. As shown in \cite{COR2} for a non-Abelian TGFT on $SU(2)$, the exact power counting is uniformly bounded by its Abelian version, which corresponds to the linearized version of the exact one around identity for all the group elements (i.e. the non-commutativity of the group variables improves the convergence of a graph amplitude compared to the Abelian version, see Section \ref{sectionOB}). \\

\noindent
The Abelian version \ref{gaugeremain} is
%\begin{align*}
%C_i^{AV}(\{\vec{q}_j\},\{\vec{q}^{\prime}_j\}) &\leq K M^{(3d-2)i}\int_{\mathcal{S}_2} d\vec{e}_k\int_{U(1)^3} d\vec{\lambda}\\\nonumber
%&\int_{0}^{2\pi} \prod_{j=1}^{d} d\theta_j^{\vec{e}_k} e^{-\delta M^i\sum_{j=1}^{d}|\vec{q}_j-\vec{q}^{\prime}_j+\vec{\lambda}+\theta_j^{\vec{e}_k}\vec{e}_{k}|},
%\end{align*}
\begin{equation}
\int \prod_{e\in L(G)} d\vec{\lambda}_e dk_e \int\prod_{f}d\theta_{k_e}^{c(f)} e^{-\delta M^{i(f)}|\sum_e\epsilon_{ef}(\vec{\lambda}_e+\theta_{k_e}^{c(f)}\vec{e}_{k_e})|}\label{gauge}.
\end{equation}
Where $\vec{e}_k$ is the unit $3d$ vector associated to the unit Lie algebra element $k$ and $|\vec{q}|:=\sqrt{\sum_{j=1}^d q_j^2}$ is the $\mathbb{R}^3$ norm (we use here explicitly the trivial isomorphism between the elements of the Lie algebra $\mathfrak{su}(2)$, and the vectors of $\mathbb{R}^3$). %In particular, the $3d$ norm corresponds, up to a normalization, to the norm given by the Killing 2-form on the Lie group algebra).\\

\noindent
%Using the measure invariance, we can choose for each faces the value of the component of $\vec{q}$ along $\vec{e}_k$ to be zero. This result, 
%The measure translating invariance allows to choose one face for each line so that the associated $\theta$ variable compensate exactly one component of $\vec{\lambda}$. \\
Integrating over a selected tree $\mathcal{T}_2$ of faces, such that the number of faces in this set equals the rank of the incidence matrix, and in an optimal way, in the sense that the faces of this set proceed recursively from the leaves to the root of the Gallavotti-Nicol\'o tree, the integral \eqref{gauge} over the $\vec{\lambda}$ and $\theta$ variables gives the power counting contribution
\begin{equation}\label{gaugeint}
\prod_i\prod_{\rho=1}^{\rho(i)}M^{-3R(G_i^\rho)},
\end{equation}
and the remaining integration
\begin{align}
&\int \prod_{e\in L(G)} dk_e \int\prod_{f\in F/\mathcal{T}_2}d\theta_{k_e}^{c(f)} e^{-\delta M^{i(f)}|\sum_{e\in\partial f}\epsilon_{ef}\theta_{k_e}^{c(f)}|}=\int\prod_{f\in F/\mathcal{T}_2}\int \prod_{e\in \partial f} d\theta_{k_e}^{c(f)}dk_ee^{-\delta M^{i(f)}|\sum_{e\in\partial f}\epsilon_{ef}\theta_{k_e}^{c(f)}|},
\end{align}
gives, up to a positive constant,
\begin{align}\label{gaugetheta}
\prod_{f\in F/\mathcal{T}_2}M^{-i(f)}&=\prod_i\prod_\rho\prod_{f\in F/\mathcal{T}_2(G_i^\rho)}M^{-1}=\prod_{i,\rho}M^{-|F(G_i^\rho)|+R(G_i^\rho)} \quad ,
\end{align}
from which we deduce  the bound on the amplitude $A_{G, \mu}$:
\begin{align}
|\mathcal{A}_{G,\mu}| &\leq K^{|L(G)|}\prod_i\prod_{\rho=1}^{\rho(i)}M^{-2|L(G_i^\rho)|+2|F(G_i^\rho)|-2R(G_i^\rho)}
= K^{|L(G)|}\prod_i\prod_{\rho=1}^{\rho(i)}M^{\omega(G_i^\rho)}.
\end{align}
\begin{flushright}
$\square$
\end{flushright}
Note that for the same model without simplicity constraint, the power counting involves a $3$ and not a $2$ in front of the contribution $F-R$ (see \cite{COR2}). It seems that it is just the dimension of the manifolds, $3$ for $SU(2)$ and $2$ for $SU(2)/U(1)$ which differs at this point. The next Section provides a confirmation of this intuition. 

\subsection{Abelian power counting for a constrained models on $[SO(D)]^{\times d}$}

It is not hard to extend the previous analysis to the homogeneous space $[SO(D)/SO(D-1)]^d \simeq \mathcal{S}_{D-1}$, allowing to obtain a preliminary classification of potentially just-renormalizable models, for various choices of $D$ and $d$. Of course, such classification is valid only to the extent in which the Abelian power counting captures in fact the exact power counting of these non-Abelian models. This is however not straightforward, and we have actually reasons not to believe it, as we are going to discuss in the following. \\

\noindent
We obtain the following result:
\begin{theorem}
The Abelian superficial divergence degree of any Feynman graph $G$ associated to a field theory on $[SO(D)/SO(D-1)]^{\times\,d}$ with closure constraint is given by
\begin{equation}
\omega(G) = -2|L(G)|+(D-1)\big[|F(G)|-R(G)\big]\label{dideg} \quad .
\end{equation}
\end{theorem}
The proof is the exact generalization of the previous one, and we will only give the main steps. The previous bound \eqref{boundprop} for the propagator becomes, on $SO(D)$:
\begin{align}\label{boundprop2}
C_i(\{g_j\},\{g^{\prime}_j\}) \leq K M^{(d\frac{D(D-1)}{2}-2)i}\int dk\int_{SU(2)} dh\int_{[SO_k(D-1)]^d} \prod_{j=1}^{d} dl_j e^{-\delta M^i\sum_{j=1}^{d}|g_jhl_jg^{\prime\,-1}_j|} \quad .
\end{align}
After integration over group variables $g_i$, the product \eqref{firstbound} becomes
\begin{equation}\label{firstbound2}
\prod_i\prod_{\rho=1}^{\rho(i)}M^{-2|L(G_i^\rho)|+\frac{D(D-1)}{2}|F(G_i^\rho)|}
\end{equation}
and the remaining integration \eqref{gauge}, in the Abelian approximation, which corresponds to the linearized version of \eqref{gaugeremain}, becomes
\begin{align}\label{gauge2}
\int \prod_{e\in L(G)} d{\lambda}_e dk_e \times\int\prod_{f}d\theta_{k_e}^{c(f)} e^{-\delta M^{i(f)}\big|\sum_e\epsilon_{ef}\big({\lambda}_{e,\mu\nu}+\theta_{k_e,\mu'\nu'}^{c(f)}\mathcal{P}_{\mu'\mu}^{k_e}\mathcal{P}_{\nu'\nu}^{k_e}\big)L_{\mu\nu}\big|} \quad .
\end{align}
The integration over the $\lambda_e$ variables replaces \eqref{gaugeint} by:
\begin{equation}\label{gaugeint2}
\prod_i\prod_{\rho=1}^{\rho(i)}M^{-\frac{D(D-1)}{2}R(G_i^\rho)}
\end{equation}
and the remaining integration over the $\theta_{k_e}^{c(f)}$ gives, instead of \eqref{gaugetheta},
\begin{equation}\label{gaugetheta2}
\prod_{i,\rho}M^{-\frac{(D-1)(D-2)}{2}\big[|F(G_i^\rho)|-R(G_i^\rho)\big]} \qquad .
\end{equation}
Combining the results \eqref{firstbound2}, \eqref{gauge2} and \eqref{gaugetheta2}, we obtain the divergence degree \eqref{dideg}. 

\subsection{Discussion: Optimal bound and Abelian power counting}
\label{sectionOB} 
In this section we want to examine further the validity of the Abelian power counting for our non-Abelian model. To this end, we will study the  behaviour of the integral \eqref{gaugeremain}\footnote{Our analysis is close to the one of \cite{matteovalentin} for the $SU(2)$ case.}. In order to simplify the reasoning, we choose the orientations of faces and lines such that $\epsilon_{ef}\geq 0$. A moment of reflection shows that it is always possible to do so: one just has to exploit the bipartite structure of the Feynman graphs, and choose the orientation of lines from black to white vertices, for instance. Hence, we will study the behaviour in $\Lambda$ of the simpler integral:\begin{equation}
\mathcal{I}_{\Lambda}=\int\prod_e dh_e dk_e\prod_{f}dl_{k_e}^{c(f)} e^{-\Lambda^2\big|\prod_{e\in\partial f}h_el_{k_e}^{c(f)}\big|^2},
\end{equation}
in the large $\Lambda$ limit. Because of the normalization of each integration measure, $\mathcal{I}$ goes to zero when $\Lambda \to \infty$, and we expect a behavior of the type $\Lambda^{-\Omega(G)}$. The aim is therefore to find $\Omega$, or, at least, an optimal bound for it. In addition, note that the integral is absolutely convergent, and trivially bounded by a constant. \\

\noindent
The large $\Lambda$ limit enforces the relations:
\begin{equation}\label{vaccum}
\prod_{e\in \partial f}h_el_{k_e}^{c(f)}=\mathbb{I},
\end{equation}
and the strategy is to expand the exponent in the vicinity of these solutions, and integrating around them, by the Laplace method. Let $x=\{\bar{h}_{e},\bar{l}_e^{c(f)}\}$ a point in the space of solutions of \eqref{vaccum}, expected to be a manifold with {\it a priori} many connected parts, eventually of null dimension (a single point). We define $\mathcal{A}=\{h_e\}$ the set of group variables attached to each line and $H_f$ the map from $SO(D)^{\times |L|}$ to $SO(D)^{\times |F|}$ defined by :
\begin{equation}
H_f(\mathcal{A})=\prod_{e\in \partial f}h_el_{k_e}^{c(f)}
\end{equation}
whose differential around $x$ is:
\begin{equation}
dH_{f}(x)=\sum_{e\in\partial f}\ad_{\big\{\prod_{e'\in\partial f|e'<e}\bar{h}_{e'}\bar{l}_{e'}^{c(f)}\big\}}\big[\hat{\delta}_e+\bar{h}_e \hat{\delta}_{e}^{f}\bar{h}_e^{-1}\big]=:\sum_{e\in \partial f} L_{fe}\big[\hat{\delta}_e+\bar{h}_e \hat{\delta}_{e}^{f}\bar{h}_e^{-1}\big],
\end{equation}
where $\hat{\delta}_e$ and $\hat{\delta}_e^{f}$, living in the Lie Algebra $\mathfrak{su}(2)$, are the right variations of $h_e$ and $l_e^{c(f)}$ respectively. Note that, we fix the gauge by setting $h_e=\mathbb{I}$ along a spanning tree before linearizing, otherwise, the gauge orbit becomes non-compact and leads to spurious divergences. Defining $\hat{\delta}_e^{f\,\sharp}=\bar{h}_e \hat{\delta}_{e}^{f}\bar{h}_e^{-1}$, we obtain, around $x$:
\begin{equation}\label{sumx}
\mathcal{I}_{\Lambda}(x)=\int\prod_{e,f}d\hat{\delta}_ed\hat{\delta}_e^{f}\prod_f e^{-\Lambda^2|\sum_{e\in\partial f}L_{fe}(\hat{\delta}_e+\hat{\delta}_e^{f\,\sharp})|^2},
\end{equation}
where only the lines in the complementary of the gauge tree are selected. In order to integrate it, we introduce the quantities $\hat{\delta}_f$ and $\hat{\delta}^{\sharp}_f$ as:
\begin{equation}
\hat{\delta}_f=\sum_{e\in\partial f}L_{fe}\hat{\delta}_e \qquad \qquad \hat{\delta}^{\sharp}_f=\sum_{e\in\partial f}L_{fe}\delta^{f\,\sharp}_e \quad ,
\end{equation}
and the notations $\hat{\delta}_{f\,\|}$ and $\hat{\delta}_{f\,\bot}$, designating respectively the components parallel and orthogonal to $\hat{\delta}^{\sharp}_f$. Inserting this in \eqref{sumx}, we find
\begin{equation}
\mathcal{I}_{\Lambda}(x)=\int\prod_{e,f}d\hat{\delta}_ee^{-\Lambda^2|\hat{\delta}_{f\,\bot}|^2} \int d\hat{\delta}_e^{f}\prod_f e^{-\Lambda^2|\hat{\delta}_{f\,\|}+\hat{\delta}^{\sharp}_f|^2} \quad ,
\end{equation}
which behaves as
\begin{align}
\mathcal{I}_{\Lambda}(x)&\lesssim \big|\det[L]_{\ker(L)_\bot}\big|^{-1}\Lambda^{-\Big\{\dim\big[SO(D)\big]-\dim\big[SO(D-1)\big]\Big\}\rk[L]}\Lambda^{-\dim\big[SO(D-1)\big]|F|}\\\nonumber
&=\big|\det[L]_{\ker(L)_\bot}\big|^{-1}\Lambda^{-(D-1)\rk[L]}\Lambda^{-\frac{(D-1)(D-2)}{2}|F|} \quad ,
\end{align}
where $\rk[L]$ is the rank of $L$, and the notation $\det[L]_{\ker(L)_\bot}$ indicates the determinant over the complementary space $\ker(L)_{\bot}$ of $\ker(L)$. Because the rank $\rk[L]$ is at least equal to the rank of $\epsilon_{ef}$, the previous bound in always bounded by its Abelian version. The sum over $x$, however can eventually spoil this result. As explained before, the support of this sum splits into continuous and discrete components, and the integral over the continuous component can be ill-defined. However, these singularities occur when the determinant vanishes, and because the integral is absolutely convergent, it is a snag of the Laplace method. Moreover, for these points, the co-dimension of the kernel of $L$ becomes bigger than the co-dimension of the kernel on the other points. Hence, presumably these singularities do not affect the conclusion. \\

This result is important for the rest of this paper, because it allows to find some just-renormalizable models only from the Abelian divergent degree. However, it seems to indicate that the abelian power counting is a pessimistic one, and as a result, that the list of just-renormalizable models obtained using the Abelian divergent degree is certainly incomplete. In addition, the flatness condition \eqref{vaccum} is different of the one obtained in standard TGFT, which is $\prod_{e\in\partial f}h_e=\mathbb{I}$. We will return on this subtlety in the following. 

\section{Abelian classification of just-renormalisable models}\label{ren}

This section is devoted to a detailed analysis of the divergence degree given by \eqref{dideg}. The aim is to determine for which values of $d$ and $D$, and for which value of the maximal degree $v_{max}$ of interactions, the theory is just-renormalizable (obviously, a stronger degree of convergence would indicate super-renormalizability). Recently, an analysis of this type has been made in \cite{COR2}  for TGFT with gauge invariance (but no other constraints) on group manifolds, for which a classification table has been obtained. We make here the same work for our models on the homogeneous space $[SO(D)/SO(D-1)]^{\times d}$. This work can also be taken as preliminary step towards a similar analysis for TGFT models for quantum gravity, obtained by constraining models of quantum BF theory, the additional constraints there having a similar effect as the projection to a homogeneous space (and in the Barrett-Crane-type models being exactly such projections on $SO(4)/SO(3)$). However, we emphasize in advance that the difficult issue in applying this classification to $SO(D)/SO(D-1)$ models lies in showing that the exact power counting is well captured by the Abelian one. We will return to this point in the following. \\

\subsection{Basics on colored graphs}

This section give some definitions and properties of colored graphs. Most of these properties are well-known in tensor model literature, so we simply adopt them and refer to, say, \cite{jimmyrazvan} for their proof. From \cite{COR2}, we adopt the following definitions.
\begin{definition}\label{def1}
(contraction operation). Let $G$ be a Feynman graph and $L_0=\{l_i\}\subset L(G)$ an ordered subset of dotted (i.e. propagation) lines in $G$ (including tadpole lines). The graph $G/L_0$ is obtain from $G$ by the following steps:\\

considering the dotted line $l_i\in L_0$:\\

\noindent
$\bullet$ deleting the line $l_i$ and its two (black and white) end vertices and all the colored lines joining these two vertice;\\
$\bullet$ identifying the colored line linked to the deleted black vertex with the corresponding line linked to the white vertex; \\
$\bullet$ repeating the same steps for $l_{i+1}$, and so on. 
\end{definition}
\begin{definition}
For a connected graph $G$ with $|L|$ lines, $|V|$ vertices and a spanning tree $\mathcal{T}\subset G$, we call \textit{tensorial rosette} or simply \textit{rosette}, the contracted graph $G/\mathcal{T}$ with one vertex and $|L|-|V|+1$ lines.
\end{definition}
The figure \ref{fig3} below illustrates the definition 1 in a simple example. 
\begin{center}
\includegraphics[scale=0.9]{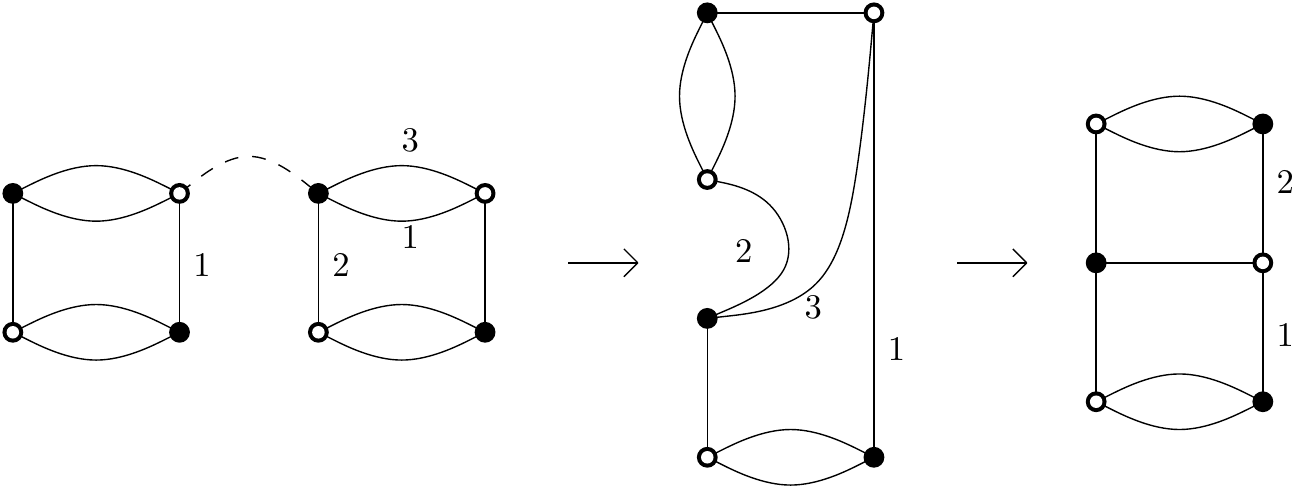} \\
\captionof{figure}{Contraction of a dotted line between two vertices}\label{fig3}
\end{center}
From this definition, we have the following lemma: 
\begin{lemma}\label{lemma1}
Let a connected graph $G$, with $|F|$ and $R$ respectively its number of faces and the rank of the incidence matrix $\epsilon_{ef}$. Under contraction of a spanning tree $\mathcal{T}$, $|F|$ and $R$ do not change.  
\end{lemma}
\textit{\textbf{Proof}}: Because $\mathcal{T}$ is a spanning tree, its lines bound faces with a number of boundary lines bigger or equal to two, so their number does not change under contraction. In the same way, because of the gauge invariance of the Feynman amplitude, allowing to fix the gauge such that $h_e=\mathbb{I}$ along an arbitrary spanning tree, we deduce that the rank does not change. 
\begin{flushright}
$\square$
\end{flushright}
\begin{definition}
Consider a Feynman graph $G$. The colored extension $G_c$ of this graph is the bipartite regular graph for which:\\ 

\noindent
$\bullet$ the vertices are partitioned in the form $\mathcal{V}(G_c)=V\cup\bar{V}$, where $V$ (respectively $\bar{V}$) is the set of black (respectively white) vertices;\\
$\bullet$ the set of lines $\mathcal{E}(G_c)$ is formed by all the lines (colored plus dotted) joining any pair $\{v,\bar{v}\}\in V\times\bar{V}$; by definition, the dotted lines have color $0$;\\
$\bullet$ the set of faces is of the form $\mathcal{F}(G_c)=F\cup F_c^{\neq 0}$, where $F$ is the set of faces in $G$, i.e. the set of faces of the form $f_{0i}$ with boundary lines of color $0$ and $i$ ($i \neq 0$), and $F_c^{\neq 0}$ is the set of faces of the form $f_{ij}$ with boundary lines of color $i$ and $j$($i\neq j; i,j\neq 0$); 
\end{definition}
\begin{definition}
Consider a colored extension $G_c$. A $k$-dipole $d_k$ is a set of $k$ colored lines necessarily including the color $0$ and linking two vertices $v$ and $\bar{v}$. An example is depicted on the Figure \ref{fig4} below. 
\end{definition}
\begin{center}
\includegraphics[scale=0.9]{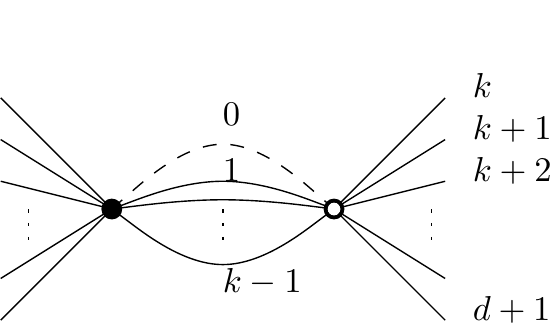} 
\captionof{figure}{Example of $k$-dipole}\label{fig4}
\end{center}
In addition, we recall the following three definitions about colored graphs:
\begin{definition}
\textbf{(jacket)} Consider a colored extension ${G}_{c}$ in dimension $d$. A \textit{jacket} $\mathcal{J}$ is a 2-subcomplex of ${G}_{c}$, labeled by a $(d+1)$-cycle $\tau$, such that $\mathcal{J}$ has the same number of lines and vertices as ${G}_c$, but only a subset of its faces: $\mathcal{F}_{\mathcal{J}}=\left\lbrace f\in \mathcal{F}_{G_{c}}|f=({\tau}^q(0),{\tau}^{q+1}(0)), q\in \mathbb{Z}_{D+1}\right\rbrace $.
\end{definition}
A jacket is a ribbon graph, corresponding to a sub-manifold of dimension 2 and of Euler-Poincaré characteristic  given by $\chi(\mathcal{J})=|\mathcal{F}_{\mathcal{J}}|-|\mathcal{E}_{\mathcal{J}}|+|\mathcal{V}_{\mathcal{J}}|=2-2g_{\mathcal{J}}$, where $g_{\mathcal{J}}$ is the genus of the surface. 

\begin{definition}
\textbf{(Degree)} The degree $\varpi({G}_{c})$ of a colored extension $G_c$ is the sum over all the genus of its jackets:
\begin{equation*}
\varpi({G}_{c})=\sum_{\mathcal{J}}g_{\mathcal{J}} \quad \Rightarrow \quad \varpi({G}_{c})\geq 0
\end{equation*}
\end{definition}
\begin{definition}\label{defmelons}
The graphs whose degree is equal to zero are called {\it melonic} graphs.
\end{definition}
In addition to these definitions, we have the three following lemmas:
\begin{lemma}
The melonic graphs are dual to a $d$-dimensional sphere. 
\end{lemma}
\begin{lemma}\label{propdeg}
In dimension $d$, the degree $\varpi({G}_{c})$ is related to the number of bi-colored faces and to the number of black (or white) vertices $p$, by the following two relations:
\begin{equation*}
|\mathcal{F}|=\dfrac{d(d-1)}{2}p+d-\dfrac{2}{(d-1)!}\varpi({G}_{c})
\end{equation*}
\begin{equation*}
\varpi({G}_{c})=\dfrac{(d-1)!}{2}(p+d-\mathcal{B}^{[d]})+\sum_{i;\rho}\varpi(\mathcal{B}^{\hat{i}}_{(\rho)}).
\end{equation*}
In addition, we can show that $p+d-\mathcal{B}^{[d]} \geq 0$.
\end{lemma}
Note that, in this lemma, the sum over $i$ in the second relation includes the color $0$. In addition, $\mathcal{B}^{\hat{i}}_{(\rho)}$ is the connected component $\rho$ of the sub-graph obtained from $G_c$ by deleting all the lines with color $i$ (including the color $0$). $\mathcal{B}^{[d]}$ is the number of these connected sub-graphs. These sub-graphs are the so-called $d$-bubbles. From this lemma, we easily deduce the following proposition:

\begin{proposition}
Under any $1$-dipole contraction, the degree of a graph is unchanged.
\end{proposition}
\noindent
With this material at hand, we now move on to the renormalizability analysis, which is the object of the next section. 
\bigskip
\subsection{Power counting Renormalizability}
We have seen that, for these models, the divergence degree of a graph grows with the number of faces.
The first question is: which are the graphs that have a maximum number of faces? To answer this question, consider a vacuum graph $G$ and its colored extension $G_c$. We can choose a tree $\mathcal{T}$ in $G$ and build the rosette $G/\mathcal{T}=\hat{G}$, $\hat{G}_c$ being its colored extension. Then, from the lemma \ref{propdeg}, we have:
\begin{align}\label{boundfaces}
|F(\hat{G})|=(d-1)|L(\hat{G})|+1-\Delta(\hat{G}) \quad ,
\end{align}
where we have used the fact that $L(\hat{G})=p$ in the lemma \ref{propdeg}, and where
\begin{equation*}
\Delta(\hat{G}):=\dfrac{2}{(d-2)!}\Big[\dfrac{1}{d-1}\varpi(\hat{G})-\varpi(\hat{G}^{0})\Big]\quad ,
\end{equation*}
where $\hat{G}^{0}$ is the $d$-bubble of color $0$ obtained from $\hat{G}$ by deleting all the lines of color $0$. Note that because the rosette $\hat{G}$ have only one vertex, $\hat{G}^{0}$ have only one connected component. Because one can prove that $\varpi(\hat{G})\geq d \varpi(\hat{G}^{0})$, we easily deduce that $|F|$ is bounded by:
\begin{equation*}
(d-1)p+1-\frac{2}{(d-1)!}\varpi(\hat{G})\leq |F| \leq (d-1)p+1-\frac{2}{(d-1)!}\varpi(\hat{G}^{0}).
\end{equation*}
The number of faces is then maximal when $\varpi(\hat{G})=0$, implying  $\varpi(\hat{G}^{0})=0$ from the lemma \ref{propdeg}. Hence we deduce that the number of faces is maximal for the melonic (colored extension) graphs. This result is actually a key one for all the TGFT models that have been studied to date. In addition to the formal Definition \ref{defmelons}, the melonic graphs have an iterative definition. From the simplest melon with $p=1$, the so-called \textit{supermelon}, pictured in Figure \ref{supermelon} below,
\begin{center}
\includegraphics[scale=1]{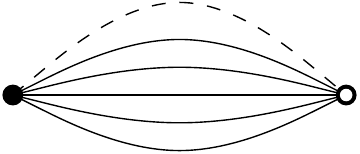} 
\captionof{figure}{The supermelon graph}\label{supermelon}
\end{center}
\noindent
we obtain the refined melons of order $p$ by replacing an edge by a $d$-dipole as in the Figure \ref{insert} below. 
\begin{center}
\includegraphics[scale=1]{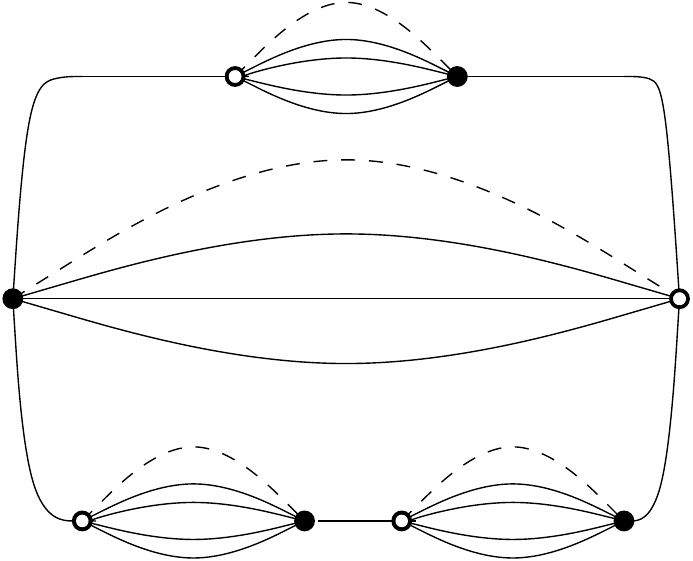} 
\captionof{figure}{Melonic recursion: $d$-dipole insertion.}\label{insert}
\end{center}
Note that for a rosette, this recursion procedure excludes the line of color $0$ because the $d$-bubble $\hat{G}^{0}$ has just one connected component.\\
Now we can turn to the analysis of the rank, the other main contribution to the divergence degree. It is obvious from the previous recursion that for a melonic rosette of order $p$, the rank is just equal to $p$, the number of lines ${L}(\hat{G})$ in $\hat{G}$. Hence, the rank is maximal for the same melonic graphs, and each insertion of a $d$-dipole increases $|F|-R$ by $d-2$. It follows that, for a melonic rosette graph:
\begin{equation}\label{boundmelon}
|F(\hat{G})|-R(\hat{G})=(d-2)|L(\hat{G})|+1.
\end{equation}
It is also the optimal bound for $|F|-R$ for arbitrary graphs. A statement that we can prove easily by recursion. Starting from the order $p=1$, the unique connected vacuum graph $\mathcal{M}_1$ is the supermelon in Figure \ref{supermelon}. For the next order $p=2$, we wish to add one black vertex and one white vertex, or, in other words, a new dotted line. Each line carries at least $d$ faces $f^{0i}\,i\neq 0$ of length one, and can increase the rank at least of $+1$. Because of the connectivity constraint, it seems that one colored line must be sacrificed, and bound a common face for the two dotted lines. Hence, the maximal number of faces is $2d-1=2(d-1)+1$, in accordance with the formula (\ref{boundmelon}). Concerning the rank, if we wish to minimize this variation in the step $p=1\rightarrow p=2$, the only possibility is to exclude the creation of a $k$-dipole for $k> 1$, and so to create a new face $f^{0i}$. Hence, we lose $d-1$ faces, and, if $d>2$, this possibility does not correspond to the leading order. Privileging the graphs with the maximum number of faces is then more advantageous, and the connectivity constraint implies that the only possibility is a melonic graph $\mathcal{M}_2$ as depicted in the Figure \ref{nextorder}.\\

The same argument survives at order $p$. Starting with a melonic graph of order $p$, $\mathcal{M}_p$, we move from the order $p$ to the order $p+1$ by adding a dotted line. This dotted line can carry at least $d$ faces, but one is necessarily common with another dotted line, ensuring the graph connectivity. Hence, adding a new line increases at least by $d-1$ the number of faces, and the optimal graph corresponds to the melon $\mathcal{M}_{p+1}$. As to the rank, it can at least increase by $1$. It is clear that the optimal graphs for the rank and the faces are incompatible, because to minimize the rank variation, so $\Delta R =0$, it is necessary that no $k$-dipole (for $k>1$) and no face is created by the new line. Then, we lose $d-1$ lines compared to the melonic graphs, and this solution does not correspond to a leading order graph.
\begin{center}
\includegraphics[scale=1]{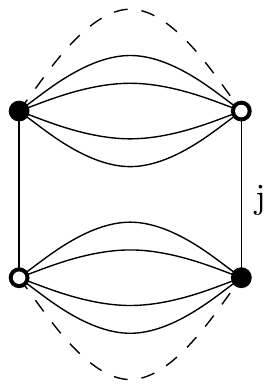} 
\captionof{figure}{The most divergent graph at the order $p=2$.}\label{nextorder}
\end{center}
\noindent
\bigskip
Hence, the melonic rosettes correspond to the most divergent graphs, i.e. the leading order of the perturbative expansion. Because $L(\hat{G})=|L(G)|-|V(G)|+1$, we deduce the following result:
\begin{proposition}\label{th2}
Let $G$ be a vacuum Feynman graph. Its divergence degree is bounded by 
\begin{align}
\omega(G)\leq \omega_{melo}(G)
\end{align} 
with
\begin{align}
\omega_{melo}(G):=\big[&(D-1)(d-2)-2\big]|L(G)|-(D-1)(d-2)(|V(G)|-1)+1 \qquad ,
\end{align}
which correspond to the divergence degree of a melonic rosette graph. 
\end{proposition}
As a result, it follows that the leading order graphs are also melonic. Indeed, it follows from another elementary result concerning the degree $\varpi(G_c)$, i.e. that it is invariant under $1$-dipole contraction \cite{jimmyrazvan, COR2}. Yet, it is obvious that the tree contraction given by the rosette in the previous proposition is a succession of $1$-dipole contractions, then:

\begin{corollary}
The leading order graphs are melonic: their degree $\varpi$ vanish. 
\end{corollary}
\begin{corollary}
Only the melonic interactions contribute to the leading order graphs.
\end{corollary}
It remains to consider the non-vacuum graphs. And we have the following proposition:

\begin{proposition}\label{prop4}
Let $G_{N}$ be a non-vacuum graph with $N$ external lines. Its divergence degree is bounded as:
\begin{align}
\omega(G_N)\leq\big[&(D-1)(d-2)-2\big]|L(G)|-(D-1)(d-2)(|V(G)|-1)
\end{align}
with equality for melonic graphs. In addition, an external mono-color face connects all the external black and white vertices.
\end{proposition}
\bigskip
\textit{\textbf{Proof}}:
Let $\hat{G}$ be a vacuum graph containing $G_N$ in the following sense. $G \subset \hat{G}$ means that $\hat{G}$ contains all the vertices of $G_N$ and has the same connectivity, and that any face of $\hat{G}$ is either an internal face of $G_N$, either splitting in some external faces of this graph. In addition, this inclusion supposes that $G_N$ can be obtained from $\hat{G}$ by cutting some internal lines. Because of proposition \ref{th2}, it follows that $\hat{G}$ is a leading order graph if and only if it is melonic. Starting from this graph, we wish to build the most divergent graph with the same number of external lines as $G_N$. Suppose that $G_N$ has $2$ external lines. From $\hat{G}$, we begin by selecting a spanning tree $\mathcal{T}$ and we construct the rosette $\bar{\hat{G}}$. Now, the simplest way to obtain a non-vacuum graph from the rosette is to cut an internal dotted line. An internal line is necessarily a dipole line, and carries $d$ faces. In addition, the rank has to be decreased by $1$, and $|F|-R$ decreases by $d-1=(d-2)+1$. Another (more complicated) way to build a $2$-points graph is to cut $n$ internal dotted lines, to select $2$ half dotted lines, and to reconnect the $2(n-1)$ remaining half dotted lines in an optimal way. But observe that cutting these $n$ lines decreases $|F|-R$ at most by $(d-2)n+1$. Hence, we must at least reconstruct $(d-1)(n-1)$ faces and increase the rank by $(n-1)$, obtaining in a more involved way exactly the same result. It is then obvious, from the recursive definition of the melonic graphs and their inherited connectivity, that any reconnecting procedure, which does not correspond to the cutting of a singular dipole, increases the length of the $d$ externals faces, and necessarily reduces the number of internal faces. Hence, cutting a single dipole is the more face-economic way to obtaine a $2$-point graph which respects the melonicity condition, and thus the maximization of the divergence degree. Therefore, from the rosette $\bar{\hat{G}}$, we obtain the leading order $2$-points graphs $\bar{G}_2^{melo}$ by cutting one dipole, and the divergence degree of $G_2$ is bounded by:
\begin{align*}
\omega_{melo}(G_2)\leq -2|L(G_2)|+(D-1)(d-2)\big[|L(\bar{\hat{G}})|-1\big]=-2|L(G_2)|+(D-1)(d-2)|L(\bar{{G}}_2)|
\end{align*}
which bounds also the divergence degree of any $2$-point graph. \\

Now, from the $2$-point graph, we would like to build the $4$-point graph. As previously, we start from a rosette graph $\bar{G}_2^{melo}$. The same argument as before shows that the leading order graphs are obtained by cutting a new dipole. But a new subtlety appears. Indeed, because of the special connectivity of the melonic graphs, two given lines in $\bar{G}_2^{melo}$ can be the boundary of one and only one internal face. If two lines do not have common faces, they are said to be \textit{face disconnected}\footnote{We recall the general definition. For a given graph $G$, the incidence matrix $\epsilon_{ef}$ can be rectangular block-factorized. And the subset of lines of a such rectangular block form a \textit{face-connected sub-graph}.}, and if we cut two ``face-disconnected" lines, we lose $2d$ faces against $2(d-1)+1$ if they are face-connected. Hence, it follows that the leading order graphs with $4$ external lines are melonic with an external face of length upper than $2$ in the colored extension graph. The same argument can be applied with $N$ external lines. Then, from the complete graph $\bar{\hat{G}}$, $|F|-R$ decreases by $(N/2)(d-1)+1$ when we cut face-connected lines. Hence, 
\begin{align*}
\omega_{melo}(G_N)\leq& -2|L(G_N)|+(D-1)(d-2)\big[|L(\bar{\hat{{G}}})|-N/2\big]\\
&=-2|L(G_N)|+(D-1)(d-2)|L(\bar{{G}}_N)|,
\end{align*}
and an external mono-color face connects all the external black and white vertices. 
\begin{flushright}
$\square$
\end{flushright}
%\begin{corollary}
%Let a non-vacuum graph $G$ with $N$ external lines. 
%\end{corollary}

From proposition \ref{prop4}, we can easily deduce a criterion for just-renormalizability. Remember that a field theory is said to be \textit{just-renormalizable} if its divergence degree does not increase with the number of vertices. Because of the following topological relationship:
\begin{equation}
|L(G)|=\sum_{k=1}^{k_{max}}kn_k(G)-N_{ext}/2\qquad |V(G)|=\sum_{k=1}^{k_{max}}n_k(G),
\end{equation}
where $n_k$ is the number of vertices of degree $k$ in $G$ (with $k$ black (or white) vertices in their corresponding bubble interaction vertex), we deduce from the previous theorem \ref{th2}:
\begin{align*}
\omega_{melo}&(G):=(D-1)(d-2)-\big[(D-1)(d-2)-2\big]\frac{N_{ext}}{2}\\\nonumber
&+\sum_{k=1}^{k_{max}}\big(\big[(D-1)(d-2)-2\big]k-(D-1)(d-2)\big)n_k(G).
\end{align*}
Hence, renormalizability is ensured if, and only if, the maximal value $k_{max}$ for the degree of the interactions does not exceed
\begin{equation}
k_{R}= \dfrac{(D-1)(d-2)}{(D-1)(d-2)-2}.
\end{equation}
This result allows to classify the just- and super-renormalizable TGFT models, on the basis of the Abelian power counting. 
\begin{center}
\begin{tabular}{|L{0.7cm}||C{0.5cm}|C{0.5cm}|C{0.6cm}|C{3cm}|}
\hline Type & d &  D &  $k_{R}$&$\omega_{melo}$\\
\hline  A & 3 & 4 & 3& $3-N/2-2n_1-n_2$\\
\hline  B & 4 & 3 & 2 & $4-N-2n_1$\\
\hline  C & 5 & 2 & 3 & $3-N/2-2n_1-n_2$\\
\hline  D & 6 & 2 & 2 & $4-N-2n_1$\\
\hline 
\end{tabular}\label{table1}
\end{center}
\captionof{table}{Table of power-counting Abelian-just-renormalizable theories}
\medskip
The super-renormalizable models are those for which $k_{max}<k_{R}$, such that the divergence degree decreases with the number of vertices, implying that only a finite number of graphs needs to be renormalized. It is only when the divergence degree does not depend on the order of the pertubative expansion, i.e. when the higher degree $k_{max}$ equals $k_{R}$, that the theory is said to be just-renormalizable, and that the divergences can be taken care of by a renormalization procedure, implying the definition of a finite number of counter-terms.
The table \ref{table1} below lists some power-counting Abelian just-renormalizable TGFT models\footnote{This classification inherit of the pessimistic nature of the Abelian Power counting.}, in the class we have been considering and on the basis of the Abelian power counting only. \\

%\begin{tabular}{|R{2cm}||C{1cm}|L{1.5cm}|L{1.5cm}|}
%\hline Type\quad & d &  D &  k_{max} 4  \\
%\hline  A & 4 & 4 & 6  \\
%\hline 
%\end{tabular}

\medskip
\noindent
Note that some promising models for quantum gravity are absent of this table. This is the case, for example, of models on $SO(4)/SO(3)$ in dimension $4$, the TGFT counterpart of the simplicial ones studied in \cite{DP-F-K-R, P-R, BCrevisited} and which have been a source of inspiration for this paper. Their absence is due to the rather \lq pessimistic\rq Abelian divergence degree, which, as discussed in section \ref{sectionOB}, is always higher than the exact divergence degree. Hence, if all the models in the table are certainly just-renormalizable, this classification certainly does not exhaust the class of just-renormalizable models. To emphasize this point, we adopt for the following definition:
\begin{definition}
Any model which is power-counting just-renormalizable on the basis of the Abelian power counting only is said to be Abelian just-renormalizable. Moreover, any power counting Abelian just-renormalizable model is also power counting just-renormalizable.
\end{definition}

Note that models that are not Abelian renormalizable can be understood to be actually renormalizable by using a different scaling of the coupling constants and working with a different set of interactions. See \cite{LCO}, in preparation.  \\

\noindent
In the rest of this paper, however, we address the issue of perturbative renormalization for a localizable rank $4$ just-renormalizable quartic melonic model over $SU(2)$. Note that localizability, which means that one can define a contraction procedure and a set of counter-terms is an essential ingredient of the renormalization procedure, not obvious in a theory with non-local interactions. 

\section{Renormalization of the $T^4_4$ model over $SU(2)$}
\label{sectionren}
\subsection{Definition and basic properties}

The previous analysis shows that the special model in dimension $4$ with group $SU(2)$ and quartic melonic interaction \label{sec4.4}is just-renormalizable. The interactions of this model are of the form depicted in the Figure \ref{fig8} below. There are exactly four interactions of this type, one for each choice of the color of the intermediate lines between the two $3$-dipoles. In the following, each interaction bubble $b_i$ will be labeled by the color of these intermediate lines.
\begin{center}
\includegraphics[scale=0.7]{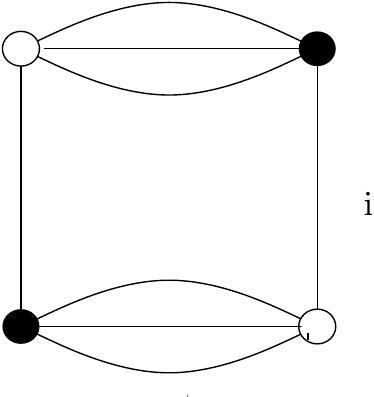} 
\captionof{figure}{The melonic $\phi^4$ interaction $b_i$}\label{fig8}
\end{center}
Normally, each bubble $b_i$ can appear in the interaction part of the action with its own coupling but in the following we will limit our attention to the simplest case where all the interaction bubbles have the same coupling. Hence :
\begin{equation}
S_{int}=\lambda \sum_{i=1}^{4} \Tr_{b_i}\big(\bar{\psi},\psi\big).
\end{equation}
As for the Gaussian measure, it is given by the formula (\ref{prop2}) (or \eqref{propagatorgaugebis}) in the first formulation), and allows to write the regularized generating functional as:
\begin{equation}\label{generfunct}
\mathcal{Z}_{\Lambda}[J,\bar{J}]:=\int d\mu_{C_{\Lambda}} e^{-S_{int}(\bar{\psi},\psi)+\langle \bar{J},\psi\rangle+\langle \bar{\psi},J\rangle},
\end{equation}
where:
\begin{equation}
\langle \bar{J},\psi\rangle:=\int [dg]^4 \bar{J}(g_1,g_2,g_3,g_4)\psi(g_1,g_2,g_3,g_4) \quad .
\end{equation}
In its minimal prescription, the aim of the renormalization procedure is to give a sense to the limit $\Lambda \rightarrow \infty$ perturbatively. This issue will be considered in the following two sections. \\

From the previous section, a question remains: if it is now clear that the most divergent graphs are melons, it is not obvious that all the divergent graphs are melonic. In other words, we have not proved that the melonic graphs contain all the divergences occurring in the graph expansion of correlation functions, and we dedicate the end of this section to the answer to this question. \\

The form of the interaction bubbles in Figure \ref{fig8} allows to use a very useful representation of the theory, the \textit{intermediate field representation}, from which the problem can be translated into a simple recursion. The building rules of the intermediate field representation are the following. From the basic properties of the Gaussian integration, the generating functional (\ref{generfunct}) can be formally rewritten as:
\begin{equation}
\mathcal{Z}_{\Lambda}[J,\bar{J}]=\int d\mu_{C_{\Lambda}}\prod_{i=1}^4 d\mu_1{(\sigma_i)} e^{i\sqrt{2\lambda}\langle \bar{\psi},\Sigma\psi\rangle+\langle \bar{J},\psi\rangle+\langle \bar{\psi},J\rangle} \quad ,
\end{equation}
where $d\mu_1{(\sigma_i)}:=e^{-\tr(\sigma^2)}d\sigma$ is the Gaussian measure for Hermitian matrices and:
\begin{equation}
\Sigma=\sum_{i=1}^4\mathbb{I}^{\otimes(i-1)}\otimes \sigma_i\otimes\mathbb{I}^{\otimes(4-i)} \quad ,
\end{equation} 
where $\mathbb{I}$ is the identity operator for a single variable function: $\int dg \mathbb{I}(g,g^\prime)\phi(g^\prime)=\phi(g)$. Now, the Gaussian integration over $\psi$ and $\bar{\psi}$ can be performed, leading to the following effective multi-matrix model:
\begin{equation}\label{intermediate}
\mathcal{Z}_{\Lambda}[J,\bar{J}]=\prod_{i=1}^4 d\mu_1{(\sigma_i)}e^{-\Tr\ln(1-i\sqrt{2\lambda}C\Sigma)+\langle \bar{J},\mathfrak{R}J\rangle},
\end{equation}
where : $\mathfrak{R}:=(1-i\sqrt{2\lambda}C\Sigma)^{-1}C$. The Feynman rules are the following: expanding the logarithm, we generate interactions with one, two,... n-external points, and a typical Feynman graph is composed of several of these vertices, connected to each other by matrix lines. 
\begin{center}
\includegraphics[scale=0.8]{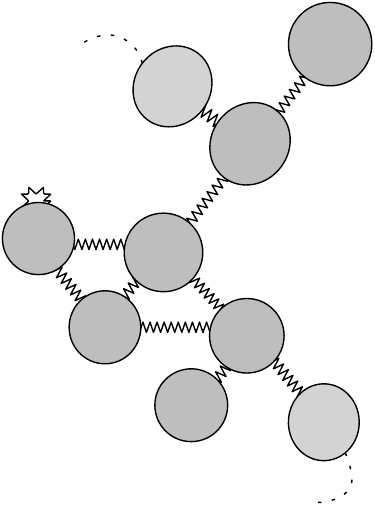} 
\captionof{figure}{A Feynman graph in the intermediate field representation, with two ciliated vertices.}\label{fig8.2}
\end{center}
These matrix lines, of color 1 to 4, are depicted by a wavy line, and the vertices, to which they are hooked, by a grey disk, as in Figure \ref{fig8.2}. In this figure, one of the grey disks has a dotted arrow line (a cilium). This grey disc does not correspond to an interaction generated by the logarithm expansion, but comes from the expansion of the $R$ operator defined before (\ref{intermediate}), and a ciliated disk with $n$ external wavy lines correspond to the term of degree $n$ in the expansion in powers of $\Sigma$.  

This representation has been studied in detail in several recent papers, e.g. \cite{constructiveTGFT}. An important result about this representation is that the leading order graphs, the melons of the original representation, appear as trees in the intermediate field representation. More precisely:
\begin{proposition}
In the intermediate field representation, the melonic graphs of the divergent sector are trees with one or two external wavy lines, all of the same color, and connected to the same external face. 
\end{proposition}
This can be easily proven by recursion. Now, we will use this property to see if all the divergent graphs are melons. Starting from a leading order graph (a tree) with $l$ wavy lines (to each wavy line corresponds a vertex of the original representation, and the number of these wavy lines is equivalent to the power of $\lambda$ associated to the graph), we will investigate the different ways to build a graph with $l+1$ wavy lines, and the possibility that one of these gives a non-melonic divergent graph. Two of these ways are depicted in Figure \ref{fig8.3}. They correspond to the addition of a tadpole graph over a grey disk, or to the replacement of a wavy line by a $2$-points grey disk. But neither of them affects the tree structure of the starting graph, which remains a melon.  
\begin{center}
\includegraphics[scale=0.5]{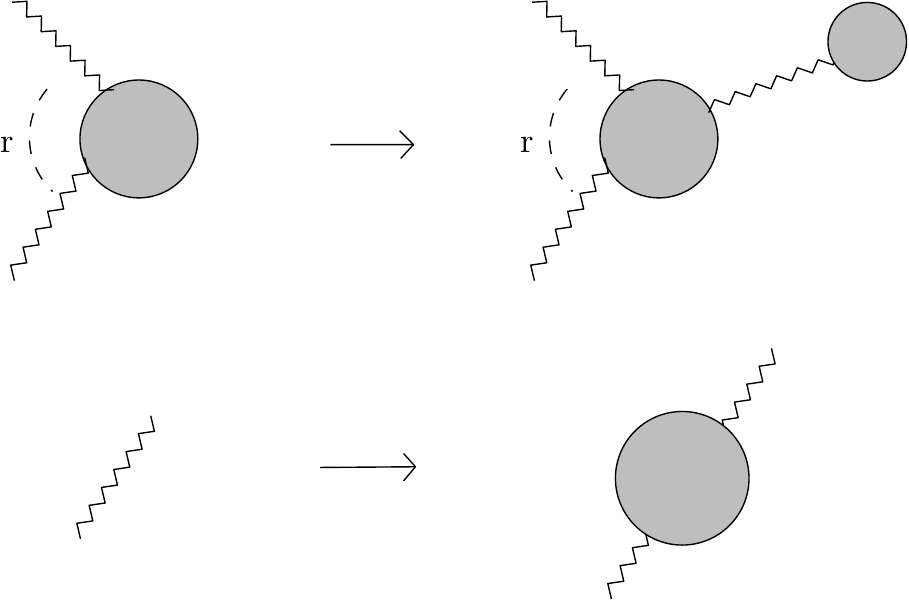} 
\captionof{figure}{Addition of a tadpole over a grey disk and the replacement of a wavy line by a $2$-points grey disk.}\label{fig8.3}
\end{center}
The two operations depicted in Figure \ref{fig8.4} are more promising. They are both a deviation from the melonicity, because both affect the tree structure of the starting graph (because of the connectivity of the starting graph, the second operation necessarily builds a loop). As a result, only these two operations can give us a non-melonic divergent graph, and we will examine these two possibilities.
\begin{center}
\includegraphics[scale=0.5]{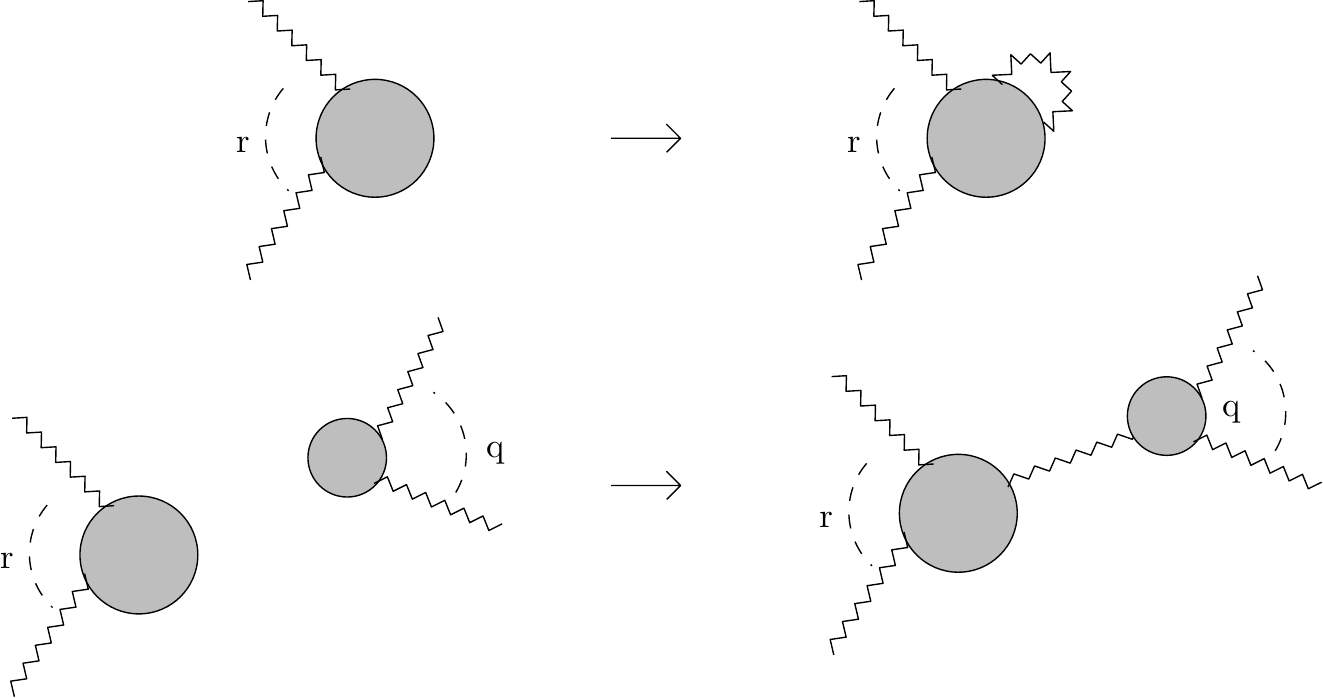} 
\captionof{figure}{Addition of a self-loop and of a wavy line between two disks.}\label{fig8.4}
\end{center}
In the two cases, we increase the number of field lines by two (the field lines are the dotted lines of the original formulation, which are hidden inside the disks in the intermediate field representation). In the worst case, the number of faces increases only by one, but it is exactly compensated by the variation of the rank. Therefore, the total variation of the divergence degree is:
\begin{equation}
\delta \omega\leq \delta\omega_{Abelian} =-2\delta |L|+2\delta(|F|-R)< -2 \qquad ,
\end{equation}
where $\omega_{Abelian}$ is the Abelian divergence degree computed previously. Because of the Abelian divergence degree is $4-N$ for a graph with $N$ external lines, any graph with $N\geq 2$ becomes superficially convergent. However, the previous result seems to show that a vacuum divergent subgraph can support one intermediate field loop. And we have proven the following result:
\begin{proposition}
All the divergent non-vacuum graphs of Abelian the melonic $\phi^4$ model are melonic.
\end{proposition}
Our conclusion about the Abelian just-renormalizable model $\phi^4$ in $d=4$ is fundamental for the following reason. The recent literature on TGFT renormalization has shown that melonic graphs are interesting for two (closely related) essential reasons. The first one is that the Abelian power counting become exact for these graphs, and the second one is that they are said to be \textit{tracial}, meaning that any connected graph remains connected under contraction of a melonic subgraph - a property which is essential for renormalization. The renormalization procedure can be defined in a worst-case-scenario, in which all the Abelian divergent graphs are regarded as \textit{dangerous}. 

\subsection{One-loop renormalization and asymptotic freedom}

In this section we compute the divergent parts of the one particle irreducible (1PI) melonic $2$- and $4$-point functions\label{oneloop}, providing a simple illustration of the renormalization procedure that we will generalize at all order in the next section. As an interesting consequence for our models, we will deduce the so called beta function of the running coupling constant $\lambda_{eff}$, and deduce that the theory is asymptotically free in the vicinity of the Gaussian fixed point.

\subsubsection{The 2-points function}

At one-loop order, the divergences are due to the melonic tadpole diagrams, an example of which is pictured in Figure \ref{fig12}, and corresponds to the following Feynman amplitude, written in the time gauge:
\begin{align}\label{oneloop}
\nonumber\mathcal{A}_{\mathcal{M}}^{(4)}&(\mathbf{g}_{t},\mathbf{g}_{s}) = \int_{1/\Lambda^2}^{+\infty} d[\alpha]^3 e^{-(\alpha+\alpha_1+\alpha_2) m^2}\int dldl_1dl_2 \int dk\int_{U_{k}(1)^{\times 12}}[dh_1]^4[dh_2]^4[dh]^4\\
&\prod_{i=1}^3 K_{\alpha}\big(lh_i\big)K_{\alpha_1+\alpha_2}\big(g_{ti}l_1h_{1\,i}l_2h_{2\,i}g_{si}^{-1}\big)K_{\alpha_1+\alpha_2+\alpha}(g_{t4}l_1h_1^{(4)}lh_4l_2h_2^{(4)}g_{s4}^{-1})\quad  ,
\end{align}
where the subscripts $t$ and $s$ meaning ``target" and ``source" label the boundary group variable. 
\begin{center}
\includegraphics[scale=1]{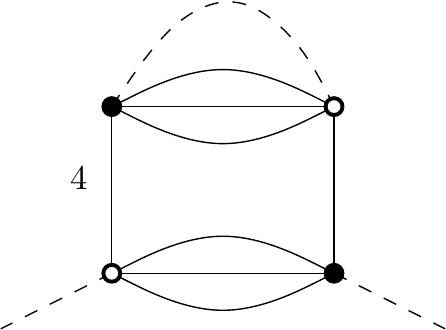} 
\captionof{figure}{Tadpole contribution to the one-loop 1PI 2-point function}\label{fig12}
\end{center}
\medskip
As the divergences occur in the vicinity of $\alpha =0$, we can make use of the corresponding approximation for the heat kernels appearing in (\ref{oneloop})
\begin{equation}\label{heatlim}
K_{\alpha}(g=e^{iX})\underset{\alpha \to 0}{\longrightarrow}\big(4\pi \alpha\big)^{-3/2}e^{-\frac{\langle X,X\rangle}{4\alpha}} \quad ,
\end{equation}
where $\langle.,.\rangle:\mathrm{su}(2)\rightarrow \mathbb{R}$ is the (normalized) Killing form on the Lie algebra. The vicinity of $\alpha = 0$ forces us, in the saddle point approximation, to evaluate only the fluctuations around the identity for each group arguments :
\begin{align}\label{saddlepoint}
e^{X_l}e^{Y_{h_i}} \approx \mathbb{I}\quad where \quad l=:e^{X_l}\,\,,h_i=:e^{Y_{h_i}} \quad .
\end{align}
Hence, the integral in (\ref{oneloop}) behaves as
\begin{align}
\nonumber\Big(\frac{1}{4\pi\alpha}\Big)^{9/2}\int_{\mathbb{R}^2} d^{2}x\int_{\mathbb{R}}\prod_{i=1}^3dy_i&e^{-\frac{3x_1^2+3x_2^2+(x_3-y_1)^2+(x_3-y_2)^2+(x_3-y_3)^2}{4\alpha}}\sim \alpha^{-2} \quad .
\end{align}
Note that the power of $\alpha$ coincides with the divergence degree computed previously. Hence, we confirm its validity in this simple example.\\

\noindent
In order to extract the divergences of the expression \eqref{oneloop}, we begin by fixing $k$ along the $Oz$ axis. Then, defining: 
\begin{align}
\nonumber\mathcal{I}_k(l,g^{-1}_{4}g_{4}^{\prime}l):=\int_{U(1)_k^{\times 4}}[dh]^4\times \int dg_4dg_4^{\prime}K_{\alpha}\big(lh_1\big)K_{\alpha}\big(lh_2\big)K_{\alpha}\big(lh_3\big)K_{\alpha}\big(g^{-1}_{4}g_{4}^{\prime}lh_{4}\big)\quad ,
\end{align}
and integrating over the $h_i$ gives:
\begin{align}
\nonumber\mathcal{I}_k(l,g^{-1}_{4}g_{4}^{\prime}l):=\sum_{\{l_i\}\in\mathbb{N}^4}\sum_{m_1=-l_1}^{l_1}\prod_{i=1}^3(2l_i+1)e^{-4\alpha l_i(l_i+1)}D_{00}^{(l_i)}(l)(2l_4+1)e^{-4\alpha l_4(l_4+1)}D^{(l_4)}_{m_40}(l)D^{(l_4)}_{0m_4}(g^{-1}_{4}g_{4}^{\prime})\quad.
\end{align}
The first product is nothing that $ \big[4\pi K_{\alpha}^{\mathcal{S}_2}(|x|)\big]^3 $ , where $x:=\pi(l)$ is the path over $\mathcal{S}_2 $ corresponding to $l\in SU(2)$. With our definitions, using the decomposition: $l=e^{i\varphi \sigma_z/2}e^{i\theta \sigma_y/2}e^{i\gamma\sigma_z/2}$, one find explicitly : $|x|=\theta$ i.e. the geodesic length from the north pole to $\vec{n}=(\theta,\phi)$. As explained before, the divergences come from the negative (or null) powers of $\alpha$, allowing to make an expansion around the identity. When $\alpha\to 0$, $K_{\alpha}^{\mathcal{S}_2}(x)\approx \delta(x)$, and using the expansion of the heat kernel over the $2$-sphere in the vicinity of $\alpha =0$:
\begin{equation}
K_{\alpha}^{\mathcal{S}_2}(x)\approx \frac{e^{-|x|^2/4\alpha}}{4\pi\alpha}\quad ,
\end{equation}
where the discarded terms involve higher power of $\alpha$ and generate sub-divergent corrections for the mass parameter, which are not relevant for our discussion. Then, because of the fact that, for any $C^{2}$ function $f:\mathcal{S}_2\to\mathcal{C}$ which is regularized at the origin:
\begin{equation}
\int d^2x e^{-3|x|^2/4\alpha}f(x)=\frac{4}{3}\pi \alpha f(0)+\bigg(\frac{4}{3}\pi\alpha\bigg)\frac{\alpha}{3}\Delta_{\mathcal{S}_2}f(0)+\cdots\quad ,
\end{equation}
where $\Delta_{\mathcal{S}_2}$ denote the Laplacian over $\mathcal{S}_2$. And because $\Delta_{\mathcal{S}_2}D_{m_40}^{(l_4)}=-l_4(l_4+1)D_{m_40}^{(l_4)}$, we deduce (up to sub-divergent term for the first correction without Laplacian):
\begin{align}
\int dl\mathcal{I}_k(l,g^{-1}_{4}g_{4}^{\prime}l)=\int_{U(1)_k}dh_k\delta(g^{-1}_{4}g_{4}^{\prime}h_k)\bigg\{\frac{1}{6}\frac{1}{\alpha^2}-\frac{13}{18}\frac{1}{\alpha}l_4(l_4+1)+\cdots\bigg\}\quad.
\end{align}
where we have taking into account a normalization factor $1/8\pi$ coming from the normalized Haar measure $dl$. Then, integrating over the $\{h_{\ell}^{(i)}\}\,,\ell=1,2$ variables in \eqref{oneloop}, we find the leading divergent par of the $2$-points function:
\begin{align}
\nonumber\mathcal{A}_{\mathcal{M}}^{(4)\,\infty}(\mathbf{g}_{t},\mathbf{g}_{s}) =& \sum_{\{l_i\}\in\mathbb{N}^4}\int_{1/\Lambda^2}^{+\infty} d[\alpha]^2 e^{-(\alpha_1+\alpha_2) (m^2+4\sum_{i=1}^4l_i(l_i+1))}\int dl\\
&\times \bigg\{\frac{1}{6}\mathcal{I}_1-\frac{13}{18}\mathcal{I}_2l_4(l_4+1)+\cdots\bigg\}\prod_{i=1}^4(2l_i+1)D_{00}^{l_i}(g_{si}^{-1}g_{ti}l)\label{2div}
\end{align}
where we have defined:
\begin{equation}
\mathcal{I}_1:=\int_{1/\Lambda^2}^{+\infty} d\alpha\frac{e^{-\alpha m^2}}{\alpha^2}\quad,
\end{equation}
\begin{equation}
\mathcal{I}_2:=\int_{1/\Lambda^2}^{+\infty} d\alpha\frac{e^{-\alpha m^2}}{\alpha}.
\end{equation}
and where the subscript $\infty$ means that we retain only the divergent part. The first term of \eqref{2div} correspond to the mass renormalization and the second one to the wave function. Similar expressions can be obtained for each interaction bubble, and the divergent part of the complete 1PI $2$-points function $\Gamma_{\infty}^{(2)}(\{g_{t_i}\},\{g_{s_i}\})$ writes as (a global factor $2$ comes from the Wick-theorem):
\begin{align}
\Gamma_{\infty}^{(2)}&=-\frac{4\lambda}{3}\bigg[\mathcal{I}_1+\frac{13}{24}\mathcal{I}_2\,\sum_{i=1}^4\Delta_{\mathcal{S}_2,i}\bigg] \quad .
\end{align}\label{twopointsdiv}

\subsubsection{The 4-point function}

At one-loop order, a typical melonic contribution involves one loop between two vertices, as an example is pictured in Figure \eqref{fig13} below.

\begin{center}
\includegraphics[scale=0.9]{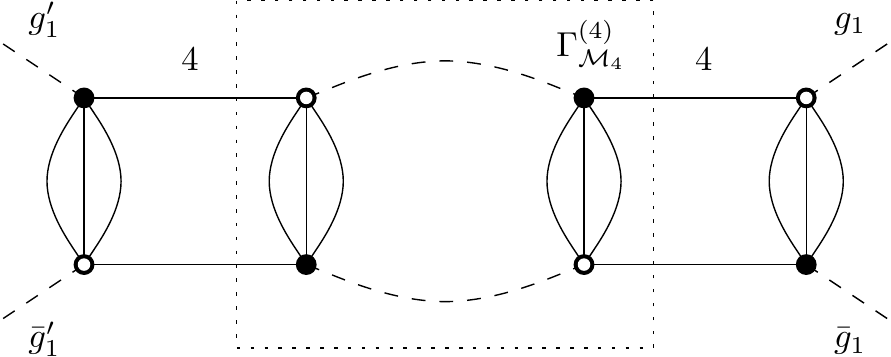} 
\captionof{figure}{Melonic contribution to the 1PI 4-points function at the one-loop order}\label{fig13}
\end{center}

The amputated Feynman amplitude, pictured on Figure \ref{fig13} can be easily obtained from the Feynman rules and from the previous calculation. From the Feynman rules - and taking into account symmetry factors:
\begin{equation}
\Gamma_{\mathcal{M}_4}^{(4)}(g_1,\bar{g}_1,g_1^{\prime},\bar{g}_1^{\prime})=8\lambda^2\sum_{\{l_i\}\in\mathbb{N}^5}\int_{1/\Lambda^2}^{\infty}d[\alpha]^2e^{-(\alpha_1+\alpha_2)m^2}\int_{SU(2)^2}dh_1dh_2\mathcal{I}_{\mathcal{M}_4,\alpha_1+\alpha_2,\{l_i\}}^{(4)}(g_1,\bar{g}_1,g_1^{\prime},\bar{g}_1^{\prime},h_1,h_2)\,,\label{amplitude4}
\end{equation}
with
\begin{align}
\nonumber\mathcal{I}_{\mathcal{M}_4,\alpha_1+\alpha_2,\{l_i\}}^{(4)}&(g_1,\bar{g}_1,g_1^{\prime},\bar{g}_1^{\prime},h_1,h_2)=\prod_{i=1}^3(2l_i+1)e^{-4(\alpha_1+\alpha_2)l_i(l_i+1)}D_{00}^{(l_i)}(h_1h_2)\\
&\times (2l_4+1)(2l_5+1)e^{-4\alpha_1 l_4(l_4+1)-4\alpha_2 l_5(l_5+1)}D_{00}^{(l_4)}(g_1^{\prime\,-1}g_1h_1)D_{00}^{(l_5)}(\bar{g}_1^{-1}\bar{g}_1h_2).\label{amplitude42}
\end{align}
Note that the amplitude is invariant under the transformation of the gauge variables: $h_1\to G_1h_1\bar{G}_1^{-1}$, $h_2\to \bar{G}_1h_2G_1^{-1}$ if $G_1\in U(1)_{\sigma_z}$, up to an irrelevant global translation of the boundary variable : $g_1,\bar{g}_1\to g_1G_1^{-1}, \bar{g}_1\bar{G}_1^{-1}$ - the $4$-points function being assumed to be a gauge invariant function i.e. $\Gamma^{(4)}_{\mathcal{M}_4}\in \mathbb{G}^4$, where $\mathbb{G}=\ker[\hat{P}-\mathbb{I}]$ has been defined in Section \ref{3.1}. Fixing the gauge such that $\bar{G}_1h_2=\mathbb{I}$, the amplitude \eqref{amplitude4} writes as:
\begin{equation}
\Gamma_{\mathcal{M}_4}^{(4)}(g_1,\bar{g}_1,g_1^{\prime},\bar{g}_1^{\prime})=8\lambda^2\sum_{\{l_i\}\in\mathbb{N}^5}\int_{1/\Lambda^2}^{\infty}d[\alpha]^2e^{-(\alpha_1+\alpha_2)m^2}\int_{SU(2)}dh\mathcal{I}_{\mathcal{M}_4,\alpha_1+\alpha_2,\{l_i\}}^{(4)}(g_1,\bar{g}_1,g_1^{\prime},\bar{g}_1^{\prime},h,\mathbb{I})\quad,\label{amplitude4}
\end{equation}
and the computation of the divergent part of $\mathcal{I}_{\mathcal{M}_4,\alpha_1+\alpha_2,\{l_i\}}$ follows the same strategy as for the computation of the mass correction in the previous section. As a matter of fact, because the diagram on Figure \ref{fig13} scale as $\ln(\Lambda)$, only the first term in the local expansion is relevant for the extraction of the divergences. Let us consider separately the Hepp sectors $\alpha_1\geq \alpha_2$ and $\alpha_2\geq \alpha_1$, and the change of variables :
\begin{align}
\alpha_1&=\alpha\\
\alpha_2-\frac{1}{\Lambda^2}&=\beta\Big(\alpha_1-\frac{1}{\Lambda^2}\Big)\quad,
\end{align}
in the first one of these sectors. We then obtain truly the same integral as for the computation of the mass correction. Because of the melonic structure of the graph, 
the local approximation at order $\lambda^2$ writes as (note that because the $4$-points function scale as $\ln(\Lambda)$, the first term of the local expansion contain all the divergences):
\begin{align}
\nonumber\mathcal{A}_{\mathcal{M}_4}^{(4)}(\mathbf{g}_1,\mathbf{g}_2,\mathbf{g}_3,\mathbf{g}_4) =\big(-4\lambda+\Gamma^{(4)}\big)\frac{1}{2}\big[\mathcal{W}^{(4)}_{\mathbf{g}_1,\mathbf{g}_2,\mathbf{g}_3,\mathbf{g}_4}+\mathbf{g}_1\longleftrightarrow \mathbf{g}_3\big] \quad ,
\end{align}\label{oneloop2}
and one finds:
\begin{align}\label{melofour}
&\Gamma^{(4)\,\infty}=\frac{4\lambda^2}{3}\mathcal{I}_2 .
\end{align}

\subsubsection{Running coupling and asymptotic freedom}
\label{asymptotic}In order to extract the dangerous part of the one-loop amplitudes, i.e. the terms involving a positive or null power of $\Lambda$, we study the behavior of the integrals $\mathcal{I}_1$ and $\mathcal{I}_2$. Firstly, observe that an integration by parts gives: $$\mathcal{I}_1=\Lambda^2e^{-m^2/\Lambda^2}-m^2\mathcal{I}_2 \quad .$$ 
Secondly, observe that the divergence of $\mathcal{I}_2$ is at most logarithmic. Hence,$$\mathcal{I}_2=A\ln(\Lambda)+\mathcal{O}(1/\Lambda) \quad .$$
By differentiating the two members of this equality, we obtain $A=2$, and finally:
\begin{align}
\mathcal{I}_1&\sim \Lambda^2-2m^2\ln(\Lambda)\\
\mathcal{I}_2&\sim 2\ln(\Lambda) \quad .
\end{align}
Hence, for the 1PI 2-point function, the dangerous part $\Gamma^{2}_{div}$ is equal to
\begin{align}\label{renwave}
&\Gamma_{\infty}^{(2)}=-\frac{4\lambda}{3}\bigg[\Lambda^2-2m^2\ln(\Lambda)+\frac{26}{24}\ln(\Lambda)\,\sum_{i=1}^4\Delta_{\mathcal{S}_2\,,i}\bigg] \quad ,
\end{align}
which fixes the divergent parts of the mass and wave function counter-terms as:
\begin{align}
\delta m^2_{div}&:=-\frac{4\lambda}{3}\big[\Lambda^2-2m^2\ln(\Lambda)\big]\\
\delta Z_{div}&:=\frac{13\lambda}{9}\ln(\Lambda) \quad .
\end{align}
Similarly, from the expression (\ref{melofour}), we deduce that the divergence is exactly compensated by an interaction of the initial form with an intermediate line of color $4$, if this interaction is proportional to the counter-term $\delta\lambda$, with:
\begin{equation}\label{rencoupling}
\delta\lambda=\dfrac{\lambda^2}{3}\ln(\Lambda).
\end{equation}
This result allows to obtain the dependence of the effective coupling at scale $\Lambda$. Indeed, the effective coupling includes the effect of the wave-function renormalization. Hence:
\begin{equation}
\lambda_{eff}(\Lambda):=\dfrac{\lambda+\delta\lambda}{Z^2} \qquad .
\end{equation}
By differentiating the two terms, and using the relations (\ref{renwave}) and (\ref{rencoupling}), we find:
\begin{equation}
\Lambda\dfrac{d\lambda_{eff}}{d\Lambda}=-\frac{23}{9}\lambda_{eff}^2 \quad ,
\end{equation}
where the minus sign means that the model is perturbatively asymptotically free in the deep UV. As a result, we expect that no-Landau pole occurs in the high energy limit, and that the theory can be properly defined beyond the perturbative regime using constructive methods. 

\subsection{Divergent graphs and renormalized amplitude}
The table \ref{table1} shows\label{sectionren1} that in the present case, the divergence degree of a Feynman graph $G$ with $N$ external lines is bounded by $4-N$ %(with equality for melonic graphs)%
. Hence, {\it a priori}, only the $2$ and $4$-points functions are potentially dangerous. Then, we adopt the following definitions:

\begin{definition}\label{divgraph}
Consider a Feynman graph $G$ and let $h\subseteq G$ be a melonic subgraph of $G$ with $n_h$ external lines. The subgraph $h$ is said to be:\\

$\bullet$ superficially convergent if $\omega_{Abelian}(h)< 0$.\\
$\bullet$ superficially divergent or dangerous if $\omega_{Abelian}(h)\geq 0 \Rightarrow N(h)\leq 4$. 
\end{definition}

\begin{definition}\label{divergent forest}
Consider a Feynman graph ${G}$. A Zimmermann forest $\mathbb{F}$ is a forest of connected divergent subgraphs $\{h\subseteq \mathcal{G}|\omega(h)\geq 0\}$. Here the word \emph{forest} 
should be understood in the sense of inclusion relations. It simply means that
taking two elements $h_1, h_2\in {\mathbb{F}}$, they are either line and vertex disjoint or included one into the other. The set of all Zimmermann forests
of ${G}$ is noted  $D({G})$
\end{definition}
The definition \ref{divgraph} is motivated by the following theorem, which states that the Feynman amplitude is finite if it does not contain any subdivergent graph in the sense of the definition \ref{divgraph} :

\begin{theorem} \textbf{\emph{(Weinberg uniform)}} \label{Weinberg}
Consider a completely convergent graph $G$, i.e. a graph with no subdivergences. Its corresponding Feynman amplitude $\mathcal{A}_{G}$ has the following bounds:
\begin{equation}
|\mathcal{A}_{G}|\leq K^{|V(G)|}, \, \, K\in \mathbb{R}^{+}.
\end{equation}
\end{theorem}
\textit{\textbf{Proof}}. The proof is standard in renormalization theory, and we will only give the main steps. The first step is to note that, when $N>4$, one has $4-N \leq -N/3$ (the graph with five external lines does not exist). Hence, for a given scale attribution $\mu$, the graph amplitude $\mathcal{A}_G$ verifies the following trivial bounds:
\begin{equation*}
|\mathcal{A}_{G \,\mu}| \leq K^{l({G})}\prod_{i,\rho}M^{-N({G}^{\rho}_i)/3} \quad .
\end{equation*}
\begin{definition}\label{linesscales}
\begin{equation*}
i_b(\mu)=\sup_{l\in L_b({G})}i_l(\mu) \qquad e_b(\mu)= \inf_{l\in L_b({G})}i_l(\mu) \quad ,
\end{equation*}
\end{definition}
where $b$ stands for a vertex bubble $b\in{G}$, and $L_b({G})$ is the set of its external lines. Note that $b$ touches a connected subgraph ${G}^{\rho}_i$ if and only if $i \leq i_b(\mu)$, and is an external vertex if $e_b < i \leq i_b$. Therefore, because each vertex touches at most $4$ subgraphs:
\begin{equation*}
\prod_{i,\rho}M^{-N({G}^{\rho}_i)/3} \leq \prod_{i,\rho}\prod\limits_{\substack{b\in {G}^{\rho}_i \\ e_b<i\leq i_b}}M^{1/12} \quad ,
\end{equation*}
and
\begin{equation*}
|\mathcal{A}_{G ,\mu}| \leq K^{l({G})} \prod_{b} M^{-\frac{|i_b(\mu)-e_b(\mu)|}{12}} \quad .
\end{equation*}
Using the fact that there are at most $4$ half-lines, and thus $6=4\times 3/2$ pairs of half-lines hooked to a given vertex, and that, for two lines $l$ and $l'$ of a bubble b, $|e_b-i_b|\geq |i_l-i_{l'}|$, we obtain:
\begin{equation*}
|\mathcal{A}_{G,\mu}| \leq K^{l({G})} \prod_{b} \prod_{(l,l')\in L_b\times L_b}M^{-\frac{|i_l-i_{l'}|}{72}} \quad .
\end{equation*}
This expression implies directly the finiteness of $A_{{G}}$. To understand why, observe that we can choose a total ordering of the lines $L(G)=\{l_1,...,l_{|L(G)|}\}$ such that $l_1$ is hooked to an external vertex $b_0$ and that each subset $\{l_1,...,l_m\}, \, m\leq |L(G)|$ is connected. Therefor, for any line $l_j$, we can choose $l_{p(j)}$ (with $p(j)<j$) a line sharing a vertex with $l_j$, from which we deduce:
\begin{equation*}
\prod_{b} \prod_{(l,l')\in L_b\times L_b}M^{-\frac{|i_l(\mu)-i_{l'}(\mu)|}{78}}\leq \prod_{j=1}^{|L({G})|}M^{-\frac{|i_{l_j}-i_{l_{p(j)}}|}{72}} \quad .
\end{equation*}
And because
\begin{equation*}
\sum_{i_{l_j}}M^{-\frac{|i_{l_j}-i_{l_{p(j)}}|}{72}} \leq \sum_{i_{l_j}\geq i_{l_{p(j)}}}M^{-\frac{|i_{l_j}-i_{l_{p(j)}}|}{72}} = \dfrac{1}{1-M^{-1/72}} \quad ,
\end{equation*}
we have finally :
\begin{equation*}
|\mathcal{A}_{G \,\mu}| \leq K^{l({G})}\sum_{\mu=\{i_1,...,i_{l({G})}\}}\prod_{j=1}^{|L({G})|}M^{-\frac{|i_{l_j}-i_{l_{p(j)}}|}{72}}\leq K'^{l({G})} \qquad .
\end{equation*}
\begin{flushright}
$\square$
\end{flushright}
When the graphs contain some dangerous subgraphs, the proof given above breaks down, and the finiteness of the sum over scale attribution is not guaranteed. Therefore, the case of the presence of these subgraphs must be considered in details.  \\

\paragraph{$\bullet$ N=2}. 
Let us consider the case of a subdivergent graph with two external lines. The situation is depicted in Figure \ref{fig9} below. The structure of the amplitude $\mathcal{A}_{G,\mu}$ for the scale attribution $\mu$ is
\begin{align}
\mathcal{A}_{G,\mu}= \int \prod_{l}d\bar{g}_{1l}d\bar{g}_{2l}dg_{1l}dg_{2l}\bar{\mathcal{A}}_{G,\mu}(\{\bar{g}_{1l}\},\{{\bar{g}}_{2l}\})C_{i_1}(\{g_{1l}\},\{\bar{g}_{1l}\})C_{i_2}(\{g_{2l}\},\{\bar{g}_{2l}\})\mathcal{M}_j(g_{11},g_{21}) \quad ,
\end{align}
where $\mathcal{M}_j$ is the 2-points subgraph pictured in Figure \ref{fig9}, $j$ its scale, i.e. the scale of its highest line, and $\bar{\mathcal{A}}_{G,\mu}$ is a completely convergent amplitude. The scale attribution is chosen such as $j>i_1,i_2$, and the subgraph $\mathcal{M}_j$ is said to be \textit{high}. This is typically the region in which this graph is potentially divergent. Obviously from the propagator structure one has $\mathcal{M}_j(g,g')=\mathcal{M}_j(gg^{\prime\,-1})$. \\

\noindent
The one-loop computations show that we have to replace the standard local expansion by an equivalence up to a right multiplication by an element of $U(1)_k$. In order to make this identification more conveniently in our analysis, we first consider the following definitions. Using the Euler parametrization, each $g\in SU(2)$ writes as : $g=e^{i\phi\sigma_z/2}e^{i\theta\sigma_y/2}e^{i\gamma\sigma_z/2}$, where $0\leq \phi <2\pi$, $0\leq \theta \leq \pi$, $0\leq \gamma <4\pi$. $(\theta, \phi, \gamma)$ are coordinates over the group manifold, with metric : $ds^2= (d\phi^2+2\cos(\gamma)d\phi d\gamma +d\gamma^2+d\theta^2)/4$. Let the map $\pi:SU(2)\to \mathcal{S}_2$ such that $\pi(g)=\vec{n}(\theta,\phi)\in\mathcal{S}_2, \,g\in SU(2)$. It provides a basic fibre bundle, say $P$ with fibre $U(1)$, acting on the right to each element $u\in P \sim SU(2)$ : $u\to u'=uh, \forall h\equiv e^{i\gamma\sigma_z/2}\in U(1)$, getting from one horizontal space to another. Moreover, we introduce the global section $s:\mathcal{S}_2\to SU(2)$, defined as $s(x)=e^{i\phi\sigma_z/2}e^{i\theta\sigma_y/2}\in SU(2),\, x\equiv \vec{n}(\theta,\phi)$, inducing a metric over $S_2$: $g_{\mathcal{S}_2}(x,x)=g_{SU(2)}(s(x),s(x))=s^*g_{SU(2)}(x,x)$. Finally, for any $g\equiv(\theta,\phi,\gamma)\in P$, let $T_g(P)\sim \mathfrak{su}(2)$ be the tangent space at $g$. We say that $X_g\in T_g(P)$ is horizontal, if and only if $d\gamma (X_g)=0 \Leftrightarrow \exists (\phi,\theta) | e^{X_g}=e^{i\phi\sigma_z/2}e^{i\theta\sigma_y/2}$. \\

Now, we define the real parameter $t\in[0,1]$ in order to interpolate between $g_{21}$ and $g_{11}$. More precisely, in the ``time gauge", choosing $k$ along $Oz$ for each lines of $\mathcal{M}_j$ and of its external lines, we have the parametrization : $G:=g_{21}g_{11}^{-1}=e^{i\phi\sigma_z/2}e^{i\theta\sigma_y/2}e^{i\gamma\sigma_z/2}$, and as explained in the Section \ref{oneloop}, $|\theta|$ correspond to the geodesic distance between the north pole and the point $\vec{n}(\theta,\phi)$. Let $X$ be the horizontal vector such that $e^X=e^{i\phi\sigma_z/2}e^{i\theta\sigma_y/2}$. We define the smooth trajectory $G(t)=e^{tX}e^{i\gamma\sigma_z/2}$, such that $\pi(G(1))=\vec{n}(\theta,\phi)$ and $\pi(G(0))=\vec{e}_z$. Therefore, we have the parametrized amplitude:
\begin{align}
\mathcal{A}_{G,\mu}(t)=\int \prod_{l}d\bar{g}_{1l}d\bar{g}_{2l}dg_{1l}dg_{2l}\bar{\mathcal{A}}_{G\mu}(\{\bar{g}_{1l}\},\{{\bar{g}}_{2l}\})C_{i_1}(\{g_{1l}\},\{\bar{g}_{1l}\})C_{i_2}(\{g_{12}(t),g_{2l\,l\neq 1}\},\{\bar{g}_{2l}\})\mathcal{M}_j(g_{11},g_{21}) \quad,
\end{align}
such that $\mathcal{A}_{G,\mu}=\mathcal{A}_{G,\mu}(t=1)$, and where $g_{12}(t):=G(t)g_{11}$. We then introduce the $*$ application $\tau_{\mathcal{M}}$ as:
\begin{equation}
\tau_{\mathcal{M}}^*\mathcal{A}_{G,\mu}(t):=\sum_{n=0}^{\omega(\mathcal{M})}\dfrac{1}{n!}\dfrac{d^n\mathcal{A}_{G,\mu}}{dt^n}(t=0) \quad ,
\end{equation} 
and the goal, motivated by the usual quantum field theory, is to prove that 
\begin{align}\label{amplituder}
\mathcal{A}_{G,\mu}^{R}=(1-\tau_{\mathcal{M}}^*)\mathcal{A}_{G,\mu}|_{t=1}=\int_{0}^{1}dt\dfrac{(1-t)^{\omega(\mathcal{M}_j)}}{\omega(\mathcal{M}_j)!}\dfrac{d^{\omega(\mathcal{M}_j)+1}\mathcal{A}_{G,\mu}(t)}{dt^{\omega(\mathcal{M}_j)+1}}
\end{align}
is finite. The first key result is the following obvious bound of the derivative of the propagator in the slice $i$:
\begin{equation}\label{boundsderiv}
|C^{(k)}_{i}(\{g_i\},\{g_i^{\prime}\})|\leq |X|^k K M^{(3d-2+k)i} \quad ,
\end{equation}
Because of this bounds, the derivative appearing in (\ref{amplituder}) increases the bound of the propagator by a factor $M^{(\omega(\mathcal{M}_j)+1)i_2}$. Furthermore, because the length $|X|$ scales as $M^{-j}$ with $M$, the power $|X|^{\omega(\mathcal{M}_j)+1}$ has a bound of the form $KM^{-(\omega(\mathcal{M}_j)+1)j}$. Taking into account all the contributions, the total exponential decay is 
\begin{equation}
\omega(\mathcal{M}_j)+(\omega(\mathcal{M}_j)+1)(i_2-j)<-1 \quad ,
\end{equation}
meaning that the subgraph becomes superficially convergent in the sense of the definition \ref{divgraph}. 
\begin{center}
\includegraphics[scale=0.8]{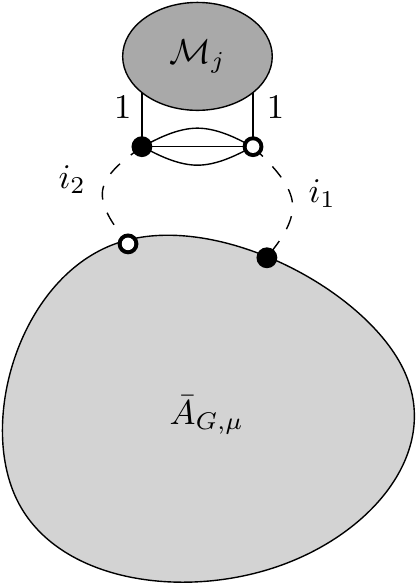} 
\captionof{figure}{Sub-divergent 2-points graph}\label{fig9}
\end{center}
\paragraph{$\bullet$ N=4} 
A typical sub-divergence of this type is depicted in Figure \ref{fig10} below, in which the graphs $\bar{\mathcal{A}}_{G,\mu}^{(i)}$ are free of sub-divergences. As in the previous case, we begin by writing the amplitude in terms of the three blocks defined in Figure \ref{fig10}. We define, with the same notations as in the case $N=2$:
\begin{align}
&\qquad\mathcal{A}_{G,\mu}(t)=\int\prod_{l=1}^{4}\prod_{k=1}^{4}dg_{lk}d\bar{g}_{lk}\mathcal{A}_{G,\mu}^{(1)}(\{\bar{g}_{1l}\},\{\bar{g}_{2l}\})\\\nonumber
&\quad\times\prod_{k=1}^{2}C_{i_k}(\{\bar{g}_{kl}\};g_{k1}(t),\{g_{kl\,l\neq 0}\})\prod_{k=3}^{4}C_{i_k}(g_{k1}(t),\{g_{kl\,l\neq 0}\};\{\bar{g}_{kl}\})\\\nonumber
&\qquad \qquad\times\mathcal{A}_{G,\mu}^{(2)}(\{\bar{g}_{3l}\},\{\bar{g}_{4l}\})\mathcal{M}_j^{(4)}(g_{11},g_{12};g_{13},g_{14}) \qquad .
\end{align}
As in the case of the $2$-point function, we introduce the $*$ operator $\tau_{\mathcal{M}^{(4)}}$ whose action is defined by the equation (\ref{amplituder}). The same argument as in the previous section can be applied to this case (there are two terms instead of one after differentiation with respect to $t$), and the conclusion is unchanged: in the domain $j>i_k\,,\forall k$, the subgraph $\mathcal{M}^{(4)}$ becomes superficially convergent, in the sense of the definition \ref{divgraph}.
\begin{center}
\includegraphics[scale=0.9]{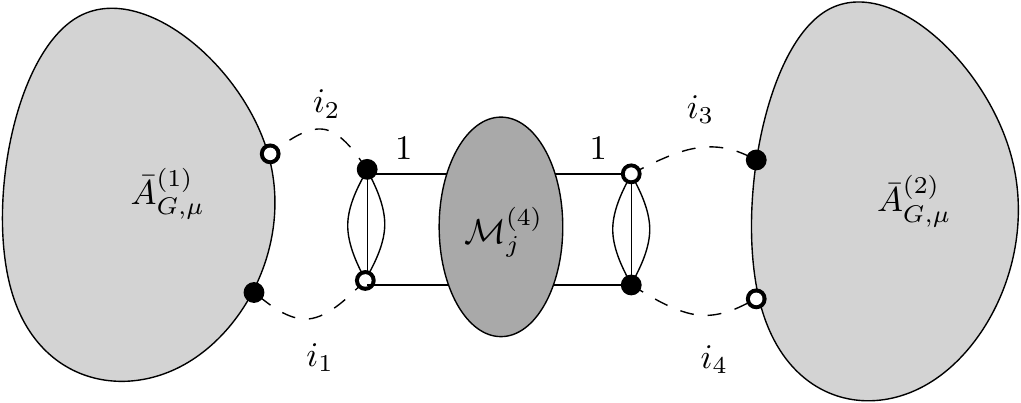} 
\captionof{figure}{Sub-divergent 4-point graph}\label{fig10}
\end{center}
\medskip
The previous analysis motivates the following definition: 
\begin{definition}\label{renampl}
The renormalized amplitude $\mathcal{A}_{G}^{R}$ associated with the graph ${G}$ is deduced from the bare amplitude $\mathcal{A}_{{G}}$ though the Zimmermann formula (or forest formula):

\begin{equation}
\mathcal{A}_{{G}}^{R}:=\sum_{\mathbb{F}\subset D({G})}\prod_{\gamma \in \mathbb{F}}(-\tau_{\gamma}^*)\mathcal{A}_{{G}},\label{renserie}
\end{equation}
where $D({G})$ is the set of Zimmerman forests. 
\end{definition}
\bigskip
The explicit form of the counter-term $\tau^*_{\mathcal{M}}\mathcal{A}_{G,\mu}$ is of interest for the next section. As in the previous paragraph, we start with the case $N=2$. From the definition (\ref{amplituder}), we have three terms, corresponding to the zeroth, first and second derivative with respect to $t$. \\

$\bullet$ Up to a change of variable, the zero derivative writes as:
\begin{equation}
\tau^{1*}_{\mathcal{M}}\mathcal{A}_{G,\mu}=\Bigg\{\int_{SU(2)} dg \mathcal{M}_j(g)\Bigg\}\mathcal{A}_{G/\mathcal{M},\mu} \quad , 
\end{equation} 
where $G/\mathcal{M},\mu$ is the graph obtained from $G$ by cutting the two lines of color $1$ linked to the melon $\mathcal{M}$, and joining to one another the two half lines of color $1$ linked to $\bar{\mathcal{A}}_{G,\mu}$. In the usual terminology, the term in square brackets corresponds to the mass renormalization. Note that this term includes presumably some sub-melonic contributions, as mentioned at the end of Section \ref{sec4.4}.\\

$\bullet$ Because in the UV $\mathcal{M}_j$ is a symmetric function of $\theta$ (that can be proved recursively in the melonic sector from the one-loop case)  $\mathcal{A}_{G,\mu}^{\prime}(0)=0$. \\

$\bullet$ Finally, the last case involve two derivative terms with respect to $t$. Note that:
\begin{equation}
\frac{d}{dt}D_{0m}^{(l)}(Gg_1)\big|_{t=0}=\mathcal{L}_{X}D_{0m}^{(l)}(g_1)=\sum_{i=1}^3X^i\mathcal{L}_{\tau_i}D_{0m}^{(l)}(g_1)
\end{equation}
where $\tau_i:=i\sigma_i/2$ and $\mathcal{L}_X$ denote the Lie derivative with respect to $X$. In the same way, the second derivative gives, for any test function $f$ of $\sum_i (X^i)^2$:
\begin{equation}
\int d^3X f(X)\sum_{i,j}X^iX^j\mathcal{L}_{\tau_i}\mathcal{L}_{\tau_j}D_{0m}^{(l)}(g_1)\propto\frac{1}{3}\int d^3X f(X)g_{SU(2)}(X,X) \Delta_{SU(2)}D_{0m}^{(l)}(g_1),
\end{equation}
where we have used to the fact that $\sum_i(\mathcal{L}_{\tau_i})^2=\Delta_{SU(2)}$. Then, taking into account that $X$ is horizontal, $g_{SU(2)}(X,X)=s^*g_{SU(2)}(x,x)$, and with: $\Delta_{SU(2)}D_{0m}^{(l)}(g_1)=4l(l+1)D_{0m}^{(l)}(g_1)$, we deduce :
\begin{align}
\tau^{2*}_{\mathcal{M}}&\mathcal{A}_{G,\mu}\propto \Bigg\{\int d^2x\mathcal{M}_j(|x|)|x|^2\Bigg\}\times \int \prod_{l}d\bar{g}_{1l}d\bar{g}_{2l}dg_{1l}dg_{2l}\bar{\mathcal{A}}_{G,\mu}(\{\bar{g}_{1l}\},\{{\bar{g}}_{2l}\}) \\\nonumber
&\qquad\qquad \times C_{i_1}(\{g_{1l}\},\{\bar{g}_{1l}\})\int_{SU(2)} dh\sum_{\{l_i,m_i\}}(-l_1(l_1+1))\prod_{i=1}^4(2l_i+1)e^{-4M^{-i_2}l_i(l_i+1)}D_{00}^{(l_i)}(g_{2i}^{-1}\bar{g}_{2i}h) \quad,
\end{align}
where the term in square brackets corresponds to the so-called wave function renormalization term, and gives the "first deviation from locality", in the sense that the combination with the Laplacian operator does not correspond exactly to an invariant trace.\\
\begin{center}
\includegraphics[scale=0.9]{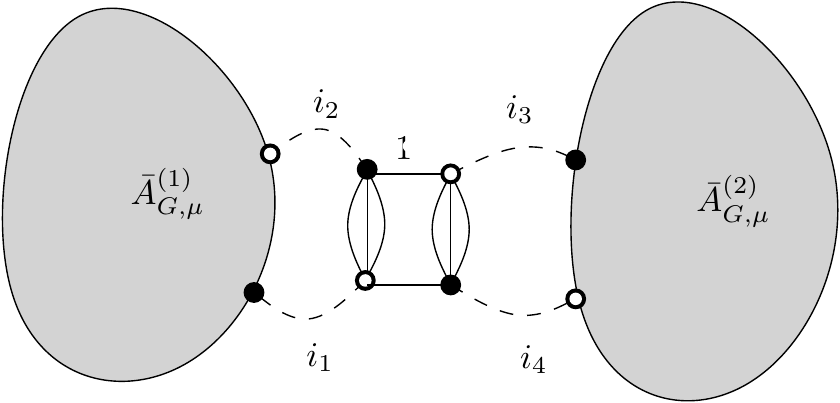}\label{fig11}
\captionof{figure}{Contraction of a 4-points subgraph}. 
\end{center}
The case of the $4$-point function follows the same pattern, but is simpler because only one term appears in the Taylor expansion: the zeroth derivative term. It follows that the divergent term can be written as 
\begin{align}
\tau^{1*}_{\mathcal{M}}\mathcal{A}_{G,\mu}=\Bigg\{\int dg dg^{\prime}\mathcal{M}^{(4)}_j(g,g')\Bigg\}\times \mathcal{A}_{G/\mathcal{M}^{(4)},\mu} \quad .
\end{align}
where $G/\mathcal{M}^{(4)}$ is the (connected) contracted graph obtained from $G$ in the procedure detailed previously, and depicted in Figure \ref{fig11} below. This counter-term gives the coupling constant renormalization.

\subsection{Bounds on the renormalized series}

The finiteness of the renormalized amplitude can be proved rigorously. In fact, we can prove that, when the graph contains some subdivergences, the renormalized amplitude $\mathcal{A}_G^R$ is finite, but increases dramatically as the factorial of the number of divergent forest. Proving this theorem requires to define precisely the dangerous and safe divergent forests:
\begin{definition}\textbf{\emph{Dangerous and safe forests}}\label{dangerousforest}
Consider a graph $G$, $\mathcal{A}_{G,\mu}$ the corresponding amplitude for the scale attribution $\mu$, and $D(G)$ the set of divergent forests. Consider then $H\subset D(G)$. We define $i_H$ and $e_H$ as:
\begin{equation*}
e_H=\sup\{i_l|l\in H/\mathbb{A}_{D(G)}(H)\}\quad i_H=\inf\{i_l|l\in L_{H}\cap \mathbb{B}_{D_G}(H)\}
\end{equation*}
where $L_H$ is the set of external lines of $H$, $\mathbb{B}_{D(G)}$ is the ancestor of $H$ in $D(G)\cup H$ and $\mathbb{A}_{D(G)}(H)$ the descendant, such as $\mathbb{A}_{D(G)}(H)=\cup_{h;H\supset h\in D(G)}h$. Moreover, $H$ is said to be \textit{compatible} with $D(G)$, in the sense that, for any $H^{\prime}\subset D(G)$, $H^{\prime}\cup H$ is still a forest. \\

The safe forest $\mathbb{F}_{\mu}$ is then the complementary in $D(G)$ of the set $D_{\mu}(\mathbb{F}_{\mu})$ of dangerous or high subgraphs in $G$ with respect to the scale assignment $\mu$, defined as : $D_{\mu}=\{H\in D(G)|e_H>i_H\}$. 
\end{definition}
This definition allows to rewrite the renormalized amplitude as:
\begin{equation}
\mathcal{A}_G^R=\sum_{f\in D(G)}A_{G,f}^R \quad ,
\end{equation}
with:
\begin{equation}
A_{G,f}^R := \sum_{\mu|f\in \mathbb{F}_{\mu}}\prod_{g\in f}(-\tau_g^*)\prod_{h\in D_{\mu}(f)}(1-\tau_h^*)\mathcal{A}_{G,\mu} \quad ,
\end{equation}
or 
\begin{equation}
A_{G,f}^R := \sum_{\mu|f\in \mathbb{F}_{\mu}}\prod_{g\in f}(-\tau_g^*)\prod_{g\in f\cup \{G\}}\prod\limits_{\substack{h\in D_{\mu}(f)\\\mathbb{B}_f(h)=g}}(1-\tau_h^*)\mathcal{A}_{G,\mu} \qquad .
\end{equation}
Beginning with the contractions over the safe forest $f$, we obtain, after appropriate  organization of the successive contractions:
\begin{equation}
\prod_{g\in f}(-\tau_g^*)\mathcal{A}_{G,\mu}=\prod_{g\in f\cup \{G\}}\nu_{\mu}(g/\mathbb{A}_f(g)) \quad ,
\end{equation}
where $\nu_{\mu}(g)$ is the discarded part of the amplitude. Note that all these terms are not exactly disconnected, because the contraction of the $2$-point graph reveal a non-local operator, which acts on another contracted component. From the multiscale analysis, it follows that:
\begin{equation}
\big|\prod_{g\in f\cup \{G\}}\nu_{\mu}(g/\mathbb{A}_f(g))\big|\leq \prod_{g\in f\cup \{G\}}\prod_{i,\rho}M^{\omega\big[(g/\mathbb{A}_f(g))_i^\rho\big]} \quad .
\end{equation}
Now, observe that the contraction over the high divergent graphs only affects the components $g/\mathbb{A}_f(g)$. It follows then, from the analysis of the previous paragraph, that the decay of a renormalized graph $g$ is at most $M^{-|e_g-i_g|}$. Hence, the renormalized amplitude is bounded by:
\begin{equation}\label{renbound}
\big|A_{G,f}^R\big|\leq \sum_{\mu|f\in \mathbb{F}_{\mu}}\prod_{g\in f\cup \{G\}}\prod_{i,\rho}M^{\omega^{\prime}\big[(g/\mathbb{A}_f(g))_i^\rho\big]} \quad ,
\end{equation}
where :
\begin{equation*}
\omega^{\prime}\big[(g/\mathbb{A}_f(g))_i^\rho\big]:=\inf\big(-1,\omega\big[(g/\mathbb{A}_f(g))_i^\rho)\big]\big) \quad ,
\end{equation*}
except when $(g/\mathbb{A}_f(g))_i^\rho=g/\mathbb{A}_f(g)$, in which case 
\begin{equation*}
\omega^{\prime}\big[(g/\mathbb{A}_f(g))_i^\rho\big]=0 \quad .
\end{equation*}
From the decay factor of equation (\ref{renbound}), we can extract the factor $M^{-\delta i_{max}(\mu)}$, where $i_{max}(\mu):=\sup(\mu)$. With the rest of the decay, we can sum over each component $g/\mathbb{A}_f(g)\,\,g\in f$, as in the proof of the Weinberg theorem. Because of the following bound:
\begin{equation}
\prod_{g\in f\cup \{G\}}K^{|V(g/\mathcal{A}_f(g))|}\leq K^{\prime\,|V(G)|} \qquad ,
\end{equation} 
the sum over internal scale assignments in each $g/\mathbb{A}_f(g)$ is bounded by $K^{\prime\,|V(g)|}$. The remaining sum over $i_{max}$ is bounded by:
\begin{equation}
\sum_{i_{max}}(i_{max})^{|f|}M^{-\delta i_{max}}\leq |f|! K^{|f|} \quad ,
\end{equation}
where $|f|$ is the cardinality of the set $f$. Because the number of sub-forests in a graph $G$ can be bounded by $2^{|D_G|}$, we finally deduce the following theorem:

\begin{theorem}\emph{\textbf{(BPH uniform)}}\label{uniform1}
Consider a Feynman graph $G$ of order $|V(G)|$. The renormalized amplitude $\mathcal{A}^R_G$ has the following bound:
\begin{equation}
|\mathcal{A}_{G}^R|\leq K^{|V(G)|}|D(G)|!, \, \, K\in \mathbb{R}^{+} \qquad ,
\end{equation}
where $|D(G)|$ is the cardinality of the divergent forest set in $G$.
\end{theorem}
As announced, the amplitude is finite but arbitrarily large, increasing dramatically with the size of the divergent forest. This is the known problem of {\it renormalons}, which implies that the convergence of the renormalized series (\ref{renserie}) is not guaranteed in a perturbative approach. To prove its convergence we would need the help of the constructive theory and of Borel summability technology, which is not the focus of this paper. To solve this technical difficulty, we use the effective series, defined in the next section, which is renormalons-free. This result, added to the asymptotic freedom, proved in Section \ref{asymptotic} at the one loop order, confirms the convergence of the effective series i.e. the perturbative series expressed in terms of the effective amplitudes and effective coupling. 
\bigskip
\subsection{The effective series}

The effective series is a more physical approach of renormalization, closely related to the Wilson approach. It is a way to solve the renormalons problem and to ensure the convergence of the perturbative series in many cases, as we will see below. The basic idea is the following. Consider a graph $G$ and its bare amplitude $\mathcal{A}_{G,\mu}$ at scale attribution $\mu$, as defined above. As we have seen before, in this graph, there are some divergent graphs, which form the set $D(G)$. But in fact, only a subset of these subgraphs is potentially dangerous, the subset noted $D_{\mu}(G)$ in the previous section. The argument is that only this subset needs to be renormalized, and the effective amplitude $A^{eff}_{G,\mu}$ is defined by
\begin{equation}
A^{eff}_{G,\mu}:=\prod_{\gamma \in D_{\mu}}(1-\tau^{*}_{\gamma})\mathcal{A}_{G,\mu} \quad , 
\end{equation}
about which we have the following theorem \cite{COR2, renormalizationreview}:
\begin{theorem}\label{effective}
\textbf{\emph{(Existence of the effective expansion)}}: Consider the formal (bare) power series defined by:
\begin{equation}
S^{\Lambda}_{N}=\sum_{G, \mu} \frac{1}{s(G)}\left(\prod_{b \in \mathcal{V}(G)} \bigg(-\lambda^{(\Lambda)}_b\bigg) \right) A_{G,\mu} \quad ,
\end{equation} 
where $\mathcal{V}(G)$ is the set of vertices in $G$ including all the interactions compatible with the just-renormalizability criterion and $\lambda^{(\Lambda)}_b$ their coupling constants. This series can be rewritten in a more convenient form in terms of the effective amplitudes:
\begin{equation}
S^{\Lambda}_{N}=\sum_{G, \mu} \frac{1}{s(G)}\left(\prod_{b \in \mathcal{V}(G)} \big(-\lambda^{(\Lambda)}_{b,e_b(G,\mu)}\big)\right) A_{G,\mu}^{eff} \quad ,
\end{equation}
where the $\lambda^{(\Lambda)}_{b,e_b(G,\mu)}$ are the effective couplings, generated by the local part of the high divergent subgraphs. They obey the following inductive relation
\begin{align}
-&\lambda^{(\Lambda)}_{b,i}=-\lambda^{(\Lambda)}_{b,i+1}+\sum\limits_{\substack{(\mathcal{H},\mu,\hat{S}) \hat{S}\neq \emptyset\\ \phi_i(\mathcal{H},\mu,\hat{S})=(b,\mu,\emptyset)}}
\frac{1}{s(\mathcal{H})}\left(\prod_{b' \in \mathcal{V}(\mathcal{H})} \big(-\lambda^{(\Lambda)}_{b',i_b'(\mathcal{H},\mu)}\big) \right)\times \left( \prod_{m \in D_{\mu}^{i+1}\setminus{\hat{S}}}(1-\tau_m^{*})\right)\prod_{M \in \hat{S}} \tau^{*}_M A_{{\mathcal{H}},\mu} \quad ,
\end{align}

\medskip
with  $e_b=\sup\{\mu_l, l \,\emph{hooked\,to}\, b\}$.\\

The notation introduced above will be defined precisely in the proof, for which we give only the main steps, referring to \cite{COR2, renormalizationreview} for details. 
\end{theorem}

\textit{\textbf{Proof}} (Sketched)\\

The basic idea is to introduce an intermediate step between the bare and the effective series as follows. We consider a slice $i$ and define:
\begin{equation}\label{sum2}
S^{\Lambda}_{N}=\sum_{G, \mu} \frac{1}{s(G)}\left(\prod_{b \in \mathcal{V}(G)} (-\lambda^{(\Lambda)}_{b,sup(i,i_b(G,\mu))} \right) A_{G,\mu}^{eff,i} \quad ,
\end{equation}
where
\begin{equation}
A^{eff,i}_{G,\mu}:=\prod_{\gamma \in D_{\mu}^{i}}(1-\tau^{*}_{\gamma})A_{G,\mu}\quad ,
\end{equation}
and
\begin{equation*}
D^{i+1}_{\mu}(G)=\{m \in D(G) | i_m > i\}  \quad i_m :=\inf\{\mu_l, l \, \emph{hooked \, to}\, b\} \quad . 
\end{equation*}
It is obvious that, if $i=\rho$, where $\Lambda=M^{\rho}$, the effective series reduces to the bare one. Assuming this is true at scale $i+1$, we can prove it at scale $i$ by induction, by multiplying the effective amplitude at scale $i+1$ by a suitable form of the identity, adding and subtracting the counter-terms in $D^{i}_{\mu}(G)\setminus D^{i+1}_{\mu}(G)=\{m \in D(G) | i_m = i+1\} $, which changes $A^{eff,i+1}_{G,\mu}$ into $A^{eff,i}_{G,\mu}$,
\begin{equation*}
A^{eff,i}_{\mu}(G):=\prod\limits_{\substack{ S\subseteq D^{i}_{\mu}\setminus D^{i+1}_{\mu}\\ S \neq \emptyset}}\prod_{M \in S} (1-\tau^{*}_{M}+\tau^{*}_{M})\prod_{\gamma \in D_{\mu}^{i}}(1-\tau^{*}_{\gamma})A_{\mu}(G) \quad .
\end{equation*}
The completely subtracted piece changes $A^{eff,i+1}_{G,\mu}$ into $A^{eff,i}_{G,\mu}$, and the second one is developed as a sum over $S$ as follows:
\begin{equation*}
S^{\Lambda}_{N} = \sum\limits_{\substack{(G,\mu,S)\\ S\subseteq D^{i}_{\mu}\setminus D^{i+1}_{\mu}}}\frac{1}{s(G)}\left(\prod_{b \in \mathcal{V}(G)} (-\lambda^{(\Lambda)}_{b,sup(i+1,i_b(G,\mu))}\right)A_{G,\mu,S}^{eff,i} \quad ,
\end{equation*}
with
\begin{equation*}
A_{\mu,S}^{eff,i}:=\prod_{M\in S}(-\tau^{*}_{M})\prod_{m\in D^{i}\setminus S}(1-\tau_m^{*})A_{G,\mu} \quad ,
\end{equation*}
and in particular $A_{\mu,\emptyset}^{eff,i}=A_{\mu}^{eff,i}$. A subtlety appears in this case because the $2$-point divergent graphs (with degree $\omega = 2$) introduce two counter-terms, one for the mass and one for the wave-function. For this reason we modify the previous definition of $S$, and introduce the new definition:
\begin{equation*}
\hat{S}=\{(M,k_M)|M\in S, k_M \in {0,2}, k_M \leq \omega(M)\} \quad .
\end{equation*} 
Secondly, we introduce the \textit{collapse} $\phi_i$ which sends the triplets $(G,\mu, \hat{S})$ to its contracted version $(\mathcal{G'}, \mu',\emptyset)$, such that the previous sum can be rewritten as a sum on $\mathcal{G'}$
\begin{equation}\label{sum1}
S^{\Lambda}_{N} = \sum_{\mathcal{G'},\mu'}\sum\limits_{\substack{\{(G,\mu,S)\}=\\ \phi_i^{-1}(\mathcal{G'},\mu',\emptyset)}}\frac{A_{G,\mu,S}^{eff,i}}{s(G)}\left(\prod_{b \in \mathcal{V}(G)} (-\lambda^{(\Lambda)}_{b,sup(i+1,i_b(G,\mu))}\right) \quad .
\end{equation}
Decomposing 
\begin{equation*}
\prod_{M \in \hat{S}}(-\tau^{*}_{M})=\prod_{b' \in \mathcal{V}(G)}\left( \prod_{M \in \hat{S}, \, M\subset \phi^{-1}_i(b')}(-\tau^{*}_M)\right)
\end{equation*}
in the sum (\ref{sum1}), we find that it gives exactly the effective sum at scale $i$ given by (\ref{sum2}), if the coupling satisfies the recursive relation of the theorem. 
\begin{flushright}
$\square$
\end{flushright}

The coupling recursion defines a discrete flow, for which the initial data are, as usual in standard quantum field theory, imposed by the 1PI functions at zero momenta. \\

The main interest of the effective series is that all these amplitudes are bounded in the form \cite{COR2}
\begin{equation}
|A^{eff}_G|\leq K^{V(G)} \quad ,
\end{equation}
a result which can be directly deduced from the theorem \ref{uniform1} proved in the previous section, in the special case where the set of inoffensive forest is empty. Remarkably,
in the previous bound, renormalons do not appear. \\

Another important fact about the effective series and effective coupling constants is their relationship with the renormalized series. In fact, if we define the renormalized coupling by $\lambda_r := \lambda_{-1}$, and if we reframe the effective series in terms of the renormalized coupling, we find exactly the renormalized series.

\section{Conclusion}

We have studied the renormalization of a TGFT model on the homogeneous space $\left(SU(2)/U(1)\right)^d$, endowed with the additional gauge invariance condition, using multi-scale methods. We have proven renormalizability to all orders in perturbation theory for the model with melonic quartic interactions in $d=4$ (and, implicitly, super-renormalizability for the model in $d=3$). 
This is the first example of a renormalization analysis for a TGFT model on a homogeneous space, rather than a group manifold, and a promising step forward towards 4d gravity models, which have similar formulations.

For the same model, we have also computed both the renormalised and effective perturbative series, and established its asymptotic freedom at one-loop order, by the analysis of the $2$-point and $4$-point correlation functions. 
This is another interesting result, because it supports the view that asymptotic freedom is generic in TGFTs, and even survives stepping out of the simple group-based setup to move to homogeneous spaces. Clearly, however, more work is needed to confirm such general expectation.

Whenever possible, we have also generalised our construction and results to arbitrary homogeneous spaces of the type $SO(D)/SO(D-1)\simeq \mathcal{S}_{D-1}$. This included a general Abelian power counting, and a corresponding classification of potentially just-renormalizable models, for various choices of $D$ and $d$. However, as we pointed out, the exact power counting of such more general non-abelian models may deviate from the Abelian one, and a more detailed case-by-case analysis needs to be carried out in order to prove (or disprove) their perturbative renormalizability.

To keep moving in the direction of $4$d quantum gravity models, as defined in the spin foam context, is our next goal. In particular, the mentioned detailed analysis of divergences and exact power counting should be performed for TGFTs on the homogeneous space $\left(SO(4)/SO(3)\right)^d$, the case $d=4$ corresponding to the so-called Barrett-Crane imposition of the simplicity constraints reducing topological BF theory to gravity (see \cite{BCrevisited} and references therein), defining interesting $4$d quantum gravity models (in absence of the Immirzi parameter). We expect the results of \cite{matteovalentin} to be a good basis for such generalisation. And work in this direction is, in fact, well in progress \cite{COL}. The Lorentzian counterpart of these models would of course be the next target. After this, one would have the proper understanding and basis to tackle the deformation of such models induced by the Immirzi parameter, which brings out of the homogeneous space setting to more general sub-manifolds of the $SO(4)$ (or $SO(3,1)$) group manifold (see \cite{GFT-Immirzi}). 

It is clear that the path towards a renormalizable quantum field theory for the \lq atoms of space\rq $\,\,$is still long, but it should be also clear that we are making steady and important progress along it.
\newpage
\appendix
\section{Geometrical interpretation}

\label{Sectiongem}In this Section we list some information on geometrical interpretation of the formalism described in Section \ref{3.1}.
The closure constraint admits a `geometrical interpretation which can be easily understood with the mathematical tool of the non-commutative (group) Fourier transform. It has originated in the quantum group literature \cite{majid}, and introduced in the GFT context in \cite{GFT-noncomm}, after being first used in the spin foam context in \cite{PR-noncomm}, and developed, in particular for the case of $SU(2)$, from the more mathematical perspective in \cite{majid, majidlaurent, carlosdanielematti}. This non-commutative Fourier Transform is a functional mapping from a (usually but not necessarily) compact group $\mathbf{G}$ into its Lie algebra $\mathfrak{g}$, sending any square-integrable function on $\mathbf{G}$ to a non-commutative function on $\mathfrak{g}$.  For the group $SU(2)$, the mapping is between $SU(2)$ and the $\mathbb{R}^3$ space, dual to its Lie algebra $\mathfrak{su}(2)$. Let $\phi$ be an integrable function on $SU(2)$, its Fourier transform is defined as:
\begin{equation}
\hat{\phi}:=\int_{SU(2)}dg\phi(g)e^{\Tr(|g|x)}\qquad \quad x\in \mathfrak{su}(2),
\end{equation}
where $\Tr{}$ is the trace in the fundamental representation and $|g|=\sign{\big[\Tr(g)\big]}g$, ensuring that the basis functions $e_g:=e^{\Tr(|g|x)}$ are trivially on $SO(3)$, because : $e_g=e_{-g}$. Note that this condition also concerns the function $\phi$, assumed to be a symmetric function: $\phi(g)=\phi(-g)$, and therefore can be understood as a field on $SO(3)$ as well. \\

\noindent
The inverse Fourier Transform is formally given by:
\begin{equation}
\phi(g)=\frac{1}{\pi}\int_{\mathbb{R}^3}d^3x\big[\hat{\phi}\star e_{g^{-1}}\big](x),
\end{equation}
where the $\star$-product is dual to the convolution product on $SU(2)$:
\begin{equation*}
\hat{\phi}\star\hat{\psi}(x)=\int_{SU(2)} dg\, e_g(x)\big(\phi\circ \psi\big)(g)\,,
\end{equation*}
and is compatible with the group structure, in the sense that:
\begin{equation}
e_{g_1}\star e_{g_2}(x) = e_{g_1g_2}(x), \quad \forall g_1,g_2 \in SU(2)\,.
\end{equation}
The Fourier Transform can easily be extended to any function on $[SU(2)]^d$ as:
\begin{equation}
\hat{\phi}(x_1,...,x_d):=\int_{[SU(2)]^d}[dg]^d\, \phi(g_1,...,g_d)\prod_{i=1}^d e_{g_i}(x_i).
\end{equation}

\noindent
We now have a look at the constraints \eqref{inv1} and \eqref{inv2} successively. With the effective field $\Psi$ defined in Section \ref{3.1}, one find, with \eqref{inv1}:
\begin{equation}\label{inv3}
\int_{[SU(2)^d]}[dg]^d\, \hat{T}_l[\Psi](g_1,...,g_d)\prod_{i=1}^d e_{g_i}(x_i)=\int_{[SU(2)^d]}[dg]^d \Psi(g_1,...,g_d)\prod_{i=1}^d e_{g_i}(x_i)\,\vec{\star}\prod_{i=1}^de_l(x_i).
\end{equation}
where in the two last expressions, the $\vec{\star}$-product distributes the $\star$-product between all the $e_g$ basis functions in accordance with their respective indices. Hence, if $\Psi=\hat{T}_l[\Psi]$, integrating over $dl$ gives:
\begin{equation}
\hat{\Psi}(x_1,...,x_d)= \hat{\Psi}(x_1,...,x_d)\,\vec{\star}\,\delta_0\Bigg(\sum_{i=1}^d x_i\Bigg).
\end{equation} 
with the $\delta_0$ distributional defined as
\begin{equation}
\delta_0\Bigg(\sum_{i=1}^d x_i\Bigg):=\int_{SU(2)} dl \,\prod_{i=1}^d e_{l}(x_i),
\end{equation}
verifies, for any function of one variable $\hat{\phi}$,
\begin{equation}
\int d^3y (\delta_0\star \hat{\phi})(y)=\int d^3y(\hat{\phi}\star \delta_0)(y)=\hat{\phi}(0).
\end{equation}
Hence, the representation of the right projector $\hat{C}=\int dl\hat{T}_l$ is a simple non-commutative multiplication acting on any fields $\psi$ as
\begin{equation}
\hat{C}[\hat{\Psi}](x_1,...,x_d):=\hat{\Psi}(x_1,...,x_d)\,\vec{\star}\,\delta_0\Bigg(\sum_{i=1}^d x_i\Bigg) \quad .
\end{equation}
We now move on to the constraint \eqref{inv2}. Consider the operators $\hat{t}^{(i)}_{h_i}$, acting on the $i$-th variable of $\phi_k$. We wish to compute the Fourier Transform of $\prod_i\int_{h_i\in U(1)_k}dh_i\hat{t}^{(i)}_{h_i}[\phi_k]$, and we find that it is proportional to:
\begin{align}
\int_{[SU(2)^d]}&[dg]^d \phi_k(g_1,...,g_d)\prod_{i=1}^d e_{g_i}(x_i)\,\vec{\star}\,\prod_{i=1}^d\delta_0\Big(\frac{1}{2}\Tr(kx_i)k\Big)=\hat{\phi}_k(x_1,...,x_d)\,\vec{\star}\,\prod_{i=1}^d\delta_0\Big(\frac{1}{2}\Tr(kx_i)k\Big)\label{inv4},
\end{align}
%with:
%\begin{equation}
%\delta_0\Bigg(\frac{1}{2}\Tr(kx)k\Bigg):=\int_{U_k(1)} dh e_{h}(x)
%\end{equation}
which follows from the definition of the basis functions $e_g$, and of:
\begin{align*}
\Tr(hx)=\Tr\Big[h\frac{1}{2}\Tr(kx)k\Big]\quad h\in U(1)_k,
\end{align*}
implying:
\begin{equation}
\int_{U(1)_k} dh\exp[\Tr(hx)]=\int_{SU(2)}dh\exp\Big[\frac{1}{2}\Tr (h\Tr(kx)k)\Big].
\end{equation}
As a result, the projector $\hat{S}_k:=\prod_i\int_{h_i\in U(1)_k}dh_i\hat{t}^{(i)}_{h_i}$ acts on $\phi_k$ as:
\begin{equation}
\hat{S}_k[\phi_k]:=\hat{\phi}_k(x_1,...,x_d)\,\vec{\star}\,\prod_{i=1}^d\delta_0\Bigg(\frac{1}{2}\Tr(kx_i)k\Bigg).
\end{equation}
Note that the non-commutativity \eqref{noncom} implies:
\begin{equation}
\hat{S}_k\circ\hat{P}=\int dl\hat{T}_l\circ \hat{S}_{lkl^{-1}}.
\end{equation}
The first result \eqref{inv3} explains why the constraint \eqref{inv1} is named ``the closure constraint". The second result \eqref{inv4} means that the Fourier variables are forced to be orthogonal to the Lie algebra index $k$. In the case $d=3$, the field can be interpreted  as describing a triangle in $\mathbb{R}^3$, where group variables are associated to the boundary lines, and its Lie algebra variables being their edge vectors (see Figure \ref{triangle}). The closure constraint forces in fact these edge vectors to sum to zero, thus the corresponding edges to "close". The constraint \eqref{inv2} implies that this triangle is orthogonal to the unit $3$-vector $\vec{k}$, associated with the index $k\in \mathfrak{su}(2)$ of the field. 

\begin{center}
\includegraphics[scale=0.8]{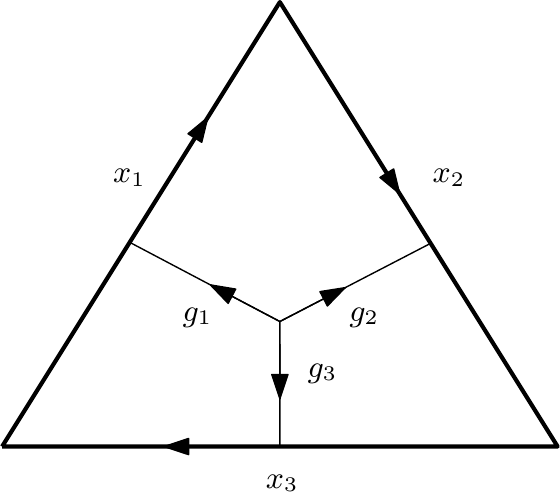} 
\captionof{figure}{Geometrical interpretation of the fields and their variables}\label{triangle}
\end{center}

%\end{multicols}


\newpage
\begin{thebibliography}{99}
\bibitem{GFT}  
D.~Oriti, ``The microscopic dynamics of quantum space as a group field theory,''
in  {\sl Foundations of space and time}, 
G. Ellis, et al. (eds.) (Cambridge University Press, Cambridge UK, 2012),
 arXiv:1110.5606 [hep-th];  
D.~Oriti, 
``The Group field theory approach to quantum gravity,''
  in {\sl Approaches to quantum gravity},  D. Oriti (ed.) 
(Cambridge University Press, Cambridge UK, 2009),
  [gr-qc/0607032]; 
D.~Oriti,
  ``The Group field theory approach to quantum gravity: Some recent results,''
in {\sl The Planck Scale},  J. Kowalski-Glikman, et al. (eds) 
AIP: conference proceedings (2009),  arXiv:0912.2441 [hep-th];

 A.~Baratin and D.~Oriti,
  ``Ten questions on Group Field Theory (and their tentative answers),''
  J.\ Phys.\ Conf.\ Ser.\  {\bf 360}, 012002 (2012)
  [arXiv:1112.3270 [gr-qc]];

T.~Krajewski,
 ``Group field theories,''
  PoS QGQGS {\bf 2011}, 005 (2011)
  [arXiv:1210.6257 [gr-qc]].

\bibitem{LQG} T. Thiemann, {\it Modern canonical quantum General Relativity} 
(Cambridge University Press, Cambridge UK, 2007); 
A. Ashtekar and J. Lewandowski,   
``Background independent quantum gravity: A status report,''
 Class. Quant. Grav {\bf 21}, R53-R152 (2004);  
C. Rovelli, {\it Quantum Gravity} (Cambridge University Press, 2006); A.~Perez,
  ``The Spin Foam Approach to Quantum Gravity,''
  Living Rev.\ Rel.\  {\bf 16}, 3 (2013)
  [arXiv:1205.2019 [gr-qc]];
 C.~Rovelli,
  ``Zakopane lectures on loop gravity,''
  PoS QGQGS {\bf 2011}, 003 (2011)
  [arXiv:1102.3660 [gr-qc]

\bibitem{tensor}   V.~Rivasseau,
  ``Quantum Gravity and Renormalization: The Tensor Track,''
  AIP Conf.\ Proc.\  {\bf 1444}, 18 (2011)
  [arXiv:1112.5104 [hep-th]];
  %%CITATION = ARXIV:1112.5104;%%
  %38 citations counted in INSPIRE as of 26 May 2014
 %   V.~Rivasseau,
  ``The Tensor Track: an Update,'' Symmetries and Groups in Contemporary Physics, 63-74, World Scientific 2013.
  arXiv:1209.5284 [hep-th];
  %%CITATION = ARXIV:1209.5284;%%
  %25 citations counted in INSPIRE as of 26 May 2014
  %  V.~Rivasseau,
  ``The Tensor Track, III,'', Fortschr. Phys. {\bf 62}, No. 1, 1-27 (2013),
  arXiv:1311.1461 [hep-th];  V.~Rivasseau,
  ``The Tensor Theory Space,''
  Fortsch.\ Phys.\  {\bf 62}, 835 (2014)
  [arXiv:1407.0284 [hep-th]]
\bibitem{GFT-LQG} D.~Oriti,
  ``Group field theory as the 2nd quantization of Loop Quantum Gravity,''
  arXiv:1310.7786 [gr-qc]; 
 D.~Oriti,
  ``Group Field Theory and Loop Quantum Gravity,''
  arXiv:1408.7112 [gr-qc]; 
 D.~Oriti, J.~P.~Ryan and J.~Th\"urigen,
  ``Group field theories for all loop quantum gravity,''
  arXiv:1409.3150 [gr-qc].
\bibitem{mike-carlo}  M.~P.~Reisenberger and C.~Rovelli,
 ``Space-time as a Feynman diagram: The Connection formulation,''
  Class.\ Quant.\ Grav.\  {\bf 18}, 121 (2001)
  [gr-qc/0002095]
\bibitem{GFT-noncomm} A.~Baratin and D.~Oriti,
  ``Group field theory with non-commutative metric variables,''
  Phys.\ Rev.\ Lett.\  {\bf 105}, 221302 (2010)
  [arXiv:1002.4723 [hep-th]]; A. Baratin, B. Dittrich, D. Oriti, J. Tambornino, Class.Quant.Grav. 28 (2011) 175011, arXiv:1004.3450 [hep-th];

\bibitem{GFT-Immirzi} A.~Baratin and D.~Oriti,
 ``Group field theory and simplicial gravity path integrals: A model for Holst-Plebanski gravity,''
  Phys.\ Rev.\ D {\bf 85}, 044003 (2012)
  [arXiv:1111.5842 [hep-th]].
\bibitem{LQG-quantgeom} J. C. Baez, J. W. Barrett, Adv. Theor. Math. Phys. 3, 815 (1999), gr-qc/9903060; J. Barrett, R. Dowdall, W. Fairbairn, H. Gomes, F. Hellman, J. Math. Phys. \textbf{50}, 112504 (2009), [arXiv:0902.1170 [gr-qc]]; J. Barrett, R. Dowdall, W. Fairbairn, F. Hellman, R. Pereira, [arXiv:0907.2440 [gr-qc]]; F. Conrady, L. Freidel, Phys. Rev. D \textbf{78}, 104023 (2008), [arXiv:0809.2280]; Y. Ding, C. Rovelli, Class.Quant.Grav. 27 (2010) 165003, arXiv:0911.0543 [gr-qc]; V. Bonzom, E. Livine, [arXiv:0812.3456]; V. Bonzom, Class. Quant. Grav. \textbf{26}, 155020 (2009) [arXiv:0903.0267]; V. Bonzom, [arXiv:0905.1501]
\bibitem{emergence} D.~Oriti,
  ``Disappearance and emergence of space and time in quantum gravity,''
  Stud.\ Hist.\ Philos.\ Mod.\ Phys.\  {\bf 46}, 186 (2014)
  [arXiv:1302.2849 [physics.hist-ph]]
\bibitem{jimmyrazvan}   R.~Gurau and J.~P.~Ryan,
  ``Colored Tensor Models - a review,''
  SIGMA {\bf 8}, 020 (2012)
  [arXiv:1109.4812 [hep-th]]; V.~Bonzom, R.~Gurau, A.~Riello and V.~Rivasseau,
  ``Critical behavior of colored tensor models in the large N limit,''
  Nucl.\ Phys.\ B {\bf 853}, 174 (2011)
  [arXiv:1105.3122 [hep-th]];  R.~Gurau,
  ``Universality for Random Tensors,''
  arXiv:1111.0519 [math.PR]
\bibitem{tensorscaling} R.~Gurau,
  ``The 1/N expansion of colored tensor models,''
  Annales Henri Poincare {\bf 12}, 829 (2011)
  [arXiv:1011.2726 [gr-qc]];  
  ``The complete 1/N expansion of colored tensor models in arbitrary dimension,''
  Annales Henri Poincare {\bf 13}, 399 (2012)
  [arXiv:1102.5759 [gr-qc]];
  %%CITATION = ARXIV:1102.5759;%%
  %53 citations counted in INSPIRE as of 08 Apr 2013
  %%CITATION = ARXIV:1011.2726;%%
  %54 citations counted in INSPIRE as of 08 Apr 2013
%\bibitem{expansion2} 
  R.~Gurau and V.~Rivasseau,
  ``The 1/N expansion of colored tensor models in arbitrary dimension,''
  Europhys.\ Lett.\  {\bf 95}, 50004 (2011)
  [arXiv:1101.4182 [gr-qc]];
R.~Gurau,
  ``The Double Scaling Limit in Arbitrary Dimensions: A Toy Model,''
  Phys.\ Rev.\ D {\bf 84}, 124051 (2011)
  [arXiv:1110.2460 [hep-th]];
W.~Kami\`nski, D.~Oriti and J.~P.~Ryan,
  ``Towards a double-scaling limit for tensor models: probing sub-dominant orders,''
  New J.\ Phys.\  {\bf 16}, 063048 (2014)
  [arXiv:1304.6934 [hep-th]];  
 S.~Dartois, R.~Gurau and V.~Rivasseau,
  ``Double Scaling in Tensor Models with a Quartic Interaction,''
  JHEP {\bf 1309}, 088 (2013)
  [arXiv:1307.5281 [hep-th]]; 
 V.~Bonzom, R.~Gurau, J.~P.~Ryan and A.~Tanasa,
  ``The double scaling limit of random tensor models,''
  JHEP {\bf 1409}, 051 (2014)
  [arXiv:1404.7517 [hep-th]];  A.~Tanasa,
  ``Multi-orientable Group Field Theory,''
  J.\ Phys.\ A {\bf 45}, 165401 (2012)
  [arXiv:1109.0694 [math.CO]];
  %%CITATION = ARXIV:1109.0694;%%
  %17 citations counted in INSPIRE as of 14 Nov 2014
 %\cite{Dartois:2013he}
%\bibitem{Dartois:2013he} 
  S.~Dartois, V.~Rivasseau and A.~Tanasa,
  ``The $1/N$ expansion of multi-orientable random tensor models,''
  Annales Henri Poincare {\bf 15}, 965 (2014)
  [arXiv:1301.1535 [hep-th]];
  %%CITATION = ARXIV:1301.1535;%%
  %13 citations counted in INSPIRE as of 14 Nov 2014 
  %\cite{Raasakka:2013eda}
%\bibitem{Raasakka:2013eda} 
  M.~Raasakka and A.~Tanasa,
  ``Next-to-leading order in the large $N$ expansion of the multi-orientable random tensor model,''
  arXiv:1310.3132 [hep-th];


\bibitem{laurent-razvan-daniele} L. Freidel, R. Gurau and D. Oriti, Phys. Rev. D \textbf{80}, 044007 (2009), [arXiv:0905.3772]
\bibitem{constructiveTGFT}   T.~Delepouve and V.~Rivasseau,
 ``Constructive Tensor Field Theory: The $T^4_3$ Model,''
  arXiv:1412.5091 [math-ph];  T.~Delepouve, R.~Gurau and V.~Rivasseau,
 ``Borel summability and the non perturbative $1/N$ expansion of arbitrary quartic tensor models,''
  arXiv:1403.0170, to appear in Annales Henri Poincar\'e, Probablit\'es;
  %%CITATION = ARXIV:1403.0170;%%
  %5 citations counted in INSPIRE as of 16 Sep 2014
  %\cite{Nguyen:2014mga}
%\bibitem{Nguyen:2014mga} 
  V.~A.~Nguyen, S.~Dartois and B.~Eynard,
  ``An analysis of the intermediate field theory of $T^4$ tensor model,''
  arXiv:1409.5751 [math-ph];     D.~O.~Samary,
  ``Closed equations of the two-point functions for tensorial group field theory,''  Class. Quant. Grav. {\bf 31}, 185005 (2014)
  [arXiv:1401.2096 [hep-th]];   D.~O.~Samary, C.~I.~P\'erez-S\'anchez, F.~Vignes-Tourneret and R.~Wulkenhaar,
  ``Correlation functions of just renormalizable tensorial group field theory: The melonic approximation,''
  arXiv:1411.7213 [hep-th]; V. Lahoche, D. Oriti, V. Rivassehau, \lq\lq Renormalization of an Abelian Tensor Group Field Theory: Solution at Leading Order\rq\rq, JHEP 1504 (2015) 095, arXiv:1501.02086 [hep-th]
\bibitem{TGFT-FRG} D.~Benedetti, J.~B.~Geloun and D.~Oriti,
  ``Functional Renormalisation Group Approach for Tensorial Group Field Theory: a Rank-3 Model,''
  arXiv:1411.3180 [hep-th]; T.~Krajewski and R.~Toriumi,
 ``Polchinski's equation for group field theory,''
  Fortsch.\ Phys.\  {\bf 62}, 855 (2014); J. Ben Geloun, R. Martini, D. Oriti, \lq\lq Functional renormalization group analysis of a tensorial group field theory on a non-compact group manifold\rq\rq, to appear;
  
  \bibitem{vincent-joseph}   J.~Ben Geloun and V.~Rivasseau,
  ``A Renormalizable 4-Dimensional Tensor Field Theory,''
  Commun.\ Math.\ Phys.\  {\bf 318}, 69 (2013)
  [arXiv:1111.4997 [hep-th]];
 % J.~Ben Geloun and V.~Rivasseau,
  ``Addendum to 'A Renormalizable 4-Dimensional Tensor Field Theory',''
  Commun.\ Math.\ Phys.\  {\bf 322}, 957 (2013)
  [arXiv:1209.4606 [hep-th]]; J.~Ben Geloun,
  ``Renormalizable Models in Rank $d\geq 2$ Tensorial Group Field Theory,'' 
Commun. Math. Phys. {\bf 332}, 117--188 (2014)
  [arXiv:1306.1201 [hep-th]]
\bibitem{joseph}   J.~Ben Geloun and D.~O.~Samary,
  ``3D Tensor Field Theory: Renormalization and One-loop $\beta$-functions,''
  Annales Henri Poincare {\bf 14}, 1599 (2013)
  [arXiv:1201.0176 [hep-th]]
\bibitem{COR}   S.~Carrozza, D.~Oriti and V.~Rivasseau,
  ``Renormalization of Tensorial Group Field Theories: Abelian U(1) Models in Four Dimensions,''
  arXiv:1207.6734 [hep-th].
  %%CITATION = ARXIV:1207.6734;%%
  %38 citations counted in INSPIRE as of 01 Mar 2014
 \bibitem{fabien} D.~O.~Samary and F.~Vignes-Tourneret,
 ``Just Renormalizable TGFT's on $U(1)^{d}$ with Gauge Invariance,''
  arXiv:1211.2618 [hep-th]

\bibitem{COR2} S.~Carrozza, D.~Oriti and V.~Rivasseau,
  ``Renormalization of an SU(2) Tensorial Group Field Theory in Three Dimensions,''
  arXiv:1303.6772 [hep-th].
  %%CITATION = ARXIV:1303.6772;%%
  %18 citations counted in INSPIRE as of 01 Mar 2014 
S. Carrozza, Tensorial methods and renormalization in Group Field Theories, Springer
Theses, 2014, XV, 226 p, arXiv:1310.3736
\bibitem{TGFT-beta}   J.~Ben Geloun,
  ``Two and four-loop $\beta$-functions of rank 4 renormalizable tensor field theories,''
  Class.\ Quant.\ Grav.\  {\bf 29}, 235011 (2012)
  [arXiv:1205.5513 [hep-th]];
  %%CITATION = ARXIV:1205.5513;%%
  %27 citations counted in INSPIRE as of 01 Mar 2014
%  \bibitem{BG2} 
 J.~Ben Geloun,
  ``Asymptotic Freedom of Rank 4 Tensor Group Field Theory,''
  arXiv:1210.5490 [hep-th];  S.~Carrozza,
  ``Discrete Renormalization Group for SU(2) Tensorial Group Field Theory,''
  arXiv:1407.4615 [hep-th]
\bibitem{josephBoulatov}  J.~Ben Geloun and V.~Bonzom,
  ``Radiative corrections in the Boulatov-Ooguri tensor model: The 2-point function,''
  Int.\ J.\ Theor.\ Phys.\  {\bf 50}, 2819 (2011),
  arXiv:1101.4294; J. Ben Geloun, ``On the finite amplitudes for open graphs in Abelian dynamical colored Boulatov-Ooguri models,'' J.\ Phys.\ A {\bf 46}, 402002 (2013),  arXiv:1307.8299 [hep-th]

\bibitem{4dRadiative} J.~Ben Geloun, R.~Gurau and V.~Rivasseau, ``EPRL/FK Group Field Theory'', Europhys.\ Lett.\  {\bf 92}, 60008 (2010), arXiv:1008.0354 [hep-th]
\bibitem{BCrevisited} A. Baratin, D. Oriti, New J. Phys. 13 (2011) 125011, arXiv:1108.1178 [gr-qc]
\bibitem{DP-F-K-R} R. De Pietri, L. Freidel, K. Krasnov, C. Rovelli, Nucl. Phys. B \textbf{574}, 785 (2000), [arXiv: hep-th/9907154]
\bibitem{FK} L. Freidel, K. Krasnov, J.Math.Phys. 41 (2000) 1681-1690, [arXiv: hep-th/9903192]
\bibitem{P-R} A. Perez, C. Rovelli, Nucl. Phys. B \textbf{599}, 255 (2001), [arXiv: gr-qc/0006107]; 
\bibitem{gluing}D. Oriti, R. Williams, Phys. Rev. D \textbf{63}, 024022 (2001), [arXiv: gr-qc/0010031]
\bibitem{alexandrov} E. R Livine, Class. Quant. Grav. 19 (2002) 5525-5542, gr-qc/0207084; S. Alexandrov, Phys.Rev. D82 (2010) 024024, arXiv:1004.2260 [gr-qc]; S. Alexandrov, Phys. Rev. D 78, 044033 (2008), [arXiv: 0802.3389 [gr-qc]]
\bibitem{majid} E. Batista, S. Majid, J.Math.Phys. 44 (2003) 107-137, arXiv: hep-th/0205128
%\bibitem{schroers}
\bibitem{PR-noncomm} L. Freidel, E. Livine, Class. Quant. Grav. 23, 2021(2006),  [arXiv: hep-th/0502106]

\bibitem{majidlaurent} L. Freidel, S. Majid, Class. Quant. Grav. \textbf{25}, 045006 (2008), [arXiv:hep-th/0601004]; E. Joung, J. Mourad, K. Noui, J. Math. Phys. \textbf{50}, 052503 (2009), [arXiv:0806.4121 [hep-th]]; E. Livine, Class.Quant.Grav. 26 (2009) 195014, [arXiv:0811.1462 [gr-qc]]

\bibitem{carlosdanielematti} C. Guedes, D. Oriti, M. Raasakka, J.Math.Phys. 54 (2013) 083508, arXiv:1301.7750 [math-ph]
\bibitem{camporesi} R. Camporesi, Phys.Rept. 196 (1990) 1-134
\bibitem{matteovalentin} V. Bonzom, M. Smerlak, Lett.Math.Phys. 93 (2010) 295-305, arXiv:1004.5196 [gr-qc]; V. Bonzom, M. Smerlak, arXiv:1008.1476 [math-ph], V. Bonzom, M. Smerlak, arXiv:1103.3961 [gr-qc]
\bibitem{renormalizationreview} R. Gurau, V. Rivasseau, A. Sfondrini,  arXiv:1401.5003 [hep-th]

\bibitem{LCO}
S.~Carrozza, V.~Lahoche and D.~Oriti, ``A just-renormalizable Group Field Theory with $Spin(4)$ Barrett-Crane constraints", in preparation;

 \bibitem{Rivasseau:1991ub} 
  V.~Rivasseau,
``From perturbative to constructive renormalization,''
  Princeton, USA: Univ. Pr. (1991) 336 p. (Princeton series in physics)
\end{thebibliography}
\end{document}